\documentclass[iop,apj,tighten,twocolappendix,numberedappendix]{emulateapj}
\usepackage{apjfonts}
\usepackage{graphicx}
\usepackage{amsmath,amssymb}
\usepackage{color}
\usepackage{bm}
\usepackage[breaklinks,colorlinks,citecolor=blue,linkcolor=red]{hyperref} 
\usepackage[all]{hypcap} 
 
\def\LaTeX{L\kern-.36em\raise.3ex\hbox{a}\kern-.15em 
    T\kern-.1667em\lower.7ex\hbox{E}\kern-.125emX}

\newcommand{\Gpc}{\mathrm{Gpc}}
\newcommand{\Msun}{\mathrm{M}_{\odot}}
\newcommand{\Hz}{\mathrm{Hz}}
\newcommand{\LSO}{{\mathrm{LSO}}}
\newcommand{\CC}{c_0}
\newcommand{\yr}{\mathrm{yr}}

\begin{document} 

\title[Parameter Estimation Accuracy of Eccentric GW sources]{Accuracy of Estimating Highly Eccentric
 Binary Black Hole Parameters With Gravitational-Wave Detections}

\author{L\'aszl\'o Gond\'an\altaffilmark{1}, Bence Kocsis\altaffilmark{1}, P\'eter Raffai\altaffilmark{1},
 and Zsolt Frei\altaffilmark{1}}.
  \affil{$^1$E\"otv\"os University, Institute of Physics, P\'azm\'any P. s. 1/A, Budapest, Hungary 1117} 

\label{firstpage}

\begin{abstract} 
 Mergers of stellar-mass black holes on highly eccentric orbits are among the targets for ground-based 
 gravitational-wave detectors, including LIGO, VIRGO, and KAGRA. These sources may commonly form through
 gravitational-wave emission in high velocity dispersion systems or through the secular Kozai-Lidov 
 mechanism in triple systems. Gravitational waves carry information about the binaries' orbital parameters
 and source location. Using the Fisher matrix technique, we determine the measurement accuracy with which 
 the LIGO-VIRGO-KAGRA network could measure the source parameters of eccentric binaries using a matched 
 filtering search of the repeated burst and eccentric inspiral phases of the waveform. We account for 
 general relativistic precession and the evolution of the orbital eccentricity and frequency during the 
 inspiral. We find that the signal-to-noise ratio and the parameter measurement accuracy may be significantly 
 higher for eccentric sources than for circular sources. This increase is sensitive to the initial pericenter 
 distance, the initial eccentricity, and component masses. For instance, compared to a $30 \, \Msun - 30 \, 
 \Msun$ non-spinning circular binary, the chirp mass and sky localization accuracy can improve for an 
 initially highly eccentric binary by a factor of $\sim 129$ ($38$) and $\sim 2$ ($11$) assuming an 
 initial pericenter distance of $20 \, M_\mathrm{tot}$ ($10 \, M_\mathrm{tot}$).
 \end{abstract}


\maketitle

\section{Introduction} 
\label{sec:Intro} 

The Advanced Laser 
 Interferometer Gravitational Wave Observatory\footnote{\url{http://www.ligo.caltech.edu/}} (aLIGO) 
 detectors \citep{Aasietal2015} and Advanced Virgo\footnote{\url{http://www.ego-gw.it/}} (AdV) \citep{Acerneseetal2015} 
 have made the first six detections of GWs from approximately circular inspiraling binaries
 \citep{Abbottetal2016c,Abbottetal2016b,Abbottetal2017,LIGOColl2017,Abbottetal2017_2,Abbottetal2017_3}, 
 and opened a new window through which to observe the universe. These advanced gravitational-wave (GW)
 detectors together with upcoming instruments KAGRA\footnote{\url{http://gwcenter.icrr.u-tokyo.ac.jp/en/}} 
 \citep{Somiya2012} and LIGO-India\footnote{\url{http://www.gw-indigo.org/}} \citep{Iyeretal2011,Abbottetal2016a} 
 are expected to continue to detect GW sources in the upcoming years \citep{Abbottetal2016a}. The orbital
 eccentricity was neglected in the analysis of the detected GW sources, but a preliminary upper limit was 
 claimed to be $e \lesssim 0.1$ at 10 Hz \citep{Abbottetal2016e,Abbottetal2016d,Abbottetal2017,Abbottetal2017_4}.
 In this paper we estimate the future potential of the aLIGO-AdV-KAGRA network of advanced GW detectors
 to measure the orbital eccentricity and other physical parameters of initially highly eccentric sources.
 
 Initially highly eccentric black hole (BH) binaries are inspiraling systems, which have orbital eccentricities
 beyond $e_0 \geq 0.9$ when their peak GW frequency \citep{Wen2003} enters the sensitive frequency band of 
 advanced Earth-based GW detectors. The orbital eccentricity decreases in the inspiral phase from this value 
 until the last stable orbit (LSO) \citep{Peters1964}. Such systems can form in multiple ways including 
 single-single encounters due to GW emission \citep{Kocsisetal2006,OLearyetal2009,Gondanetal2017} in 
 dense, high-velocity-dispersion environments; dynamical multibody interactions
 \citep{Gultekinetal2006,OLearyetal2006,Kushniretal2013,AmaroSeoaneChen2016,AntoniniRasio2016,Rodriguezetal2017}; 
 the secular Kozai-Lidov mechanism
 \citep{Wen2003,Thompson2011,Aarseth2012,AntoniniPerets2012,Antogninietal2014,Antoninietal2014,Antoninietal2016,Breiviketal2016,Rodriguezetal2016a,Rodriguezetal2016b,VanLandinghametal2016,Hoangetal2017,PetrovichAntonini2017,SilsbeeTremaine2017,RandallXianyu2018} 
 in hierarchical triples, or the binary-single interaction \citep{Samsingetal2014,SamsingRamirezRuiz2017,Samsing2017,Samsingetal2017}.
 Eccentric BH binaries offer promising new detection candidates.
 
 Previous parameter estimation studies of stellar-mass compact binaries have mostly focused
 on circular binaries 
 (see \citealt{Finn1992,FinnChernoff1993,Markovic1993,CutlerFlanagan1994,JaranowskiKrolak1994,
 Kokkotasetal1994,Krolaketal1995,PoissonWill1995} for the first papers and 
 \citealt{Chatziioannouetal2014,Favata2014,Mandeletal2014,OShaughnessyetal2014,Rodriguezetal2014,Canizaresetal2015,Veitchetal2015,Berryetal2015,Milleretal2015,Farretal2016,Mooreetal2016,Langeetal2017,Vitaleetal2017} for recent developments) 
 due to their predicted high detection rates. The current detections constrain the merger rate 
 density of BH-BH mergers in the Universe to $12-213 \, \Gpc^{-3} \yr^{-1}$ \citep{Abbottetal2017}, 
 which corresponds to a detection rate between $400-7,000 \, \yr^{-1}$ for a typical $2 \, \Gpc$ 
 detection range for aLIGOs design sensitivity. Furthermore, see \citet{Abadieetal2010} for a partial
 list of historical compact binary coalescence rate predictions, and
 \citet{Dominiketal2013,Kinugawaetal2014,Abbottetal2016e,Abbottetal2016h,Abbottetal2016f,Abbottetal2016g,Abbottetal2016j,Belczynskietal2016,Rodriguezetal2016a,Bartosetal2017,McKernanetal2017,Hoangetal2017,Stoneetal2017} and references therein 
 for recent rate estimates.
 
 However, several theoretical studies have shown that the detection rates of highly eccentric BH 
 binaries may be non-negligible. For sources formed by GW-emission in galactic nuclei (GNs), the expected
 aLIGO detection rate at design sensitivity may be higher than $\approx 100 \rm \, yr^{-1}$ if the BH 
 mass function extends to masses above $25 \, \Msun$ \citep{OLearyetal2009,KocsisLevin2012}. Recently, 
 such heavy BHs have been observed in several LIGO/VIRGO detections 
 \citep{Abbottetal2016h,Abbottetal2017,Abbottetal2017_2}. Additionally, the expected merger rate densities
 in the Kozai-Lidov channel are $1-1.5 \,{\rm Gpc}^{-3} \, {\rm yr}^{-1}$ for BH binaries forming in nuclear
 star clusters without supermassive BHs (SMBHs) through multi-body interactions \citep{AntoniniRasio2016} 
 and $0.14-6.1 \,{\rm Gpc}^{-3} \, {\rm yr}^{-1}$ in isolated triple systems \citep{SilsbeeTremaine2017}. 
 Of order $1-5\,{\rm Gpc}^{-3} \, {\rm yr}^{-1}$ merger rate density is expected for BH 
 binaries forming via the Kozai-Lidov mechanism in globular clusters  \citep{Antoninietal2014,Antoninietal2016,Rodriguezetal2016a}
 and in GNs \citep{AntoniniPerets2012,Hoangetal2017}, and non-spherical nuclear star clusters may 
 produce BH binary merger rates of up to $15\, {\rm Gpc}^{-3} \, {\rm yr}^{-1}$ \citep{PetrovichAntonini2017}.
 Smaller size GNs with intermediate mass BHs may produce higher rates \citep{VanLandinghametal2016}.
 Binary-single gravitational interactions may greatly
 increase the rates \citep{Samsingetal2014,SamsingRamirezRuiz2017,Samsing2017,Samsingetal2017}. 
 In a companion paper \citep{Gondanetal2017}, we have shown that GW capture sources in 
 galactic nuclei, which appear to be circular to within $e<0.2$ near the LSO, may be highly 
 eccentric at the beginning of the detected waveform at $10 \, \mathrm{Hz}$, and that heavier
 BH binaries are expected to be systematically more eccentric in this channel. The ongoing development of detectors 
 towards their design sensitivity at low frequencies may open the possibility of detecting 
 eccentricity in such systems.
 
 In this paper, we determine the expected accuracy with which a network of ground-based interferometric
 GW detectors may determine the physical parameters that describe highly eccentric BH binaries in 
 comparison to circular sources. We investigate how signal-to-noise ratios (SNRs) and parameter 
 measurement errors depend on the initial orbital parameters, particularly the initial pericenter 
 distance and eccentricity.  We examine if it is possible to measure the initial binary parameters 
 (initial eccentricity and pericenter distance) at formation for sources that form in the GW frequency 
 band of the instrument.
 
 Previous GW parameter estimation accuracy studies for eccentric waveforms were carried out for extreme
 mass ratio (EMRI) sources around SMBHs for LISA (\citealt{BarackCutler2004,PorterSesana2010};
 \,\mbox{\citealt{CornishKey2010}};\,\mbox{\citealt{Mikoczietal2012,Nishizawaetal2016}}) and for 
 low-eccentricity stellar-mass compact binaries for Earth-based GW detector network \citep{Sunetal2015}. 
 The premerger localization accuracy of eccentric neutron star binary systems was determined by 
 \citet{KyutokuSeto2014}, and the source localization accuracy was investigated for low-eccentricity 
 binaries by \citet{Maetal2017}.
 
 The parameter space of an eccentric spinning binary waveform is generally very large, 17-dimensional 
 \citep{Vecchio2004,CornishKey2010}. Therefore, state-of-the-art methods such as Monte Carlo Markov 
 Chain calculations (see \citealt{OShaughnessyetal2014} and references therein) are numerically 
 prohibitively expensive to explore the full range of source parameters for a large set of binaries. 
 For Gaussian noise and a large SNR, the posterior distribution function of the measured parameters 
 is generally well approximated by a multidimensional Gaussian, and the parameter measurement errors 
 can be estimated accurately and very efficiently using the Fisher matrix method 
 \citep{FinnChernoff1993,CutlerFlanagan1994,CutlerVallisneri2007}. Using this technique, we determine 
 the physical parameters' measurement accuracy. 

 We restrict this first study to waveforms introduced by \citet{MorenoGarridoetal1994} and 
 \citet{MorenoGarridoetal1995}, which account for part of the leading order post-Newtonian 
 correction, the GR pericenter precession (hereafter simply precession) and neglect other 
 first post-Newtonian and higher order corrections including those due to spins. Future 
 extensions of this work should include higher order post-Newtonian and merger waveforms 
 (see \citealt{Levinetal2011,Csizmadiaetal2012,Eastetal2013} for waveform generators,  \citealt{Damouretal2004,Memmesheimeretal2004,Konigsdorfferetal2005,Konigsdorfferetal2006,Yunesetal2009,TessmerSchafer2010,TessmerSchafer2011,Huertaetal2014,Mikoczietal2015,Mooreetal2016,Tanayetal2016,Boetzeletal2017,CaoandHan2017,HindererBabak2017,Huertaetal2017,Huertaetal2017_2,LoutrelYunes2017} for analytic waveform models, and \citealt{Hinderetal2008,Eastetal2012,Goldetal2012,GoldBrugmann2013,Eastetal2015,Paschalidisetal2015,Eastetal2016,Lewisetal2017} for waveforms of numerical relativity simulations of eccentric compact binary inspirals). In this paper, we focus 
 on BH-BH binaries, but the method is also applicable for neutron star-neutron star and
 neutron star-black hole binaries on highly eccentric orbits as long as tidal interactions
 and matter exchange among the components are negligible (see
 \citealt{Goldetal2012,Eastetal2015,Eastetal2016,Radiceetal2016} and references therein).
 \footnote{Eccentric neutron star (NS) binaries (NS-NS or NS-BH) will also benefit from 
 additional information if an electromagnetic counterpart is identified, which may lead 
 to smaller parameter errors \citep{Radiceetal2016}.}

 Once a large number of GW sources is detected, the correlations between the orbital eccentricity, 
 binary total mass, reduced mass, and spins may be distinctive among different astrophysical mechanisms
 leading to BH mergers \citep{OLearyetal2009,Cholisetal2016,Rodriguezetal2016a,Rodriguezetal2016b,Chatterjeeetal2017,Gondanetal2017,SamsingRamirezRuiz2017,SilsbeeTremaine2017,Kocsisetal2017}. Therefore, detections of eccentric BH binaries have a potential in constraining
 GW source populations. 
 
 However, detecting eccentric sources and recovering their physical parameters is very challenging.
 So far three search methods were developed to find the signals of stellar-mass eccentric BH 
 binaries in data streams of GW detectors \citep{Taietal2014,Coughlinetal2015,Tiwarietal2016}. All
 three methods achieve substantially better sensitivity for eccentric BH binary signals than existing 
 localized burst searches or chirp-like template based search methods. Once a source is detected, 
 different algorithms are used to recover its physical parameters. For compact binary coalescences,
 \textsc{BAYESTAR} \citep{SingerPrice2016} is an online fast sky localization algorithm that produces 
 probability sky maps, \textsc{LALInference} \citep{Veitchetal2015} is an offline full parameter 
 estimation algorithm, and \textsc{gstlal} \citep{Cannonetal2012,Priviteraetal2014} is a low-latency 
 binary BH parameter estimation algorithm. All three algorithms use waveform models of compact binaries 
 on circular orbits. In addition, for short-duration GW ''bursts'' with poorly modeled or unknown waveforms, 
 \textsc{Coherent WaveBurst} \citep{Klimenkoetal2016}, \textsc{BayesWave} \citep{CornishLittenberg2015}, 
 and \textsc{LALInferenceBurst} \citep{Veitchetal2015} pipelines produce reconstructed waveforms with 
 minimal assumptions on the waveform morphology. The development of algorithms recovering the parameters
 of compact binaries on eccentric orbits are currently underway. These algorithms will play an important
 role for the astrophysical interpretation of eccentric sources.
 
 The paper is organized as follows. In Section \ref{sec:WaveMod}, we summarize the basic 
 formulae describing the time-domain and frequency-domain eccentric waveform model. In Section
 \ref{sec:GWdetectors}, we outline the properties of advanced detectors we use in the analysis. 
 In Section \ref{sec:FisherForm}, we describe the signal parameter measurement estimation method. 
 In Section \ref{sec:ParamSpace}, we discuss which parameters of an eccentric binary can be 
 measured through the binary's waveform. We present our main results in Section \ref{sec:NumRes},
 and compare or results with previous papers. Finally, we summarize our conclusions in Section 
 \ref{sec:Conc}. Several details about our methodology is included in appendices. In Appendix 
 \ref{sec:circlim}, we consider the values of source parameters in the circular limit. Next, 
 in Appendix \ref{sec:RoGBD}, we introduce the geometric conventions we use to describe 
 how the GWs interact with ground-based detectors. In Appendix \ref{sec:TimeinDetBand}, we 
 discuss the applicability of the assumptions of neglecting the Earth's rotation around its 
 axis and Earth's motion around the Sun. In Appendix \ref{sec:NumEffSNRFisher}, we derive 
 numerically effective formulae to reduce the computational cost of numerical calculations
 of the $\mathrm{SNR}$ and the Fisher matrix. In Appendix \ref{sec:TestCodes}, we present 
 numerical comparisons to validate our codes for both precessing and non-precessing waveforms.
 
 We use $G=1=c$ units when referring to the initial orbital parameters, and when determining the 
 phases of waveforms. We work in the observer frame assuming a binary at cosmological redshift $z$.
 In this frame, all of the formulae have redshifted mass parameters \mbox{$m_z=(1+z)m$}.\footnote{Additional corrections are necessary if the binary has a peculiar velocity \citep{Kocsisetal2006a}.}

\section{Eccentric Waveform Model}
\label{sec:WaveMod}
 
 In this section, we summarize the basic formulae describing the time-domain (Section 
 \ref{sec:TimeWaveMod}) and frequency-domain (Section \ref{subsec:FreqDom}) eccentric waveform models
 including precession in the leading quadrupole-order radiation approximation using the Fourier-Bessel
 decomposition. Note that we neglect the radiation of higher multipole orders, which are typically 
 subdominant at least in cases where the initial pericenter distance is not close to a grazing or 
 zoom-whirl configuration and the initial velocity is much less than the speed of light
 \citep{Davisetal1972,Bertietal2010,Healyetal2016}.

\subsection{The waveform in time domain} 
\label{sec:TimeWaveMod}
 
 We adopt the waveform model of \citet{MorenoGarridoetal1994} and \citet{MorenoGarridoetal1995}, which 
 describes the quadrupole waveform emitted by a spinless binary on a Keplerian orbit undergoing slow 
 precession. For a fixed semi-major axis $a$ and orbital eccentricity $e$, the two polarization states
 of a GW, $h_+$ and $h_{\times}$, with component masses $m_A$ and $m_B$, and at luminosity distance 
 $D_\mathrm{L}$, can be given in the observer's time-domain as \citep{MorenoGarridoetal1995}: 
\begin{align} \label{eq:hplusBessel}
 \nonumber h_+(t) & = - \frac{ h \, \sin^2 \Theta }{ 2 } \sum_{n=1}^{\infty} A_n \cos \Phi_n(t)
  \\ 
 & + \frac{ h (1+\cos^2 \Theta) }{ 2 } \sum_{n=1}^{\infty} \left( B^+_n \cos \Phi^-_n(t)
  - B^-_n \cos \Phi^+_n(t) \right) \, ,\\
 \label{eq:hxBessel}
  h_{\times} (t) &= - h \, \cos \Theta \sum _{n=1}^{\infty}
  \left( B^-_n \sin \Phi^+_n(t) + B^+_n \sin\Phi^-_n(t)   \right) \, ,
\end{align}
where $\Theta$ is the angle between the orbital plane and the line-of-sight to the 
 observer, $\Phi_n^{\pm}$ describe the orbital phase given below for the $n^{\rm th}$ harmonic,
\begin{equation}  \label{eq:h}
  h = \frac{ 4 \, M_{{\rm tot},z} \, \mu_z }{a \, D_\mathrm{L}  }  \, , 
\end{equation} 
 where $M_{\mathrm{tot},z}= (m_{A}+m_B)(1+z)$ is the redshifted total binary mass at cosmological 
 redshift $z$, and \mbox{$\mu_z =(1+z) m_A m_B (m_A + m_B)^{-1}$} is the redshifted reduced mass. 
 We can express the luminosity distance for a flat $\Lambda$CDM cosmology as a function of $z$ as 
\begin{equation} \label{eq:CovDLz} 
  D_\mathrm{L} = \frac{ (1+z)c }{ H_0 } \int_0 ^z \frac{dz'}{\sqrt{ \Omega_\mathrm{M} 
  (1+z')^3 + \Omega_{\Lambda} }} \, , 
\end{equation} 
 where  $H_0 = 68 \mathrm{km} \, \mathrm{s}^{-1} \mathrm{Mpc}^{-1}$ is the Hubble constant, 
 and $\Omega_\mathrm{M} = 0.304$ and $\Omega_{\Lambda} = 0.696$ are the density parameters 
 for matter and dark energy, respectively \citep{PlanckCollaboration2014a,PlanckCollaboration2014b}. 
 
 The  $A_n$ and $B_n^{\pm}$ prefactors in Equations (\ref{eq:hplusBessel}) and (\ref{eq:hxBessel})
 are the linear combinations of Bessel function of the first kind, $J_n(x)$,
 \begin{equation}\label{eq:AnBn}
  A_n = J_n (ne)\,, \quad B^{\pm}_n = \frac{S_n \pm C_n }{2} \, ,
 \end{equation}
  where $e$ is the orbital eccentricity,
\begin{align} 
  S_n & = - \frac{(1-e^2)^{1/2}}{e} \frac{2}{n} J'_n (ne) + \frac{(1-e^2)^{3/2}}{e^2} 
  2 n J_n (ne) \, , \label{eq:Sn} \\ 
  C_n & = - \frac{ 2-e^2 }{ e^2 } J_n (ne) + \frac{ 2 (1-e^2) }{ e } J'_n (ne) \, , \label{eq:Cn} 
\end{align} 
 and $J'_n(ne)$ is the first derivative of $J_n(ne)$ with respect to $e$, which satisfies
\begin{equation} \label{eq:dJnde} 
  J'_n (ne) = \frac{ n }{ 2 } \left[ J_{n-1} (ne) - J_{n+1} (ne)\right] \, . 
\end{equation} 
 We will also need the second derivative of $J_n(ne)$ with respect to $e$ when calculating the 
 Fisher matrix, thus we introduce $J''_n(ne)$ as
\begin{align} \label{eq:J2na} 
 J''_n &= \frac{n^2}{4} \left[ J_{n-2}(ne)-2J_{n}(ne) + J_{n+2}(ne) \right]  \quad{\rm if~}n\geq 2 
 \,, \\
 J''_1 &=  - \frac{J_{1}(e)}{2}-\frac{1}{4} \left[ J_{1}(e)-J_{3}(e) \right] \, .
\end{align} 
 
 The phase functions $\Phi_n(t)$ and $\Phi_n^{\pm}(t)$ in Equations (\ref{eq:hplusBessel}) and
 (\ref{eq:hxBessel}) are
\begin{align} 
   \Phi_n (t) &= \Phi_c - 2 \pi n \int^{t_c}_{t} \nu(t') dt' \, , \label{eq:Phi} \\
  \Phi^{\pm}_n(t) & = \Phi_n(t) \pm 2\gamma(t) \, , \label{eq:Phipm}
\end{align} 
 where the second term in the right hand side is $n$ times the mean anomaly expressed 
 with the time-integral of the redshifted Keplerian mean orbital frequency $\nu$, $\Phi_c$ is
 the phase extrapolated to coalescence time $t=t_c$, and $\gamma$ is the azimuthal angle of the 
 pericenter relative to the $x$ axis of the coordinate system defined by the orbital plane. 
 The redshifted Keplerian mean orbital frequency may be expressed with the dimensionless 
 pericenter distance 
 \begin{equation}
  \rho_\mathrm{p}=\frac{a(1-e)}{M_{\mathrm{tot},z}}
 \end{equation}
(where $a$ is the semi-major axis in the observer frame) as
\begin{equation}  \label{eq:nuKepler} 
 \nu(e,\rho_\mathrm{p}) = \frac{ (1-e)^{3/2} }{2 \pi \rho_\mathrm{p}^{3/2} M_{\mathrm{tot},z}} \,. 
\end{equation} 
 
 For an inspiraling binary, both the eccentricity and the Keplerian orbital frequency evolve in
 time. Assuming quadrupole radiation and adiabatic evolution of orbital parameters, the equations
 of time evolution of $e$ and $\nu$, as seen at some cosmological redshift, can be given to
 leading order as \citep{Peters1964}
\begin{eqnarray} \label{eq:dedt} 
  \dot{e} &=& - \frac{304}{15} \frac{e \mathcal{M}_z^{5/3} (2 \pi \nu)^{8/3}}{(1-e^2)^{5/2}} 
   \left( 1 + \frac{121}{304} e^2 \right) \, , \\
 \label{eq:dnudt} 
 \dot{ \nu } &=& \frac{48}{5 \pi} \frac{ \mathcal{M}_z^{5/3} (2 \pi \nu)^{11/3}}{(1-e^2)^{7/2}} 
  \left(1 + \frac{73}{24} e^2 + \frac{37}{96} e^4 \right) \, , 
\end{eqnarray} 
 where $\mathcal{M}_z = \mu^{3/5}_z M_{\mathrm{tot},z}^{2/5}$ is the redshifted chirp mass, and 
 the overdot denotes a redshifted time-derivative $\dot{x} \equiv dx/dt$. The fraction of the 
 two equations
\begin{equation}  \label{eq:dnude} 
 \frac{d \nu }{de} = - \frac{18}{19} \frac{\nu}{e} \frac{\left(1 + \frac{73}{24} e^2 + 
 \frac{37}{96} e^4 \right)}{(1-e^2) \left(1 + \frac{121}{304} e^2 \right) } \, ,
\end{equation}
 may be integrated as  \citep{Peters1964,Mikoczietal2012}
\begin{equation} \label{eq:nue}
 \nu(e) = \CC/H(e)
\end{equation} 
 where we define
\begin{equation}  \label{eq:He}
  H(e) = e^{18/19}(1-e^2)^{-3/2} \left(1+\frac{121}{304} e^2 \right)^{\frac{1305}{2299}} \, ,
\end{equation}
 and $\CC$ is an integration constant set by the initial condition $\nu(e_0, \rho_{\mathrm{p}0})
 = \nu_0$ or the conditions at the LSO, \mbox{$\nu(e_{\rm LSO}, \rho_\mathrm{pLSO}) = \nu_{\LSO}$}
 (see Equations (\ref{eq:defeLSO}) and (\ref{eq:nuLSO}) below). Equation (\ref{eq:nue}) shows that 
 the product $\CC = \nu(e) H(e)$ is conserved during the evolution. Similarly, it is straightforward 
 to determine the evolution of the dimensionless pericenter distance
\begin{align}  \label{eq:rhop}
 \rho_\mathrm{p} (e) & = \frac{c_1 e^{12/19}}{ M_{{\rm tot},z}^{2/3} (1 + e)}
 \left(1 + \frac{121}{304} e^2 \right)^{\frac{870}{2299}} 
 \nonumber  \\
 & = \rho_\mathrm{p0} \frac{
 e^{12/19}(1 + e)^{-1}\left[1 + (121/304) e^2 \right]^{\frac{870}{2299}}}{
 e_0^{12/19}(1 + e_0)^{-1}\left[1 + (121/304) e_0^2 \right]^{\frac{870}{2299}}}
\nonumber\\&
 = \rho_\mathrm{pLSO} 
 \frac{ e^{12/19}(1 + e)^{-1}\left[1 + (121/304) e^2 \right]^{\frac{870}{2299}}}{ e_{\LSO}^{12/19}
 (1 + e_{\LSO})^{-1}\left[1 + (121/304) e_{\LSO}^2 \right]^{\frac{870}{2299}} } 
\end{align}
 \citep{Peters1964}, where $c_1=(2\pi \CC)^{-2/3}$ and in the second and third lines we expressed
 the evolution with the initial condition $\rho_\mathrm{p0}$ and $e_0$, or the ``final condition'' 
 at the LSO, which satisfies 
\begin{equation} \label{eq:defeLSO}
 \rho_\mathrm{pLSO} = \rho_\mathrm{p} (e_\mathrm{LSO}) = \frac{6 + 2 e_\mathrm{LSO}}{1 +
 e_\mathrm{LSO}} 
\end{equation}
 in the leading order approximation in the test mass geodesic zero spin limit \citep{Cutleretal1994}.
 This shows that the evolution may be parameterized with the single parameter $e_\mathrm{LSO}$, or the
 two parameters $\rho_{p0}$ and $e_0$. Note that for any $e$, the orbital frequency depends only on the
 single parameter $\CC$, which is set uniquely by $e_{\rm LSO}$ and $M_{\mathrm{tot},z}$ as
 \begin{equation}\label{eq:CC}
 \CC = \frac{1}{2\pi M_{{\rm tot},z} }\frac{(1-e_\mathrm{LSO})^{3/2}H(e_\mathrm{LSO})}{[\rho_
 \mathrm{pLSO}(e_\mathrm{LSO})]^{3/2}}\,.
 \end{equation}
 
 We restrict our interest to the repeated burst \citep{OLearyetal2009,KocsisLevin2012} and eccentric
 inspiral phases of the waveform model between $0 < e_\mathrm{LSO}\leq e\leq e_0\leq 1$ and neglect 
 the merger and ringdown phases in this analysis. The repeated burst phase starts when the binary is
 formed with initial eccentricity $e_0>0.9$ and initial dimensionless pericenter distance $\rho_{p0}$,
 and the eccentric inspiral phase ends when the binary reaches the LSO with eccentricity $e_\mathrm{LSO}$.
 Note that during the evolution $e$ and $\rho_{p}$ both shrink strictly monotonically in time. 
 
 Let us also note for further use, that the Keplerian redshifted orbital frequency at the end of 
 the assumed eccentric inspiral waveform (i.e. at the LSO) is given by Equations (\ref{eq:nuKepler})
 and (\ref{eq:defeLSO}) as
\begin{equation} \label{eq:nuLSO}
  \nu_\mathrm{LSO} = \nu(e_\mathrm{LSO}) = \frac{ 1 }{2 \pi M_{\mathrm{tot},z} } 
   \left( \frac{ 1 - e_\mathrm{LSO}^2 }{ 6 + 2 e_\mathrm{LSO} }  \right)^{3/2} \,. 
\end{equation}

 Precession leads to a time-dependent $\gamma$ in Equation (\ref{eq:Phipm}). Using the analysis in 
 \citet{Mikoczietal2012}, we adopt pericenter precession from the classical relativistic motion, 
 and assume that the adiabatic evolution of the orbital parameters are governed by Equations 
 (\ref{eq:dedt}) and (\ref{eq:dnudt}). The angle of precession for a single eccentric orbit in
 the test particle geodesic approximation around a Schwarzschild BH is
\begin{equation}
  \Delta \gamma = \frac{ 6 \pi M_{\rm tot} }{ a(1-e^2) } \, .
\end{equation}
 Using an adiabatic approximation, we approximate the redshifted precession rate to be constant 
 during the orbit with
\begin{equation}  \label{eq:dgammadt}
 \dot{ \gamma} \approx \frac{\Delta \gamma}{T} = \frac{3(2 \pi \nu)^{5/3} M_{{\rm tot},z}^{2/3}}
 {1-e^2} \,. 
\end{equation}
 
 The phase functions given by Equations (\ref{eq:Phi}) and (\ref{eq:Phipm}), can be calculated 
 from Equations (\ref{eq:dedt}) and (\ref{eq:nue}) as\footnote{For circular orbits the Fourier 
 phase is conveniently parameterized by $\nu$ \citep{CutlerFlanagan1994}. However, for eccentric
 inspirals, since $\nu(e)$, $\dot{\nu}(e)$, and $\dot{e}(e)$ are given analytically in the PN 
 approximation, the phase is more conveniently parameterized by $e$ 
 \citep{OLearyetal2009,Mikoczietal2012}.} 
\begin{equation}  \label{eq:Phint}
  \Phi_n (t) = \Phi_c + 2 \pi n \int^{e(t)}_0 \frac{\nu(e')}{\dot{e}(\nu(e'),e')} de' 
\end{equation}
 \citep{CutlerFlanagan1994}. The phase functions, which arise due to precession, $\Phi^{+}_n(t)$ 
 and $\Phi^{-}_n(t)$, follow from Equations (\ref{eq:Phipm}), (\ref{eq:nue}), (\ref{eq:dgammadt}) 
 and (\ref{eq:Phint})
\begin{eqnarray} \label{eq:Phinpmt}
  \Phi^{\pm}_n(t) &=& \Phi _n(t) \pm 2 \gamma_c \mp 2 \int^{t_c}_{t} \dot{\gamma} (t') dt' \\ 
  \nonumber
   &=& \Phi _c \pm 2 \gamma_c + \int^{e(t)}_{0} \frac{2\pi n\nu(e')\pm 2\dot{\gamma}(\nu(e'),e')}
   {\dot e(\nu(e'),e')}  de' \, ,
\end{eqnarray}
 where $\gamma_c$ is the angle of periapsis extrapolated to coalescence. Note that $\rho_{p0}$, 
 $t_c$, $\gamma_c$, $\Phi_c$, and $e_\LSO$ are free parameters of the waveform. Alternatively, we 
 may use the corresponding initial values $e_0$, $t_0$, $\gamma_0$, $\Phi_0$, and $\rho_{\mathrm{p}0}$.

\subsection{The waveform in frequency domain }
\label{subsec:FreqDom}
 
 Since the expressions defining the SNR and the Fisher matrix are both given in Fourier space 
 (Section \ref{sec:FisherForm}), we construct the Fourier transforms of the waveform 
 \footnote{ We find that modulations due to Earth's rotation around 
 its axis and Earth's motion around the Sun can be neglected because the signal spends relatively
 short time in the advanced detectors' sensitive frequency band (see Appendix \ref{sec:TimeinDetBand}).} 
\begin{align}
 \label{eq:FThpx} \tilde{h}_{+,\times}(f) & = \int_{-\infty}^{\infty} h_{+,\times}(t) e^{2 \pi i f} dt
                  \, , 
\end{align} 
 where $h_+(t)$ and $h_{\times}(t)$ are given in Equations (\ref{eq:hplusBessel}) and (\ref{eq:hxBessel})
 as an infinite sum over orbital harmonics $n$. In the stationary phase approximation each frequency 
 harmonic splits into a triplet due to precession (see Equation (\ref{eq:dgammadt})) $\mathbf{f} \equiv 
 (f_n, f_n^{\pm})$ \citep{MorenoGarridoetal1995,Mikoczietal2012}, where
\begin{align}
 \label{eq:fn} f_n & = n \nu \, , \\ 
 \label{eq:fnpm} f_n^{\pm} & = n \nu \pm \frac{\dot{\gamma}}{\pi} \, ,
\end{align}  
 and the Fourier transform simplifies to 
\begin{align} 
  \label{eq:SPAp}                     
     \tilde{h}_+ ( \mathbf{f} ) & = - \frac{h_0}{4} \sin^2 \Theta
    \sum _{n=1}^{\infty} A_n \Lambda_n e^{i ( \Psi_n - \pi/4)} 
     \nonumber\\&\quad
   - \frac{h_0}{4} (1 + \cos^2 \Theta)  \sum _{n=1}^{\infty} B^+_n 
            \Lambda_n ^- e^{i ( \Psi_n^- - \pi/4)}    
     \nonumber\\&\quad
   + \frac{h_0}{4} (1 + \cos^2 \Theta) \sum _{n=1}^{\infty} B^-_n \Lambda_n ^+  
   e^{i ( \Psi_n^+ - \pi/4)} \, , \\
 \label{eq:SPAx} \tilde{h}_{\times} ( \mathbf{f}) & = - \frac{h_0}{2} \, \cos\Theta
   \sum _{n=1}^{\infty}  B^-_n \Lambda_n ^+ e^{i ( \Psi_n^+ + \pi/4) } 
    \nonumber\\&\quad    
   - \frac{h_0}{2} \, \cos\Theta \sum _{n=1}^{\infty} B^+_n \Lambda_n ^- 
     e^{i ( \Psi_n^- + \pi/4 )}  \, , 
\end{align}
 where 
\begin{eqnarray} \label{eq:h0}
  h_0 &=& \frac{ 4 \mathcal{M}_z^{5/3} (2 \pi \nu)^{2/3} }{ D_\mathrm{L} } \, ,\\\label{eq:Lambda}
    \Lambda_ n  &=& \frac{1}{\sqrt{\arrowvert n \dot{\nu} \arrowvert }}\, ,\quad 
  \Lambda_ n^{\pm} = \frac{1}{ \sqrt{\arrowvert n \dot{\nu} \pm \ddot{\gamma}/ \pi 
  \arrowvert } } \, ,
\end{eqnarray}
 $\dot{\nu}$ and $\ddot{\gamma}$ are given by Equations (\ref{eq:dnudt}), (\ref{eq:dnude}) and 
 (\ref{eq:dgammadt}), and \mbox{$\mathcal{M}_z = (1+z) \mathcal{M}$}. The $ \Psi_n $ and $ \Psi_n^{\pm}$
 phases and their first ($\dot{\Psi}_n\, ,\dot{\Psi}^{\pm}_n$) and second ($\ddot{\Psi}_n\, , 
 \ddot{\Psi }^{\pm}_n$) derivatives with respect to redshifted time $t$ are \citep{Mikoczietal2012}
\begin{align}
  \Psi_n (e, f_n) & = 2 \pi f_n t_n - \Phi_n \, , \label{eq:PsiN}  \\
  \Psi^{\pm}_n (e, f^{\pm}_n) & = 2 \pi f_n^{\pm} t_n^{\pm} - \Phi_n^{\pm} \, .   \label{eq:Psipm} 
\end{align}  
 Here the $(t_n,t_n^{\pm})$ parameters of the stationary phase approximation specify the 
 times at which the orbital frequency satisfies Equations (\ref{eq:fn}) and (\ref{eq:fnpm}) 
 for given $(f_n,f_n^{\pm})$, see Appendices A and B in \citet{Mikoczietal2012} for details.
 In Equations (\ref{eq:PsiN}) and (\ref{eq:Psipm}), $(\Phi_n, \Phi_n^{\pm})$ are to be 
 substituted from Equations (\ref{eq:Phint}) and (\ref{eq:Phinpmt}). We eliminate $\nu(t)$ 
 and $e(t)$ for $(f_n,f_n^{\pm})$ using Equations (\ref{eq:nue}), (\ref{eq:CC}), and 
 (\ref{eq:dgammadt}) together with Equations (\ref{eq:fn}) and (\ref{eq:fnpm}) to obtain 
 the frequency-domain waveform.\footnote{In practice, there 
 are closed analytic expressions for the $e$-dependence of $\nu$, $\dot{\nu}$, $\dot\gamma$, 
 $\ddot\gamma$, $\Phi_{n}$, $\Phi_n^{\pm}$, and hence also for $f_n$ and $f_n^{\pm}$. We 
 must invert these relations $f_n(e)$ and $f_n^{\pm}(e)$ to obtain the waveform in frequency
 domain.} The result depends on constant parameters $t_c$, $\Phi_c$, $\gamma_c$, $e_\mathrm{LSO}$,
 $\mathcal{M}_z$, and $\mathrm{M}_{\mathrm{tot},z}$. Further, we note that if the precessing 
 eccentric BH binary forms with $\rho_{\mathrm{p}0}$ and $e_0$, then the frequency-domain 
 waveform is truncated at the corresponding minimum frequency \mbox{$(f_{n,\min}$, $f_{n,\min}
 ^{\pm})=(n\nu_0,n\nu_0\pm \dot\gamma_0/\pi)$}. Furthermore, the waveform model becomes invalid after reaching the LSO (with $\rho
 _{\mathrm{p LSO}}$ and $e_\mathrm{LSO}$), which corresponds to a maximum frequency for each 
 harmonic \mbox{$(f_{n,\max}$, $f_{n,\max}^{\pm})=(n\nu_{\mathrm{LSO}},n\nu_{\mathrm{LSO}}\pm \dot\gamma_{\mathrm{LSO}}/\pi)$} where this model is applicable. If we 
 truncate the waveform at these maximum frequencies, this respectively introduces an explicit
 $(\rho_{\mathrm{p}0}, e_0)$ and $(\rho_{\mathrm{p LSO}}, e_\mathrm{LSO})$ parameter dependence
 in the waveform model. This is shown in Equation~(\ref{eq:hPPC}) in Appendix~\ref{sec:NumEffSNRFisher}.
 Examples of the frequency-domain waveforms are shown in \citet{{KocsisLevin2012}}.
 
 In principle, the number of spectral harmonics of an eccentric binary system is infinite. Note
 however that a large fraction of the signal power is accumulated in a finite number of harmonics.
 Therefore, in order to reduce the necessary computation time, we truncate $n$ at $n_\mathrm{max}
 (e_0)$ \citep{OLearyetal2009,Mikoczietal2012}
\begin{equation}  \label{eq:nmax}
  n_\mathrm{max}(e_0) = \left\{ 5 \frac{ (1+e_0)^{1/2} }{ (1-e_0)^{3/2} } 
  \right\} \, ,
\end{equation}
 which accounts for $99\%$ of the signal power \citep{Turner1977}. Here the bracket $\{ \}$ 
 denotes the floor function. In Appendix \ref{sec:NumEffSNRFisher}, we discuss other technical
 details to optimize the calculation of the SNR and the Fisher matrix.
 
 To test our calculations, we examine the limiting cases of no precession  ($\dot{\gamma}\rightarrow
 0$) and circular orbits ($e\rightarrow 0$), respectively. In Appendix \ref{sec:NumEffSNRFisher},
 we discuss numerical and analytic tricks to optimize the calculation and discuss results for the
 precessing (\emph{Prec}) and precession-free (\emph{NoPrec}) waveform model (i.e. $\dot{\gamma} 
 \equiv 0$).

\section{GW Detectors Used in the Analysis}
\label{sec:GWdetectors}
 
 Here we summarize the GW detectors and the assumed properties of the detector noise in our analysis.
 
 The aLIGO and AdV detectors completed their first two observing run, and made the first six detections of
 GWs \citep{Abbottetal2016c,Abbottetal2016b,Abbottetal2017,Abbottetal2017_2,Abbottetal2017_3,LIGOColl2017}.
 Two additional GW detectors are planned to join the network of aLIGO and AdV; (i) the Japanese KAGRA is 
 under construction with baseline operations beginning in $2018$ \citep{Somiya2012}; while (ii) the proposed
 LIGO-India is expected to become operational in $2022$ \citep{Iyeretal2011,Abbottetal2016a}. LIGO-India was
 approved by the government of India and a study has already suggested site location and orientations of arms 
 for the detector based on scientific figures of merit \citep{Raffaietal2013}. These parameters, however, have
 not been finalized yet, and because of this we omit LIGO-India from the analysis. 

\begin{table}
\centering  
   \begin{tabular}{@{}lrrr}
     \hline
       Detector  &  East Long.  &  North Lat. &  Orientation $\psi$  \\
     \hline\hline
       LIGO H  & $ -119.4^{\circ}$ & $46.5^{\circ}$ & $-36^{\circ}$ \\
       LIGO L  &   $ -90.8^{\circ}$   &   $ 30.6^{\circ} $ & $-108^{\circ}$  \\
       VIRGO &    $ 10.5^{\circ}$  &  $ 43.6^{\circ}$ & $ 20^{\circ}$  \\
       KAGRA &    $ 137.3^{\circ}$ &   $ 36.4^{\circ}$ & $ 65^{\circ}$  \\
     \hline
   \end{tabular} 
   \caption{ \label{tab:DetCord} Locations and orientations of considered GW detectors in 
   the coordinate system defined in Appendix \ref{sec:RoGBD}. LIGO H marks the Advanced LIGO
   detector in Hanford, and LIGO L marks the Advanced LIGO detector in Livingston. }
\end{table}

 Due to the expected similarities of design sensitivities of the aLIGO, AdV, and KAGRA 
 detectors  within the frequency range of BH inspiral waveforms, for simplicity we adopt
 the design sensitivities of the two aLIGO \citep{Abbottetal2016a} for AdV \citep{Abbottetal2016a}
 and for KAGRA \citep{Somiya2012} detectors. Table \ref{tab:DetCord} gives the locations
 and orientations of these detectors, which we used to calculate the response functions.
 For each detector, we define the detector's orientation angle, $\psi$, as the angle 
 measured clockwise from North between the $x$-arm of the detector (see Appendix 
 \ref{sec:RoGBD} for the geometric conventions of detectors) and the meridian that 
 passes through the position of the detector.
 
 We assume that the noise in each detector is stationary colored Gaussian with zero mean, and 
 that it is uncorrelated between different detectors. In reality, detector noise arises 
 from a combination of instrumental, environmental, and anthropomorphic sources that are difficult 
 to characterize precisely \citep{Aasietal2012,Asoetal2013,Aasietal2015}, and non-Gaussian 
 noise transients (glitches) may arise as well \citep{Blackburnetal2008}. However, there 
 are existing techniques to identify and remove glitches from GW strain channels 
 and to reduce the level of these artifacts 
 \citep{LittenbergCornish2010,Prestegardetal2012,Biswasetal2013,Powelletal2015,Boseetal2016,Torresetal2016,Georgeetal2017,Mukundetal2017,Powelletal2017,Shenetal2017}. 
 Furthermore, correlated noise between widely separated detectors can arise from so-called 
 Schumann resonances (predicted in \citet{Schumann1952a,Schumann1952b}  and observed soon 
 thereafter \citep{SchumannKonig1954,BalserWagner1960}), as well as from other EM phenomena 
 such as solar storms, currents in the van Allen belt \citep{Rycroft2006}, and anthropogenic 
 emission \citep[see][and references therein]{Shvetsetal2010,Thraneetal2013}. Note however, that 
 Schumann resonances mostly affect the stochastic GW background searches \citep{Thraneetal2013},
 and a strategy against such noise artifact already exists \citep{Thraneetal2014}. Our 
 simplifying assumptions on uncorrelated Gaussian noise are therefore partly justified.

\section{Overview of the Fisher Matrix Formalism} 
\label{sec:FisherForm} 
 
 In this section, we provide a brief overview of the Fisher matrix method to estimate the measurement 
 errors of physical parameters characterizing a precessing eccentric BH binary source, and refer 
 the reader to \citet{Finn1992} and \citet{CutlerFlanagan1994} for further details. 
 
 The output of a GW detector, $s(t)$, is a combination of a signal, $h(t)$, and a noise term, $n(t)$; i.e.
\begin{equation}  \label{eq:st} 
  s(t) = h(t) + n(t) \, .
\end{equation} 
 We assume the noise of a detector to be stationary, zero mean, and Gaussian, where the different
 Fourier components of the noise are uncorrelated, i.e. 
 \begin{equation}  \label{eq:defSa} 
  \left\langle \tilde{n}(f) \tilde{n}^*(f') \right\rangle = \frac{1}{2} \delta(f-f') S_n(f) \,  
\end{equation}
 \citep{Nissankeetal2010}, where $\langle\cdot\rangle$ denotes the average, $S_n(f)$ is the one sided noise 
 power spectral density of the detector, the ${}^*$ superscript denotes complex conjugate. With these assumptions, 
 the probability for the noise to have some realization $n_0(t)$ is given as
\begin{equation}  \label{eq:noiseprob} 
   p\left(n \equiv n_0 \right) \propto e^{- \left( n_0 \mid n_0 \right)/2 } \, 
\end{equation} 
 \citep{Finn1992}, where $p(n)$ is the probability distribution function of the noise to assume
 a value $n$, and $(\ldots | \ldots)$ denotes the following inner product between any two functions
 of frequency, e.g. x(f) and y(f):
\begin{equation}  \label{eq:InnProd} 
  \left( x \mid y \right) \equiv 4 \int_{0}^{\infty} 
   \frac{ \tilde{x}(f) \, \tilde{y}^*(f) }{S_n(f)} \, df \, .
\end{equation} 
 The optimal SNR is given by the standard expression 
\begin{equation}  \label{eq:SNR2_1det} 
  \frac{S}{N} = \sqrt{ \left(h \mid h \right) } = \sqrt{ 4 \int_{0}^{\infty} \frac{|\tilde{h}(f)|^2 
  }{S_n(f)} df } \, . 
\end{equation} 

 Here the signal waveform, $h(f)$, depends on the parameter set $\left\{ \lambda_p \,|\, p \in\{ 1 
 \ldots P \right\} \}$, which characterizes the source. For a large SNR, the parameter estimation
 errors \mbox{$\bm{\Delta \lambda}=\left\{ \Delta \lambda_p \,|\, p \in \{1, \ldots, P \right\} \}$} 
 defined as the measured value minus the true value have the Gaussian probability distribution 
 for a given signal
\begin{equation}  \label{eq:probdistpar}
  p ( \bm{\Delta \lambda} ) = \mathcal{N} \, \exp \left( - \frac{1}{2} 
  \Gamma_{ij} \Delta \lambda_i \Delta \lambda_j \right)
\end{equation}
 \citep{Finn1992}, where $\mathcal{N}$ is a normalization constant. In Equation (\ref{eq:probdistpar}), 
 we assume summation over repeated indices, and $\Gamma_{ij}$ is the Fisher information matrix defined as
\begin{equation}  \label{eq:Fisher}
  \Gamma_{ij} \equiv \left( \partial_i h \mid \partial_j h \right) = 
  4 \int_0^{\infty} \frac{ \Re \left( \partial_i \tilde{h}^*(f) \partial_j 
  \tilde{h}(f) \right) }{ S_n(f) } \, df \, ,
\end{equation}
 where  $\partial_i h = \partial h / \partial \lambda_i$ and $\Re$ labels the real part.
 
 Following \citet{CutlerFlanagan1994}, we define the combined signal-to-noise ratio of 
 a network of detectors ($\mathrm{SNR}_\mathrm{tot}$) as an uncorrelated superposition 
 of individual SNRs 
\begin{equation}  \label{eq:SNRtot}
   \left(\frac{S}{N}\right)^2_{\mathrm{tot}} = \sum_{k=1}^{N_\mathrm{det}} \left(\frac{S}{N}\right)^2_k \, ,
\end{equation} 
 where the number of detectors in the network is denoted by $N_\mathrm{det}$, and $(S/N)_k$
 denotes the SNR in the $k^{\mathrm{th}}$ detector. 

 Similarly, for uncorrelated Gaussian noise, the Fisher matrix of a network of detectors is the
 sum of the Fisher matrices of individual detectors,
\begin{equation} \label{eq:Fishertot}
  \Gamma_{ij,\mathrm{tot}} = \sum _{k=1}^{N_\mathrm{det}} \Gamma_{ij,k} \, .
\end{equation}
 The covariance matrix is defined with the inverse of the Fisher matrix:
\begin{equation} \label{eq:CovMatrix}
  \Sigma_{ij} = \left( \Gamma_{ij,\mathrm{tot}} \right)^{-1} = 
   \langle \Delta \lambda_i \Delta \lambda_j \rangle \, ,
\end{equation}
 where the angle brackets denote an average over the probability distribution function in 
 Equation (\ref{eq:probdistpar}). The root-mean-square parameter measurement error $\sigma_i$
 in the parameters $\lambda_i$ marginalized over all other parameters is
\begin{equation}  \label{eq:Sigma}
  \sigma_i = \langle \left(\Delta\lambda_i \right)^2 \rangle^{1/2} = \sqrt{\Sigma_{ii}} \, .
\end{equation}
 The off-diagonal elements of $\Sigma_{ij}$ give the cross-correlation coefficients 
 between parameters $\lambda_i$ and $\lambda_j$.
 
 Parameters can be measured independently if the corresponding Fisher matrix $\Gamma_{ij,\rm
 tot}$ is nonsingular. Otherwise, if the Fisher matrix is singular, then the eigenvector(s) 
 corresponding to the zero-eigenvalue(s) of the Fisher matrix represent the linear combination(s)
 of the parameters, which cannot be measured by the network.
 
 We derive efficient formulae to compute the SNR and the Fisher matrix in Appendix
 \ref{sec:NumEffSNRFisher}.

\section{Measuring the Parameters of Precessing Eccentric Black Hole Binaries}
\label{sec:ParamSpace} 
 
 In this section, we identify the parameters of a precessing eccentric binary that can be extracted
 from the detected waveform for the signal model introduced in Section \ref{sec:WaveMod}. We set the
 parameters in our calculations and measure their errors as follows.
 \begin{itemize}
  \item $D_\mathrm{L}:$ 
  We set $D_\mathrm{L}=100 \,\mathrm{Mpc}$, and measure its relative error $\langle \Delta D_{\rm L}^2 
  \rangle^{1/2} / D_{\rm L} = \langle (\Delta \ln D_{\rm L})^2 \rangle^{1/2}$. This choice is arbitrary, 
  smaller than the nearest circular BH-BH merger detection to date, $340 \pm 140 \, \mathrm{Mpc}$ 
  \citep{LIGOColl2017}. The Fisher matrix method gives accurate results for the parameter measurement 
  errors for high SNR. For moderately larger distances, the errors scale as $\propto D_{\rm L}$. 
  
  \item $\theta_N$ and $\phi_N$: 
  We generate an isotropic random sample of the sky position angles $\theta_N$ and $\phi_N$ by drawing
  $\cos \theta_N$ and $\phi_N$ from a uniform distribution between $[-1,1]$ and $[0,2 \pi]$, and calculate
  the parameter estimation covariance for each sample. The errors of the sky position is described by 
  a localization ellipse. We characterize the sky localization accuracy either by the corresponding 
  proper angular length of the semi-major and semi-minor axes of the sky-localization error ellipsoid
  given by \citet{LangHughes2006}, $(a_N,b_N)$, or its proper solid angle $\Delta\Omega_N = \sqrt{\pi 
  a_N b_N}$. The calculated results are valid if $a_N\ll 1$ radian and $b_N\ll 1$ radian. 

 \item $\theta_L$ and $\phi_L$: We draw the angular momentum vector direction angles from an isotropic
 distribution and construct their error ellipsoids or solid angles similar to that given for $\theta_N$ 
 and $\phi_N$.

 \item $m_A$ and $m_B$: We fix the fiducial component masses to $m_A=m_B=30 \, \Msun$, consistent
  with the first discovered source GW150914 \citep{Abbottetal2016b}. Such high mass sources are
  expected in galactic nuclei since mass segregation helps to increase their numbers relative to 
  the lower mass binaries, and the SNR is also higher for these binaries \citep{OLearyetal2009}. 
  Since we neglect additional post-Newtonian
  corrections of the GW phase, we restrict the measurement error estimation to $\mathcal{M}_z$ 
  for calculations evaluated for comparison in which we neglect precession.
  However, generally the assumed precessing eccentric waveform model depends on two independent combinations of component 
  masses: $\mathcal{M}_z$ sets the inspiral rate, and $M_{\mathrm{tot},z}$ sets both the apsidal
  precession rate and the final frequency at the LSO. We calculate the relative errors for both
  of these mass parameters and for the precessing eccentric waveform model \mbox{$\langle \Delta 
  \mathcal{M}_z^2 \rangle^{1/2}/ \mathcal{M}_z = \langle( \Delta \ln \mathcal{M}_z)^2 \rangle^
  {1/2}$}, and similarly for $M_{\mathrm{tot},z}$. 

 \item $t_c$, $\Phi_c$, and $\gamma_0$: These parameters only enter in the complex phase of the
  waveform through $\Psi_n$ and $\Psi_n^{\pm}$ (see Equations (\ref{eq:PsiN}) and (\ref{eq:Psipm})), 
  but do not affect the SNR. Since these parameters are responsible for an overall phase shift of
  the waveform, we do not randomize their values but assume the fiducial value $t_c = \Phi_c = 
  \gamma_c = 0$ for each binary in the Monte Carlo sample.

 \item $e_\mathrm{LSO}$: The adopted eccentric inspiral waveform model depends explicitly on 
  the final eccentricity at the LSO, see Equation (\ref{eq:defeLSO}). This quantity parameterizes 
  the evolutionary path of the binary during its eccentric inspiral in the $(\rho_\mathrm{p},e)$
  plane as shown in Equations (\ref{eq:rhop}) and (\ref{eq:defeLSO}); see also Figure 3 in 
  \citet{KocsisLevin2012} for illustration. In fact, any segment of the evolutionary path 
  $\rho_\mathrm{p}(e)$ specifies the value of $e_\mathrm{LSO}$ uniquely. Conversely, $e_
  \mathrm{LSO}$ specifies $\rho_\mathrm{p}(e)$, which sets a constraint on the possible values
  of $(\rho_{\mathrm{p}0},e_0)$, if the post-Newtonian binary inspiral model is extrapolated 
  backwards in time. Indeed, in some cases this is the only indirect information we may have 
  on the formation parameters $(\rho_{\mathrm{p}0},e_0)$. In particular, $e_0\leq 1$ puts an 
  upper bound on $\rho_{\mathrm{p}0}$ for a given $e_\mathrm{LSO}$.\footnote{When studying 
  the measurement errors for non-precessing eccentric binaries, the waveform depends 
  explicitly on a single combination of $e_\mathrm{LSO}$ and $M_{\mathrm{tot},z}$ parameters
  $c_0$ (Section \ref{sec:TimeWaveMod}). Therefore, we use $c_0$ for the \emph{NoPrec} model 
  to avoid a singularity of the Fisher matrix.}
 
 \item $e_0$: We choose several $e_0$ values from the highly eccentric ($e_0 \geq 0.9$) limit when
  discussing the $e_0$ dependence of measurement errors (see Section \ref{subsec:ResParamDist}). 
  However, we restrict to $e_0 = 0.9$ for calculations of a large survey of binaries.\footnote{We 
  note that the $e_0$ dependence of the waveform is due to the truncation of the time-domain 
  waveform for times when $e<e_0$.}
 
 \item $\rho_{\mathrm{p}0}$:  We examine two values for the dimensionless initial pericenter 
  distance $\rho_{\mathrm{p}0} = \lbrace 10,20 \rbrace$, and the circular limit corresponds to 
  $\rho_{\mathrm{p}0} \rightarrow \infty$ \citep{OLearyetal2009}. These values are likely for 
  sources that form through the GW capture mechanism in high velocity dispersion environments 
  such as GN as shown in \citet{OLearyetal2009,Gondanetal2017} or the core collapsed regions of 
  star clusters without a central massive black hole \citep{Kocsisetal2006,AntoniniRasio2016}.
 \end{itemize} 

 If the peak GW frequency of the initial orbit is large enough to be in the detectors' sensitive 
 frequency band then $\rho_{\mathrm{p}0}$ and $e_0$ are directly measurable due to the truncation
 of the time-domain waveform for times when $e<e_0$ and $\rho_{\mathrm{p}}>\rho_{\mathrm{p}0}$.
 In the opposite case only a lower limit may be given for $\rho_{\mathrm{p}0}$, which corresponds
 to $e_0 \rightarrow 1$ \citep{KocsisLevin2012}.

 In summary, we use the following free parameters in the Fisher matrix analysis:
\begin{align}
   \bm{\lambda}_\mathrm{Prec}=& 
   \lbrace \mathrm{ln} (D_\mathrm{L}), \mathrm{ln} 
  (\mathcal{M}_z), \mathrm{ln} (M_{{\rm tot},z}), \theta_N,\phi_N,\theta_L,\phi_L,e_0, 
  \nonumber\\ & \quad 
  e_\mathrm{LSO}, t_c,\Phi_c,\gamma_c \rbrace \, .
\end{align}
 Given these parameters, other parameters' marginalized measurement errors may be determined
 by linear combinations of the covariance matrix based on Equation (\ref{eq:Sigma}). For example, 
 $\rho_{\mathrm{p}0}$ is given by $e_0$ and $e_\mathrm{LSO}$ using Equations (\ref{eq:rhop}) and
 (\ref{eq:defeLSO}). Its measurement error is 
\begin{align}  \label{eq:deltarhop0}
 \left\langle(\Delta \rho_{\mathrm{p}0})^2 \right\rangle
 = & \left( \frac{\partial \rho_\mathrm{p}(e_0,e_\mathrm{LSO})}{\partial e_0} \right)^2 \left\langle (\Delta e_0)^2
 \right\rangle
   \nonumber\\ 
   & + \left( \frac{\partial \rho_\mathrm{p}(e_0,e_\mathrm{LSO})}{\partial e_\mathrm{LSO}} \right)^2 
  \left\langle (\Delta e_\mathrm{LSO})^2 \right\rangle
  \nonumber\\
  & + 2 \frac{\partial \rho_\mathrm{p} (e_0,e_\mathrm{LSO})}{\partial e_0} \frac{\partial \rho_\mathrm{p}(e_0,
  e_\mathrm{LSO})}{\partial e_\mathrm{LSO}} \left\langle \Delta e_0 \Delta e_\mathrm{LSO} \right\rangle\,.
\end{align}
 The parameter estimation errors of individual component masses or the mass ratio can be estimated 
 similarly using $\Delta M_{\rm tot}$ and $\Delta \mathcal{M}$ after inverting $M_{\rm tot}
 (m_a, m_b)$ and $\mathcal{M}(m_a, m_b)$.

\section{Results}
\label{sec:NumRes}
 
 The measurement errors depend on the sky position of the source with respect to the detectors 
 and on the relative orientation of the binary. We generate random Monte Carlo samples of $\sim 
 4500$ binaries by drawing from isotropic distributions of the sky position and of the binary
 orientation normal vector. We present the results for the $\mathrm{SNR}_\mathrm{tot}$ for 
 detecting precessing highly eccentric BH binaries with the GW detector network described 
 in Table \ref{tab:DetCord} and for the expected parameter measurement errors.
 
 \begin{figure} 
    \centering 
    \includegraphics[width=79mm]{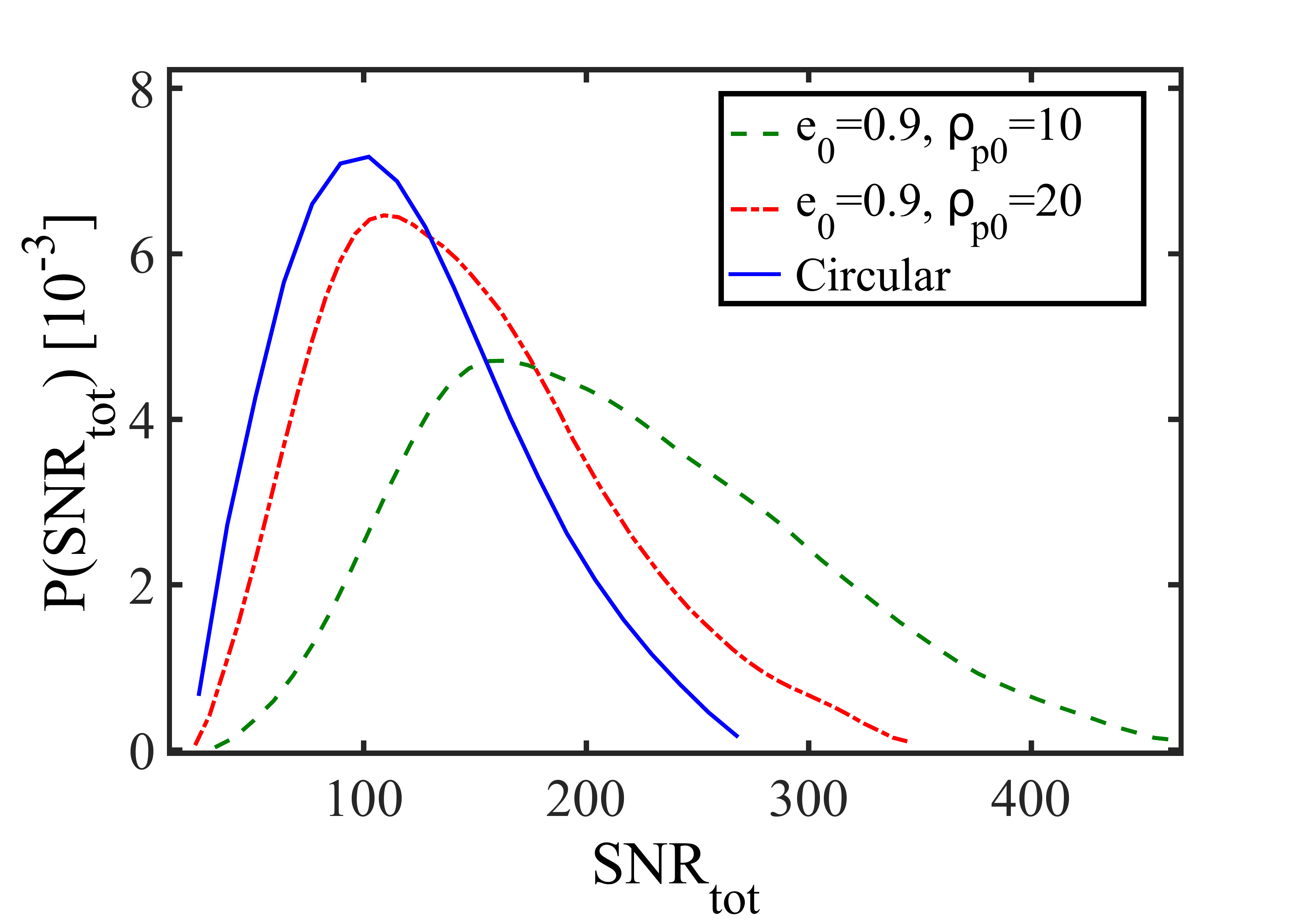} 
\caption{  Smoothed probability density function of the total network signal-to-noise ratio 
         ($\mathrm{SNR}_\mathrm{tot}$) of gravitational-wave detection from  $30 \, \Msun-30 \, 
         \Msun$ precessing eccentric BH binaries with initial eccentricity $e_0=0.9$ and dimensionless 
         pericenter distance $\rho_{\mathrm{p}0} = 10$ and $20$ (green dashed and red dash-dot) and 
         similar binaries in the circular limit (blue solid) at luminosity distance $D_\mathrm{L} = 
         100 \, \mathrm{Mpc}$ with a random source direction and orientation. Distributions correspond 
         to a Monte Carlo sample of $4500$ binaries. Parameters of the assumed detector network are 
         given in Table \ref{tab:DetCord}. The medians of $\mathrm{SNR}_\mathrm{tot}$ distributions 
         are $108.7$, $202.7$, and $137.3$ in the circular limit and for binaries with $\rho_{\mathrm{p}0} 
         = 10$ and $20$, respectively. Systematically higher $\mathrm{SNR}_\mathrm{tot}$ values for 
         precessing highly eccentric BH binaries implies that they are detectable to a larger distance 
         compared to precessing eccentric BH binaries in the circular limit. } \label{fig:SNR_Prec} 
\end{figure}

\subsection{Signal-to-noise ratio distributions} 

 Figure \ref{fig:SNR_Prec} displays the distribution of the $\mathrm{SNR}_\mathrm{tot}$ for precessing
 highly eccentric BH binaries detected with the detector network described in Table \ref{tab:DetCord}, 
 assuming binary parameters $m_A=m_B=30 \, \Msun$, $e_0=0.9$, $\rho_{\mathrm{p}0} = \lbrace 10, 20 
 \rbrace$, and similar binaries in the circular limit (see Appendix \ref{sec:circlim} for details).
 Generally, similar to the results of \citet{OLearyetal2009} (see Figure 11 therein, which corresponds
 to a single aLIGO detector), the $\mathrm{SNR}_\mathrm{tot}$ is systematically higher for binaries with
 $\rho_{\mathrm{p}0} = 10$ than for binaries with $\rho_{\mathrm{p}0} = 20$. We find that increasing 
 the initial eccentricity from $e_0=0.9$ to $0.97$ for fixed $\rho_{\mathrm{p}0}$ does not change the 
 $\mathrm{SNR}_\mathrm{tot}$ significantly (see Table \ref{tab:PrecBine0rho0Dep} below), hence we expect
 that the distribution of $\mathrm{SNR}_\mathrm{tot}$ for fixed $\rho_{\mathrm{p}0}$ converges in the $e_0 
 \rightarrow 1$ limit. This is expected since Figure 10 in \citet{OLearyetal2009} shows that a low amount
 of the $\mathrm{SNR}_\mathrm{tot}$ accumulates near $e_0 \approx 1$ for low to moderately high BH masses.  

\begin{figure}
    \center
    \includegraphics[width=79mm]{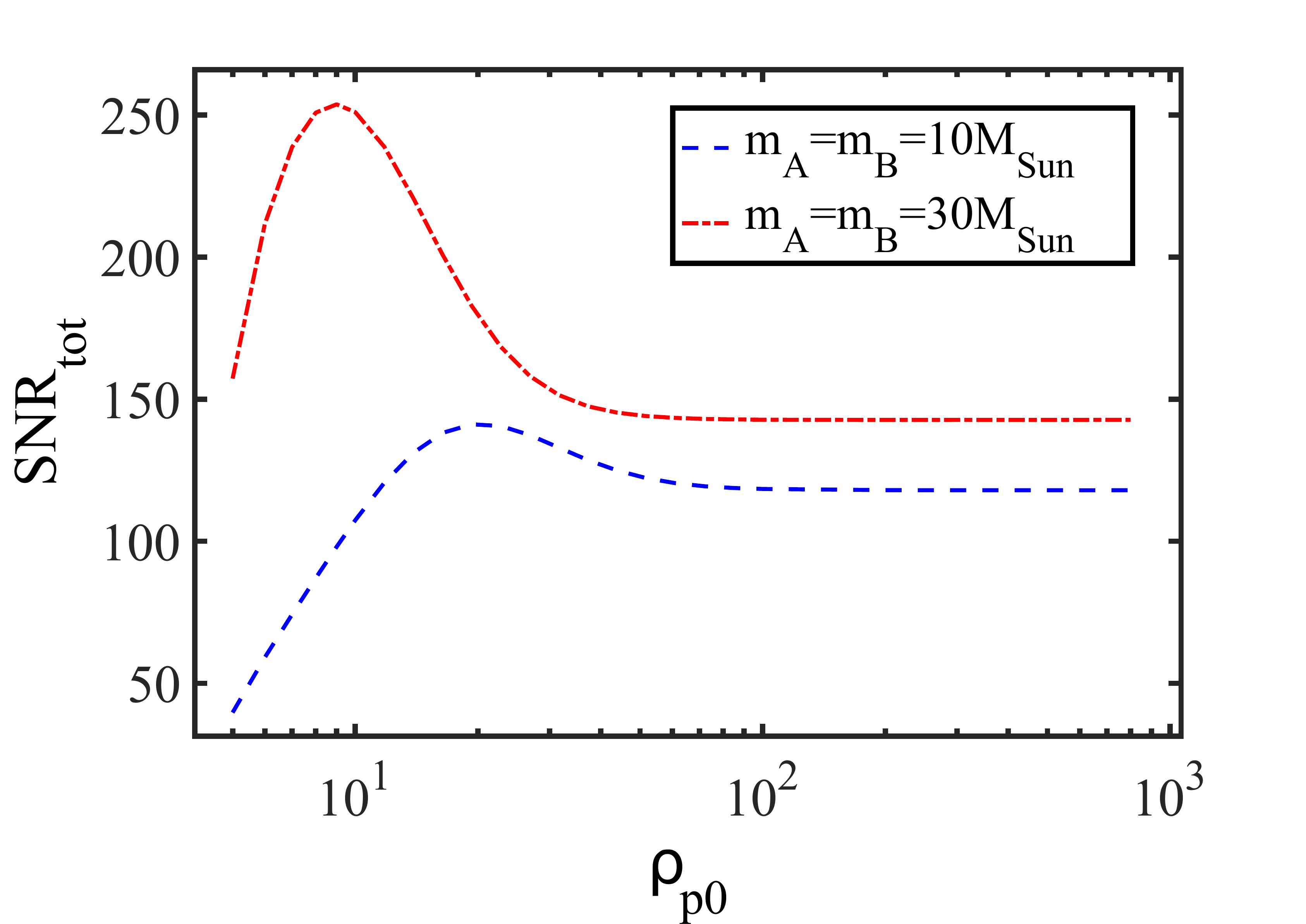}
\caption{ The $\mathrm{SNR}_\mathrm{tot}$ of precessing eccentric BH binaries as a function of 
          their initial dimensionless pericenter distance $\rho_{\mathrm{p}0}$. Parameters of
          the assumed detector network are given in Table \ref{tab:DetCord}. Here the luminosity 
          distance is $D_\mathrm{L}=100 \, \mathrm{Mpc}$, the initial eccentricity is $e_0 = 0.9$, 
          the source direction and orientation angular parameters are fixed at $\theta_N=\pi/2$, 
          $\phi_N=\pi/3$, $\theta_L=\pi/4$, and $\phi_L =\pi/5$. Depending on the binary mass, 
          the $\mathrm{SNR}_ \mathrm{tot}$ has a maximum at $\rho_{\mathrm{p}0}$ between $9$ and 
          $20$, and converges asymptotically to the value of precessing eccentric binaries in 
          the circular limit for high $\rho_{\mathrm{p}0}$. Note that we find similar trends 
          with $\rho_{\mathrm{p}0}$ for other random choices of binary direction and orientation 
          (not shown). }  \label{fig:SNRtot_rho}
\end{figure} 

 Figure \ref{fig:SNRtot_rho} shows that the $\mathrm{SNR}_\mathrm{tot}$ increases rapidly with $\rho_{
 \mathrm{p}0}$ for low $\rho_{\mathrm{p}0}$, has a maximum between $\rho_{\mathrm{p}0}\sim 9$ and 
 $20$, and decreases for higher $\rho_{\mathrm{p}0}$ approaching the circular binary limit for $\rho
 _{\mathrm{p}0} \rightarrow \infty$. These findings may be understood qualitatively as follows. 
 Within $\rho_{\mathrm{p}0}\leq 40.9 (M_{\rm tot,z}/20 \, \Msun)^{-2/3}$ the binary forms with a 
 characteristic frequency above $10 \, \mathrm{Hz}$ in the detector band \citep[see Equation 
 (59) in][]{Gondanetal2017}. The rapid decrease of the $\mathrm{SNR}_\mathrm{tot}$ for decreasing 
 $\rho_{\mathrm{p}0}<9$ is due to the fact that we neglect the GWs of the first hyperbolic encounter 
 \citep{Kocsisetal2006}. The decrease of the $\mathrm{SNR}_\mathrm{tot}$ at high $\rho_{\rm p0}$ is 
 due to the fact that part of the GW spectrum falls outside of the detectors' sensitive frequency 
 band. For very large $\rho_{\mathrm{p},0}$, the binary becomes circular by the time it enters the 
 detectors' sensitive frequency band, and a significant fraction of the $\mathrm{SNR}_\mathrm{tot}$ 
 accumulates only in the $n=2$ harmonic (Figure \ref{fig:time}), which explains the flat asymptotics 
 for high $\rho_{\mathrm{p}0}$. Decreasing $\rho_{\mathrm{p}0}$ from high to moderate values higher 
 harmonics start to contribute to the $\mathrm{SNR}_\mathrm{tot}$ (Figure \ref{fig:time}), which 
 explains the increase of the $\mathrm{SNR}_\mathrm{tot}$. A combination of these arguments leads 
 to the peak of the $\mathrm{SNR}_\mathrm{tot}$ at an intermediate $\rho_{\mathrm{p}0}$ value seen
 in Figure \ref{fig:SNRtot_rho}. However, note that in addition to neglecting the 
 initial hyperbolic encounter and the final coalescence/ringdown segments of the signal, our waveform
 model also neglects contributions of spherical moments beyond the quadrupole-order and deviations from
 a precessing Keplerian orbit \citep{Davisetal1972,Bertietal2010,Healyetal2016}. This approximation may
 not be valid for low $\rho_{\mathrm{p}0}$ (particularly for $\rho_{\mathrm{p}0} \leq 10 $). The $\mathrm{SNR}
 _\mathrm{tot}$ is expected to be underestimated in this region in Figure \ref{fig:Delta_rho}.

\begin{table*}
\centering  
   \begin{tabular}{@{}cccccccccc}
     \hline
      $\rho_\mathrm{p0}$ & $10$ &$10$& $10$ & $20$ & $20$ & $20$ & Circular& Circular& Circular \\
      quantile & $10\%$ & $50\%$ & $90\%$ & $10\%$ & $50\%$ & $90\%$ &$10\%$ & $50\%$ & $90\%$ \\
     \hline\hline
   $\Delta ( \mathrm{ln} D_\mathrm{L})$  &  $1.13\,(-2)$  &  $2.49\,(-2)$  &  $7.21\,(-2)$  &  $1.89\,(-2)$  &  $4.36\,(-2)$  &  $0.154$  &  $8.81\,(-3)$  &  $5.86\,(-2)$  &  $0.311$  \\
   $\Delta\Omega_N \,\, [\mathrm{sr}]$  &  $1.13\,(-5)$  &  $1.46\,(-4)$  &  $7.23\,(-4)$  &  $2.21\,(-4)$  &  $7.28\,(-4)$  &  $2.89\,(-3)$  &  $5.65\,(-5)$  &  $1.58\,(-3)$  &  $6.67\,(-3)$  \\
   $\Delta\Omega_L \,\, [\mathrm{sr}]$  &  $7.89\,(-5)$  &  $3.85\,(-3)$  &  $0.13$  &  $2.21\,(-5)$  &  $1.13\,(-2)$  &  $0.85$  &  $4.36\,(-4)$  &  $2.21\,(-2)$  &  $9.6$  \\
   $a_N \,\, [\mathrm{deg}]$  &  $0.14$  &  $0.62$  &  $2.42$  &  $0.57$  &  $1.36$  &  $4.07$  &  $0.28$  &  $2.01$  &  $5.16$  \\
   $b_N \,\, [\mathrm{deg}]$  &  $0.11$  &  $0.22$  &  $0.48$  &  $0.26$  &  $0.52$  &  $0.98$  &  $0.41$  &  $0.76$  &  $1.44$  \\
   $a_L \,\, [\mathrm{deg}]$  &  $0.96$  &  $2.78$  &  $20.04$  &  $1.44$  &  $4.65$  &  $45.48$  &  $0.78$  &  $6.98$  &  $2.02\,(+2)$  \\
   $b_L \,\, [\mathrm{deg}]$  &  $0.55$  &  $1.33$  &  $7.51$  &  $0.84$  &  $2.46$  &  $20.58$  &  $0.32$  &  $3.41$  &  $50.63$  \\
   $\Delta \Phi_c \,\, [\mathrm{rad}]$  &  $9.27\,(-2)$ & $0.23$ & $0.71$  &  $0.29$  &  $0.66$  &  $2.44$  &  $0.36$  &  $0.72$  &  $57.91$  \\
   $\Delta t_c \,\, [\mathrm{ms}]$  &  $4.32\,(-2)$ & $8.41\,(-2)$ & $0.181$  &  $9.28\,(-2)$  & $0.167$  & $0.311$ & $9.41\,(-2)$ & $1.582$ & $3.011$  \\
   $\Delta ( \mathrm{ln} \mathcal{M}_z )$  &  $3.53\,(-5)$  &  $6.17\,(-5)$  &  $1.71\,(-4)$  &  $1.04\,(-5)$  &  $1.81\,(-5)$  &  $3.42\,(-5)$  &  $1.36\,(-3)$  &  $2.34\,(-3)$  &  $4.27\,(-3)$  \\
   $\Delta (\mathrm{ln} M_{{\rm tot},z})$  &  $5.42\,(-4)$  &  $9.51\,(-4)$  &  $2.43\,(-3)$  &  $2.82\,(-4)$  &  $4.81\,(-4)$  &  $9.18\,(-4)$  &  $5.88\,(-3)$  &  $1.13\,(-2)$  &  $1.81\,(-2)$  \\
   $\Delta e_\mathrm{LSO}$  &  $1.18\,(-4)$  &  $2.16\,(-4)$  &  $5.83\,(-4)$  &  $2.39\,(-5)$  &  $3.19\,(-5)$  &  $5.88\,(-5)$  &  $-$  &  $-$  &  $-$  \\
   $\Delta e_0$  &  $1.44\,(-3)$  &  $2.16\,(-3)$  &  $3.95\,(-3)$  &  $1.72\,(-3)$  &  $2.91\,(-3)$  &  $5.79\,(-3)$  &  $-$  &  $-$  &  $-$  \\
   $\Delta \rho_{\mathrm{p}0}$  &  $6.64\,(-3)$  &  $1.08\,(-2)$  &  $2.28\,(-2)$  &  $1.33\,(-2)$  &  $2.29\,(-2)$  &  $4.58\,(-2)$  &  $-$  &  $-$  &  $-$  \\
   $\Delta \gamma_c \,\, [\mathrm{rad}]$  &  $0.04$  &  $0.11$  &  $0.47$  &  $0.14$  &  $0.34$  &  $1.51$  &  $-$  &  $-$  &  $-$  \\
\hline\hline
   \end{tabular} 
   \caption{ \label{tab:PrecBin} The $10\%$, $50\%$, and $90\%$ quantile of measurement errors for parameters
   of $30 \, \Msun-30 \, \Msun$ precessing eccentric BH binaries with initial eccentricity $e_0=0.9$ and
   dimensionless pericenter distance $\rho_{\mathrm{p}0}=10$ and $20$, and circular binaries at distance $D_
   \mathrm{L}=100 \, \mathrm{Mpc}$, random source sky location and orientation using the detector network 
   in Table \ref{tab:DetCord}. Here $(\Omega_N, a_N, b_N)$ are respectively the area, semi-major and semi-minor 
   axes of the 2D error ellipse corresponding to the source's sky direction, and similarly for $(\Omega_L, 
   a_L, b_L)$ describing the source's orbital plane orientation (i.e. angular momentum vector direction). 
   Note that $e_{\rm LSO}=0.187$ and $0.059$ for $\rho_{\mathrm{p}0}=10$ and $20$, respectively, if $e_0=0.9$
   is assumed. In the circular limit the binary forms outside of the sensitive frequency band of the detector 
   network, and $\Delta e_0\rightarrow \infty$ and $\Delta \rho_{\mathrm{p}0}\rightarrow \infty$. We adopt the 
   following notation in the table: $1.13\,(-2)=1.13 \times 10^{-2}$. }
\end{table*}

\begin{table*}
\centering  
   \begin{tabular}{@{}ccccccc}
     \hline
      $\rho_{\mathrm{p}0}$ & $10$ & $10$ & $10$ & $20$ & $20$ & $20$  \\
      $e_0$ & $0.9$ &$0.95$& $0.97$ & $0.9$ & $0.95$ & $0.97$  \\
      $e_\mathrm{LSO}$  &  $0.1872$  &  $0.1932$  &  $0.1956$  &  $5.89\,(-2)$   &  $6.07\,(-2)$ &  $6.15\,(-2)$ \\
      $\mathrm{SNR}_\mathrm{tot}$  &  $251.1$  &  $257.7$  &  $260.4$  &  $179.4$   &  $182.1$ &  $182.9$ \\
     \hline\hline
   $\Delta ( \mathrm{ln} D_\mathrm{L})$  &  $3.92\,(-2)$  &  $3.79\,(-2)$ &  $3.76\,(-2)$  &  $6.41\,(-2)$  
   &  $6.31\,(-2)$   &  $6.27\,(-2)$  \\
   $\Delta\Omega_N \,\, [\mathrm{sr}]$  &  $8.37\,(-5)$  &  $7.78\,(-5)$  &  $7.53\,(-5)$  &  $4.06\,(-4)$ 
   &  $3.91\,(-4)$   &  $3.85\,(-4)$  \\
   $\Delta\Omega_L \,\, [\mathrm{sr}]$  &  $9.68\,(-3)$  &  $9.07\,(-3)$  &  $8.84\,(-3)$  &  $2.74\,(-2)$ 
   &  $2.65\,(-2)$  &  $2.61\,(-2)$  \\
   $a_N \,\, [\mathrm{deg}]$  &  $0.501$  &  $0.484$  &  $0.477$  &  $0.994$  &  $0.977$  &  $0.970$  \\
   $b_N \,\, [\mathrm{deg}]$  &  $0.174$  &  $0.168$  &  $0.165$  &  $0.427$  &  $0.418$  &  $0.415$  \\
   $a_L \,\, [\mathrm{deg}]$  &  $4.31$  &  $4.18$  &  $4.13$    &  $6.96$ &  $6.85$ &  $6.81$  \\
   $b_L \,\, [\mathrm{deg}]$  &  $2.35$  &  $2.26$  &  $2.23$  &  $4.12$  &  $4.04$  &  $4.01$  \\   
   $\Delta \Phi_c \,\, [\mathrm{rad}]$  &  $0.331$ & $0.313$  &  $0.308$  & $1.102$ &  $1.049$  &  $1.031$  \\
   $\Delta t_c \,\, [\mathrm{ms}]$  &  $9.68\,(-2)$ & $9.08\,(-2)$  & $8.88\,(-2)$  & $0.198$ &  $0.192$ 
   & $0.189$   \\
   $\Delta ( \mathrm{ln} \mathcal{M}_z )$  &  $4.61\,(-5)$  &  $1.89\,(-5)$  &  $1.39\,(-5)$  &  $1.38\,(-5)$ 
   &  $6.03\,(-6)$  &  $4.33\,(-6)$  \\
   $\Delta (\mathrm{ln} M_{{\rm tot},z})$  &  $7.09\,(-4)$  &  $5.62\,(-4)$  &  $4.85\,(-4)$  &  $3.67\,(-4)$ 
   &  $2.87\,(-4)$  &  $2.38\,(-4)$  \\
   $\Delta e_\mathrm{LSO}$  &  $1.61\,(-4)$  &  $1.28\,(-4)$ &  $1.11\,(-4)$ &  $2.43\,(-5)$  &  $1.93\,(-5)$ 
   &  $1.62\,(-5)$   \\
   $\Delta e_0$  &  $1.95\,(-3)$  &  $1.74\,(-3)$  &  $1.69\,(-3)$ &  $2.25\,(-3)$  &  $2.17\,(-3)$  &  $2.14\,(-3)$   \\
   $\Delta \rho_{\rm p0}$  &  $8.33\,(-3)$  &  $7.45\,(-3)$  & $7.11\,(-3)$  &  $1.71\,(-2)$  &  $1.61\,(-2)$ 
   &  $1.56\,(-2)$   \\
   $\Delta \gamma_c \,\, [\mathrm{rad}]$  &  $0.169$  &  $0.162$ &  $0.159$  &  $0.556$   &  $0.529$  &  $0.520$    \\
\hline\hline
   \end{tabular}  
   \caption{ \label{tab:PrecBine0rho0Dep} Measurement errors for parameters of $30 \, \Msun-30 \, \Msun$ 
   precessing eccentric BH binaries with initial eccentricities $e_0=0.9$, $0.95$, and $0.97$ for initial
   dimensionless pericenter $\rho_{p0}=10$ and $20$, luminosity distance $D_\mathrm{L}=100 \, \mathrm{Mpc}$, 
   and arbitrarily fixed source direction $(\theta_N,\phi_N)=(\pi/2,\pi/3)$ and orientation $(\theta_L,
   \phi_L)=(\pi/4,\pi/5)$ for the detector network specified in Table \ref{tab:DetCord}. }
\end{table*}

\subsection{Parameter measurement errors} 
\label{subsec:ResParamDist} 
 
\begin{figure*}
    \centering 
    \includegraphics[width=79mm]{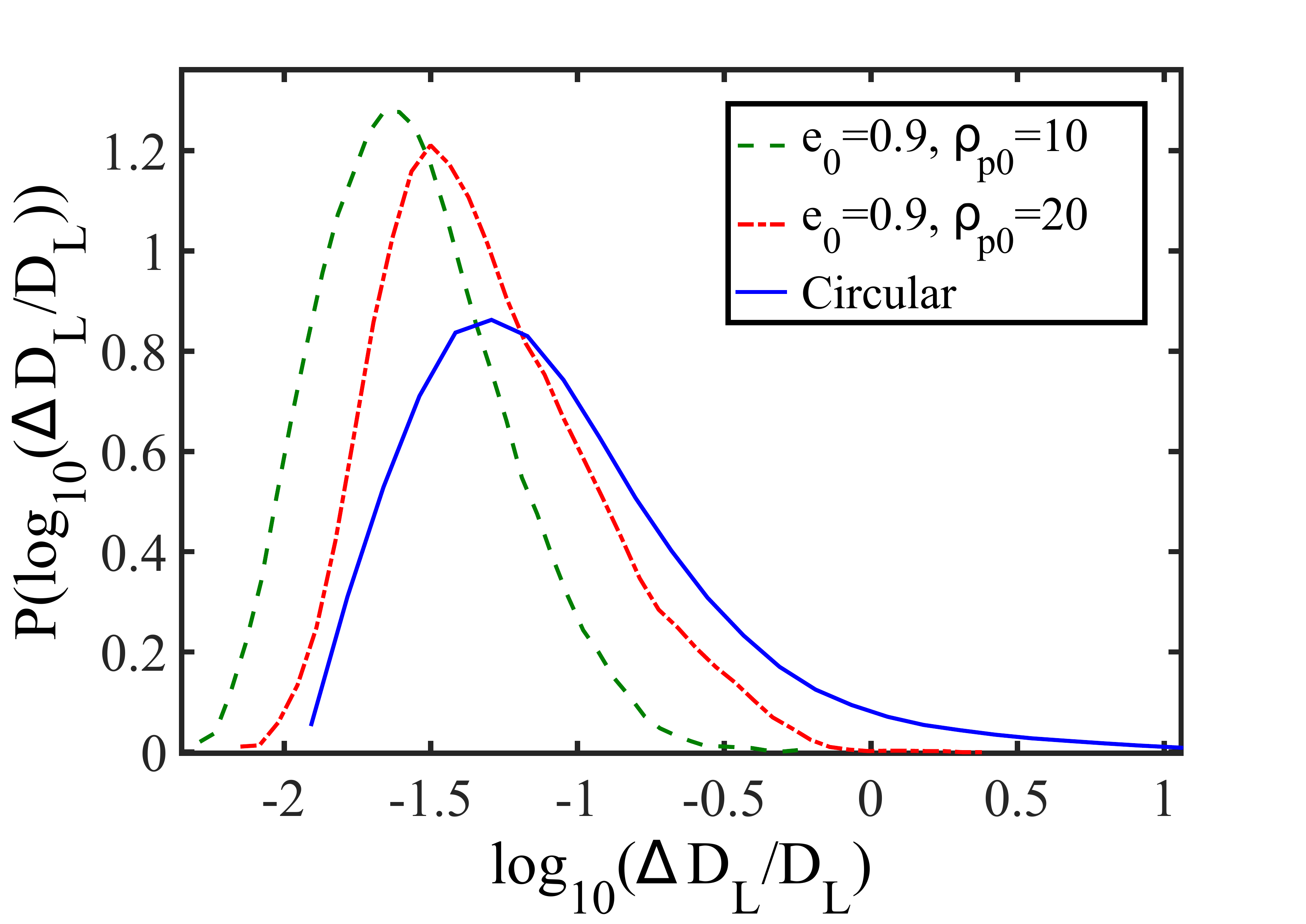} 
    \includegraphics[width=79mm]{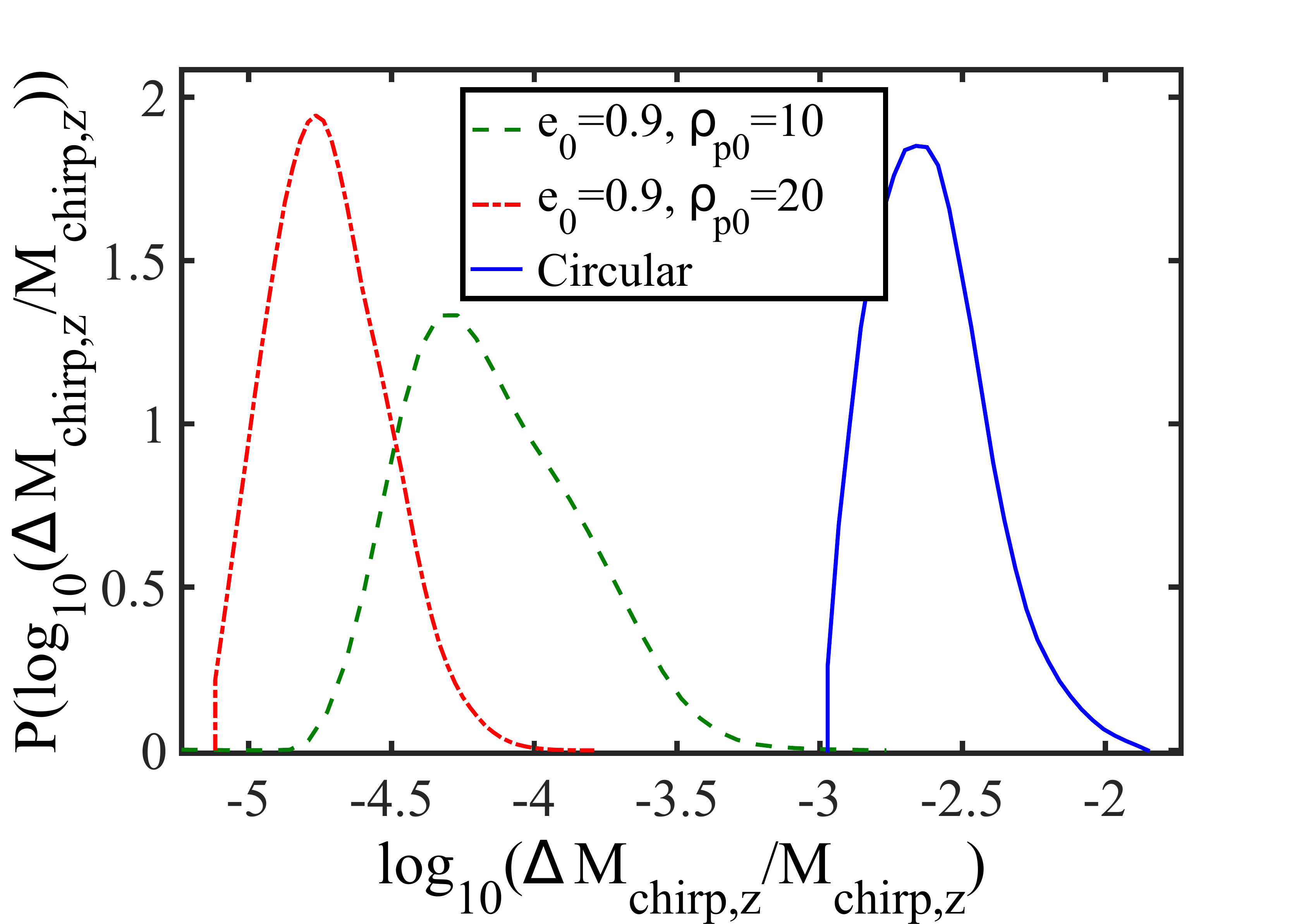}    
    \\
    \includegraphics[width=79mm]{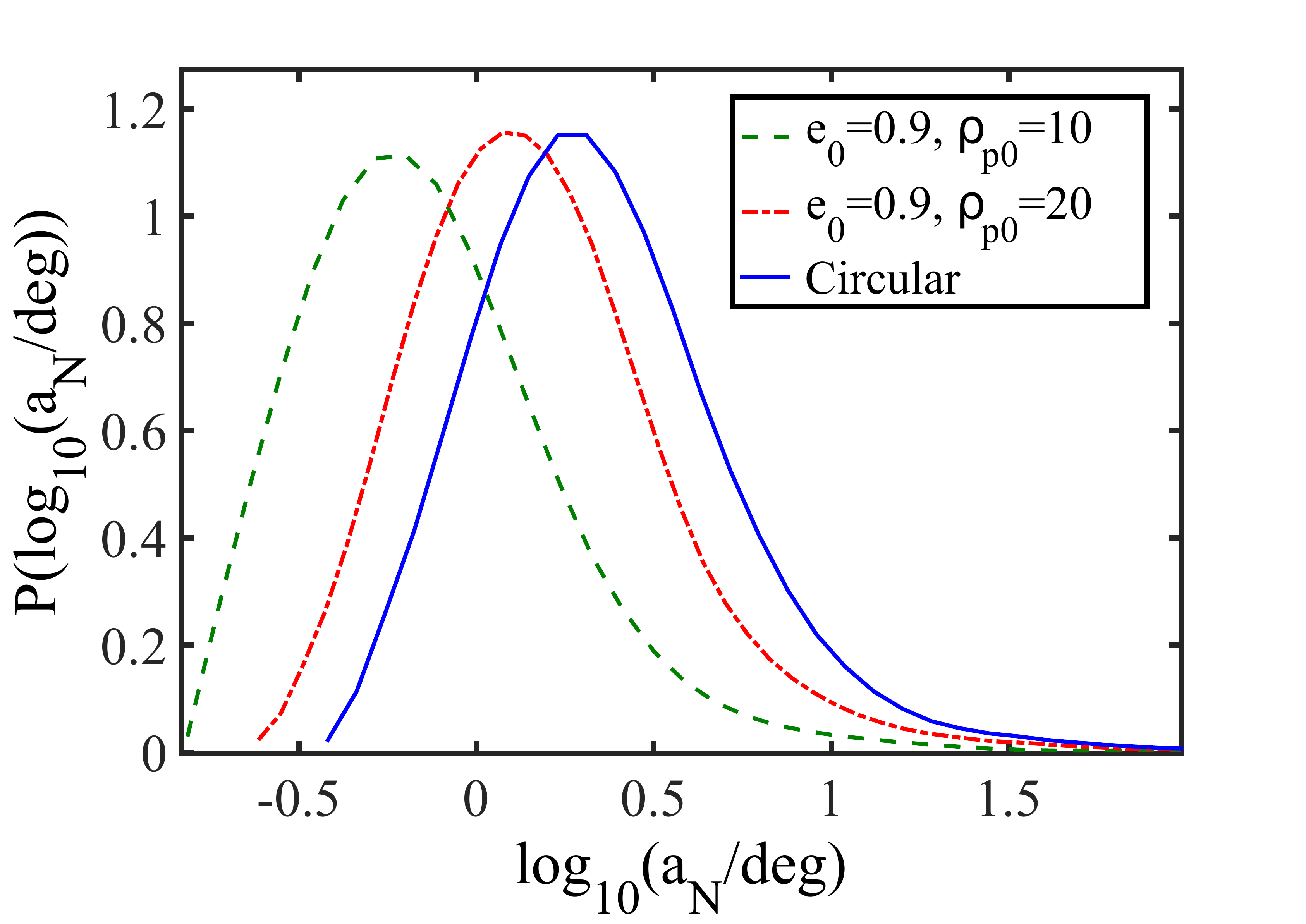}   
    \includegraphics[width=79mm]{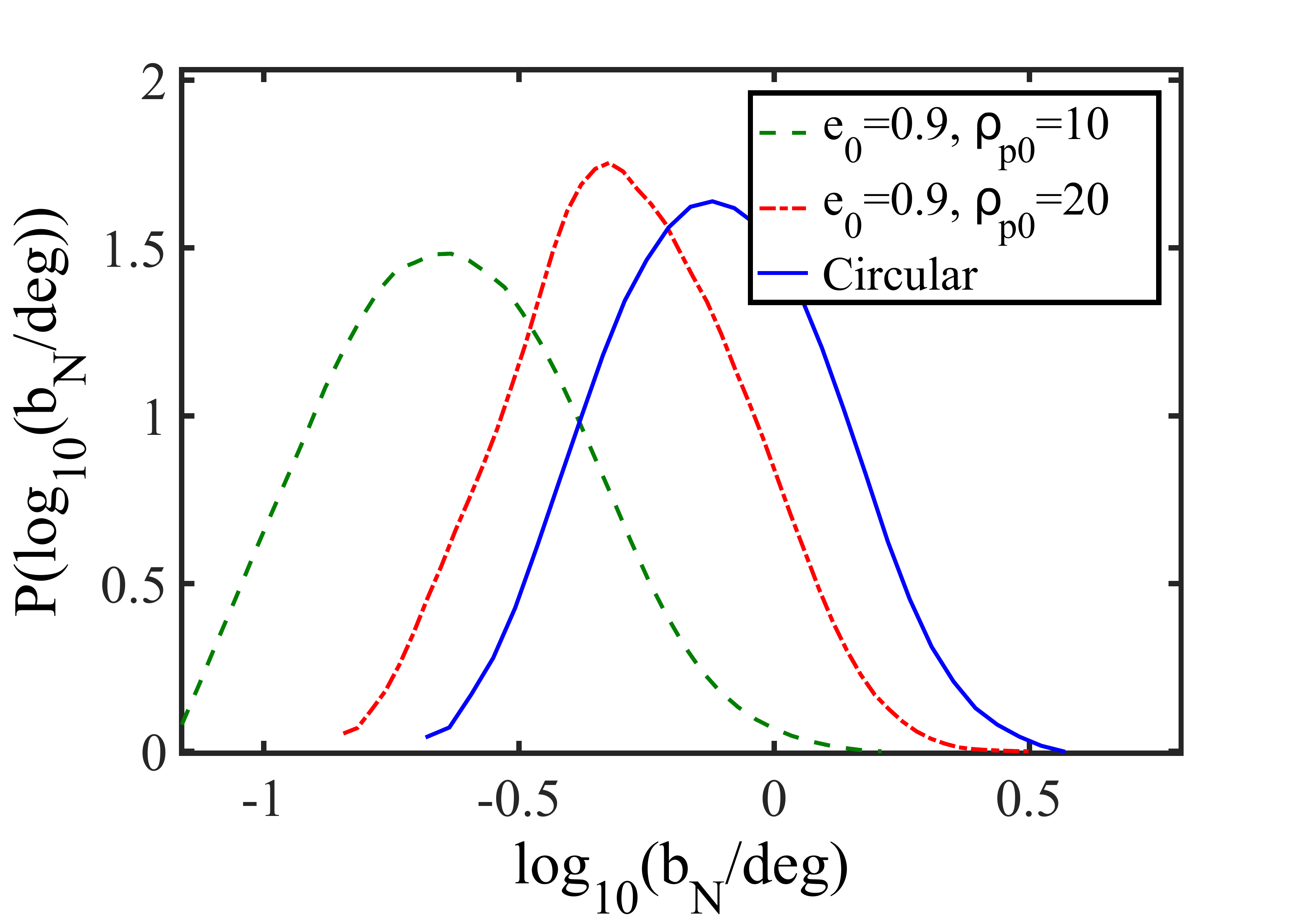}
    \\
    \includegraphics[width=79mm]{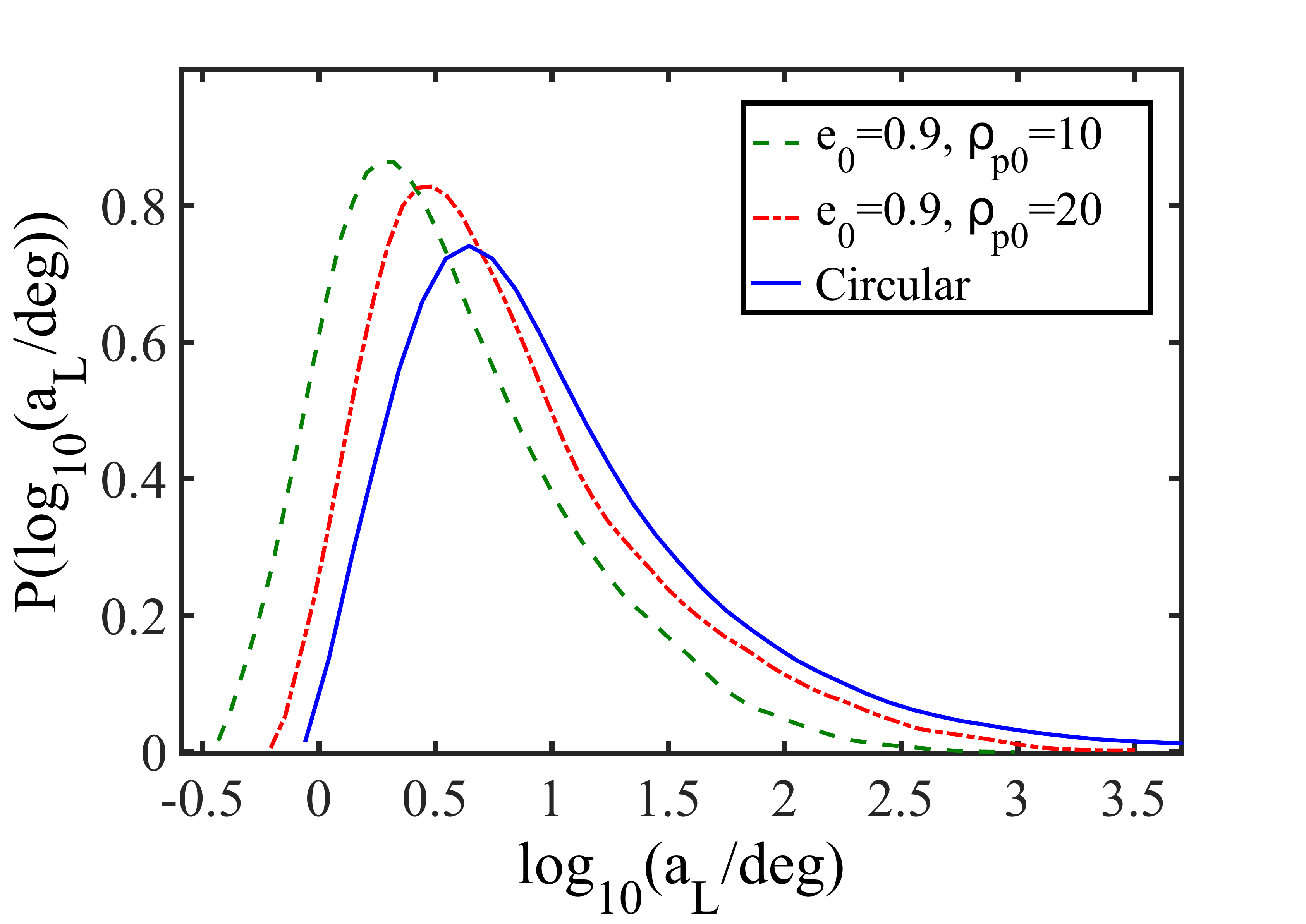}   
    \includegraphics[width=79mm]{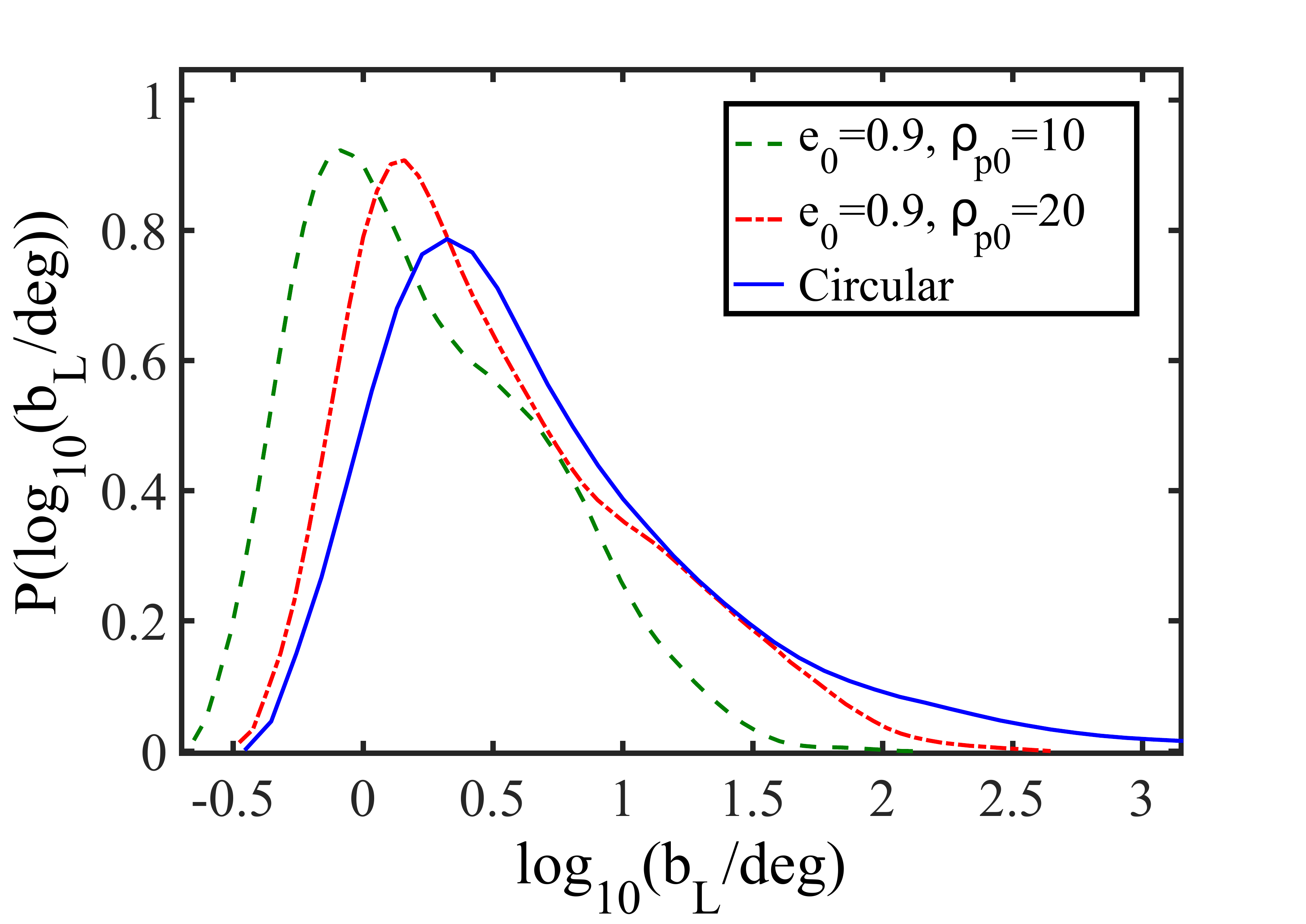}
\caption{ Smoothed distribution of the measurement errors of parameters measured for $30 \, \Msun-30 \,
 \Msun$ precessing eccentric BH binaries with initial eccentricity $e_0 = 0.9$ and dimensionless pericenter 
 distance $\rho_{\mathrm{p}0}=10$ and $20$ (green dashed and red dash-dot) and for similar binaries in 
 the circular limit (blue solid) at luminosity distance $D_\mathrm{L} = 100 \, \mathrm{Mpc}$ with a
 random source direction and binary orientation. We have assumed the detector network specified in 
 Table \ref{tab:DetCord}. \emph{Top row:} Distribution of the relative measurement error of luminosity 
 distance, $\Delta D_{\rm L} / D_{\rm L} = \Delta (\ln D_{\rm L})$, and redshifted chirp mass, $\Delta 
 \mathcal{M}_{z} / \mathcal{M}_{z} = \Delta (\ln \mathcal{M}_{z})$. \emph{Middle row:} Distribution of 
 semi-major axis of the sky position error ellipse, $a_N$, and its semi-minor axis, $b_N$. \emph{Bottom 
 row:} Distribution of the semi-major axis of the binary's orbital plane orientation error ellipse, $a_L$,
 and its semi-minor axis, $b_L$.  } \label{fig:Dist1}
\end{figure*}

\begin{figure*}
    \centering 
    \includegraphics[width=79mm]{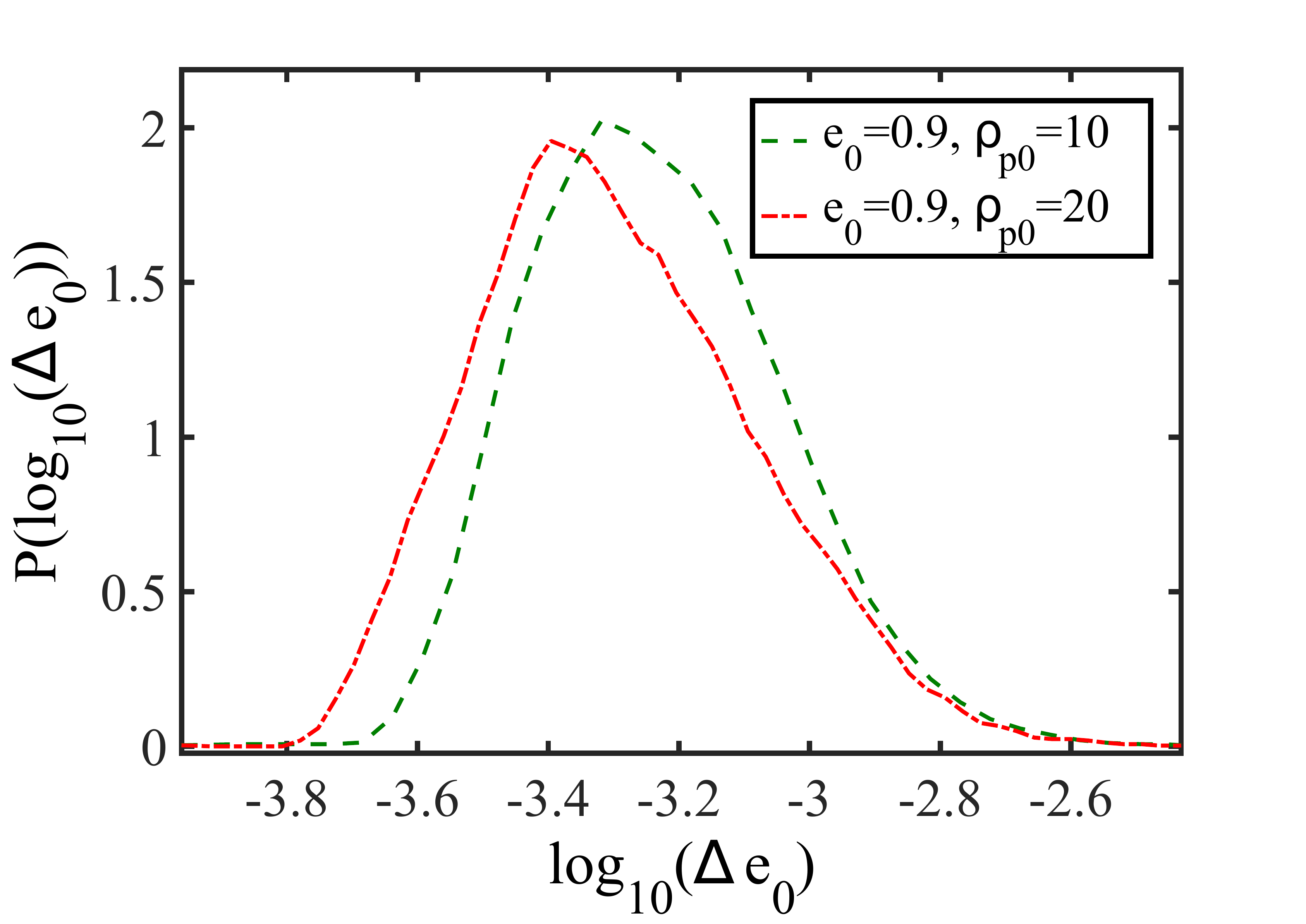}   
    \includegraphics[width=79mm]{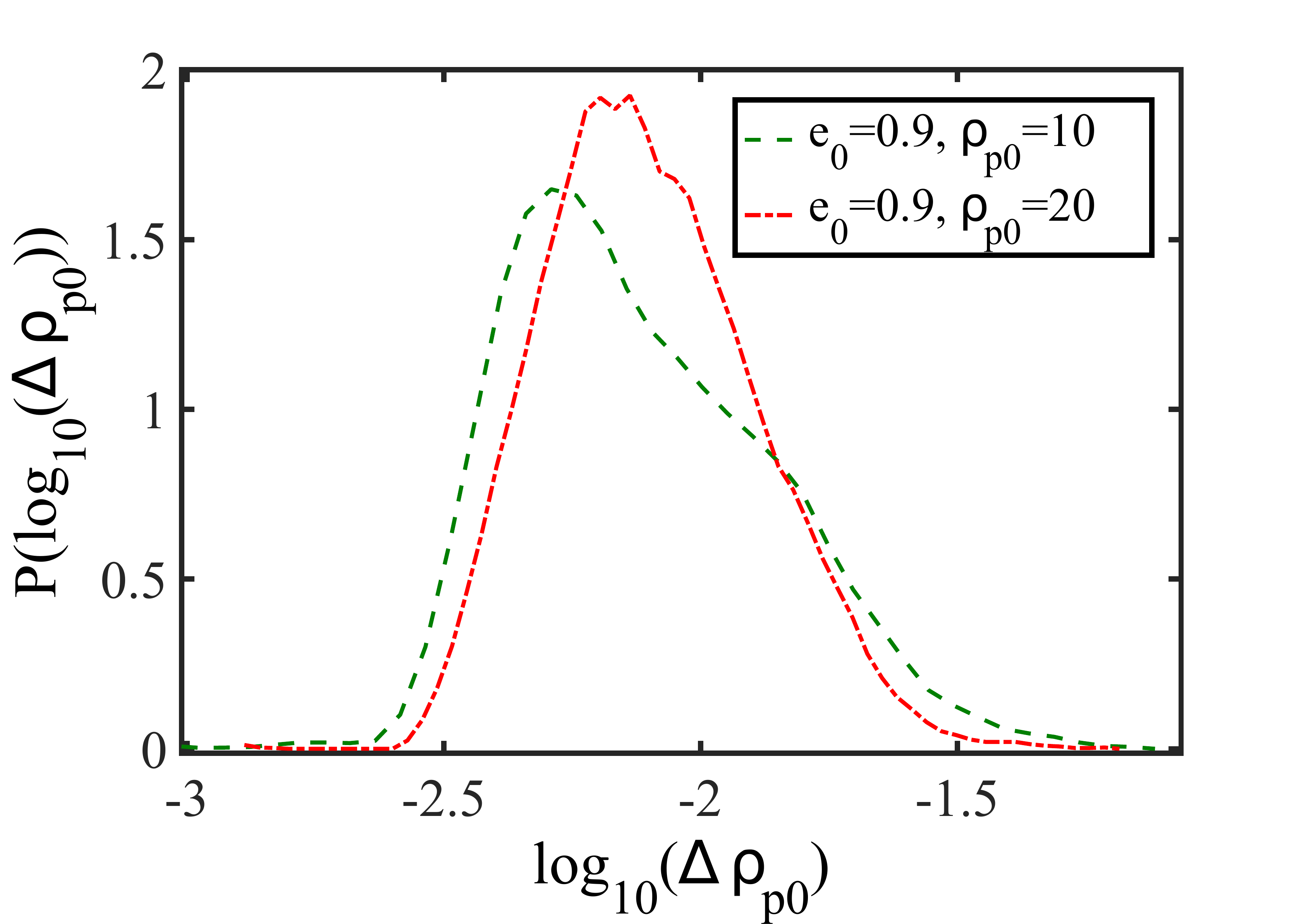}
    \\
    \includegraphics[width=79mm]{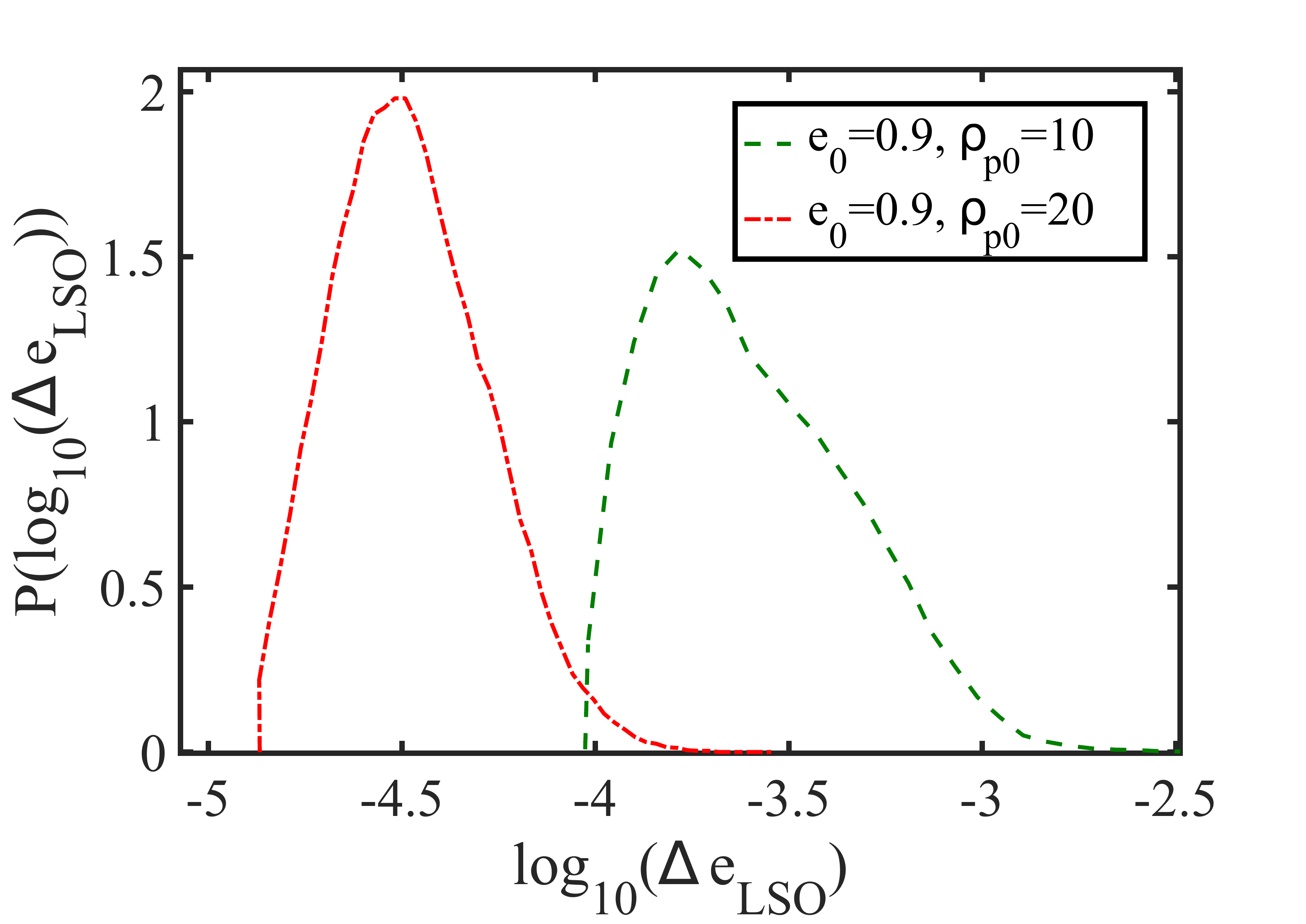}
\caption{ Smoothed distribution of the measurement errors of various parameters measured only for 
 precessing eccentric binaries. Similar to Figure \ref{fig:Dist1}, these distributions correspond
 to a Monte Carlo sample of $4500$ binaries with random source direction and binary orientation for 
 $30 \, \Msun - 30 \, \Msun$ precessing eccentric BH binaries with initial eccentricity $e_0 = 
 0.9$ and dimensionless pericenter distance $\rho_{\mathrm{p}0} = 10$ and $20$ (green dashed and red 
 dash-dot). The source distance is also fixed at $D_\mathrm{L} = 100\, \mathrm{Mpc}$, and the detector 
 network is specified in Table \ref{tab:DetCord}. \emph{Top left:} Distribution of the measurement 
 error in initial orbital eccentricity, $e_0$. \emph{Top right:} Distribution of the measurement error
 in initial dimensionless pericenter distance, $\rho_{\mathrm{p}0}$. \emph{Bottom:} Distribution of the 
 measurement error in the eccentricity at the last stable orbits, $e_{\rm LSO}$. }  \label{fig:Dist2}
\end{figure*} 

 We present the measurement accuracy for the final eccentricity at the LSO for parameters grouped as  
\begin{align} 
  \bm{\lambda}_\mathrm{slow} & = \lbrace \mathrm{ln} (D_\mathrm{L}), \theta_N, \phi_N, \theta_L, 
  \phi_L \rbrace \, ,  \\
  \bm{\lambda}_\mathrm{fast} & = \lbrace \Phi_c,  t_c, \mathrm{ln} (\mathcal{M}_z), 
  \mathrm{ln} (M_{{\rm tot},z}), e_\mathrm{LSO}, \gamma_c \rbrace \,\
\end{align} 
 \citep{Kocsisetal2007}. The $\bm{\lambda}_\mathrm{fast}$ fast parameters are related to the high 
 frequency GW phase, while the $\bm{\lambda}_\mathrm{slow}$ slow parameters appear only in the 
 slowly-varying amplitude of the GW signal. Slow parameters are mostly determined by a comparison 
 of the GW signals measured by the different detectors in the network. For the polar angles $(\theta_N,
 \phi_N)$ describing the source direction, we calculate the minor and major axes ($a_N,b_N$) of the 
 corresponding 2D sky location error ellipse and its area ($\Omega_N= \pi a_N b_N$), and we do the 
 same for the binary orientation error ellipse $(a_L,b_L)$ and its area ($\Omega_N=\pi a_L b_L$).

 Figures \ref{fig:Dist1} and \ref{fig:Dist2} show the distribution of the measurement errors for randomly
 chosen source sky position and binary orientation for $\rho_{\mathrm{p}0}=10$ and $20$, and for the circular
 limit $\rho_{\mathrm{p}0} \rightarrow \infty$ (see Appendix \ref{sec:circlim}), while Table \ref{tab:PrecBin}
 shows the $10 \%$, $50\%$, and $90\%$ quantiles of the error distributions. Compared to a $\rho_{\mathrm{p}0}
 = 20$ binary, a $\rho_{\mathrm{p}0} = 10$ binary is more eccentric throughout its evolution, which leads to
 a higher $\mathrm{SNR}_\mathrm{tot}$, and most of its measurement errors are smaller. There are, however, 
 exceptions to this finding: the fast parameters such as the mass parameters and the eccentricity have higher errors for $\rho_{\mathrm{p}0}
 = 10$ than for $\rho_{\mathrm{p}0} = 20$ (see discussion below). 
  
 Many of the binaries in galactic nuclei\footnote{particularly the heavy BHs therein \citep{Gondanetal2017}} form with very high $e_0$, close to unity, in single-single
 encounters due to GW emissions \citep{OLearyetal2009,Gondanetal2017}. However, similarly to the finding that 
 the $\mathrm{SNR}_\mathrm{tot}$ does not increase significantly for $1 > e_0 \geq 0.9$, we find that $\bm{\lambda}
 _\mathrm{slow}$ parameter errors do not improve due to the early very eccentric evolutionary period beyond $e > 
 0.9$ (repeated burst phase) compared to waveforms with $e_0=0.9$ as shown in Table \ref{tab:PrecBine0rho0Dep}. 
 However, some $\bm{\lambda}_\mathrm{fast}$ parameters' measurement errors improve more 
 significantly with $e_0 > 0.9$. In particular, the measurement errors of the mass parameters ($\mathcal{M}_z,
 M_{{\rm tot},z}$) improve by a factor of $\sim 2$, and the measurement error of $e_\mathrm{LSO}$ improves 
 by $\sim 50\%$ if increasing $e_0$ from $0.9$ for $0.97$. This difference is due to the fact that eccentricity
 modifies the GW phase significantly, which affects the determination of $\bm{\lambda}_\mathrm{fast}$ parameters
 only.

\begin{figure*}
    \centering 
    \includegraphics[width=79mm]{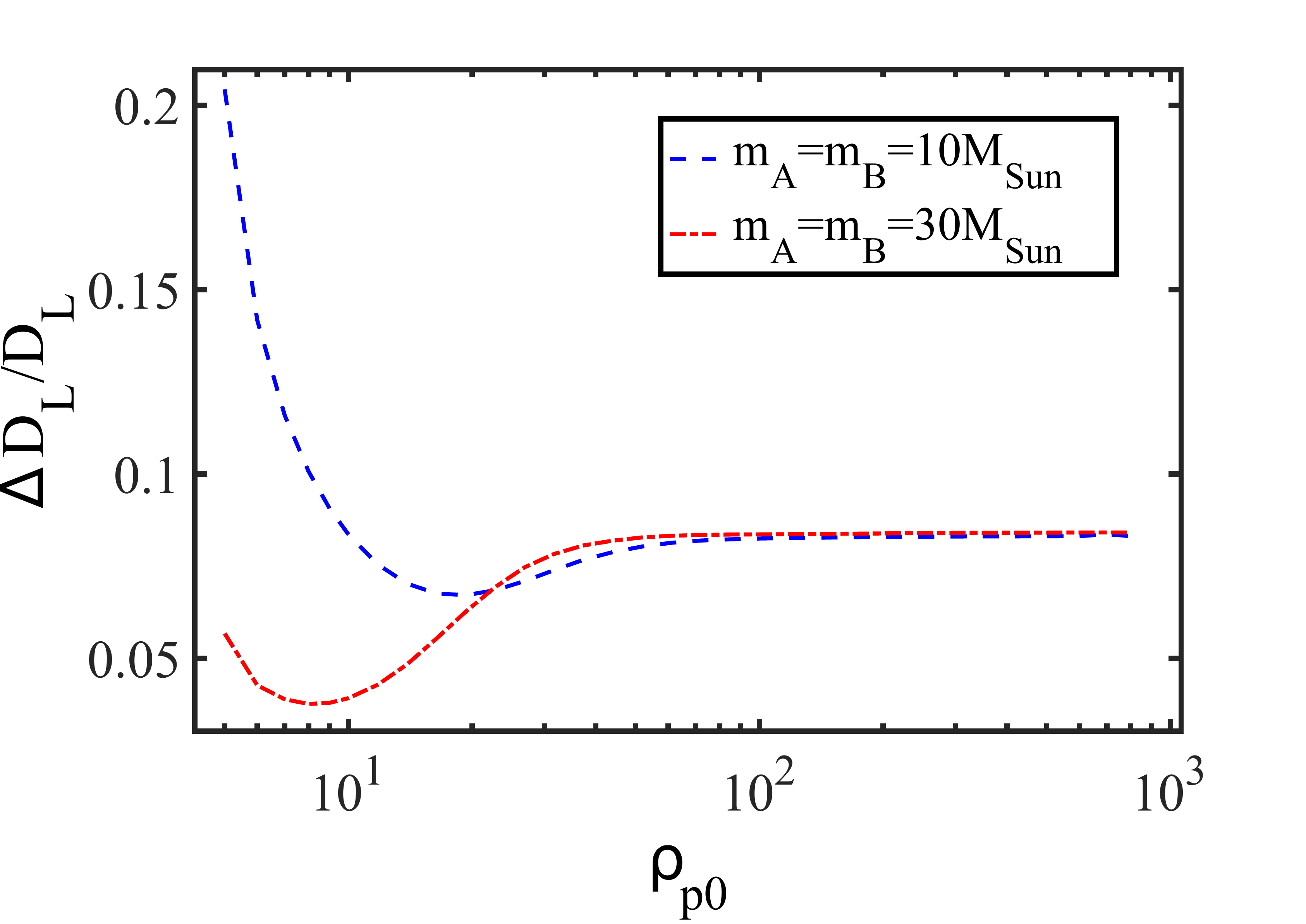}  
    \includegraphics[width=79mm]{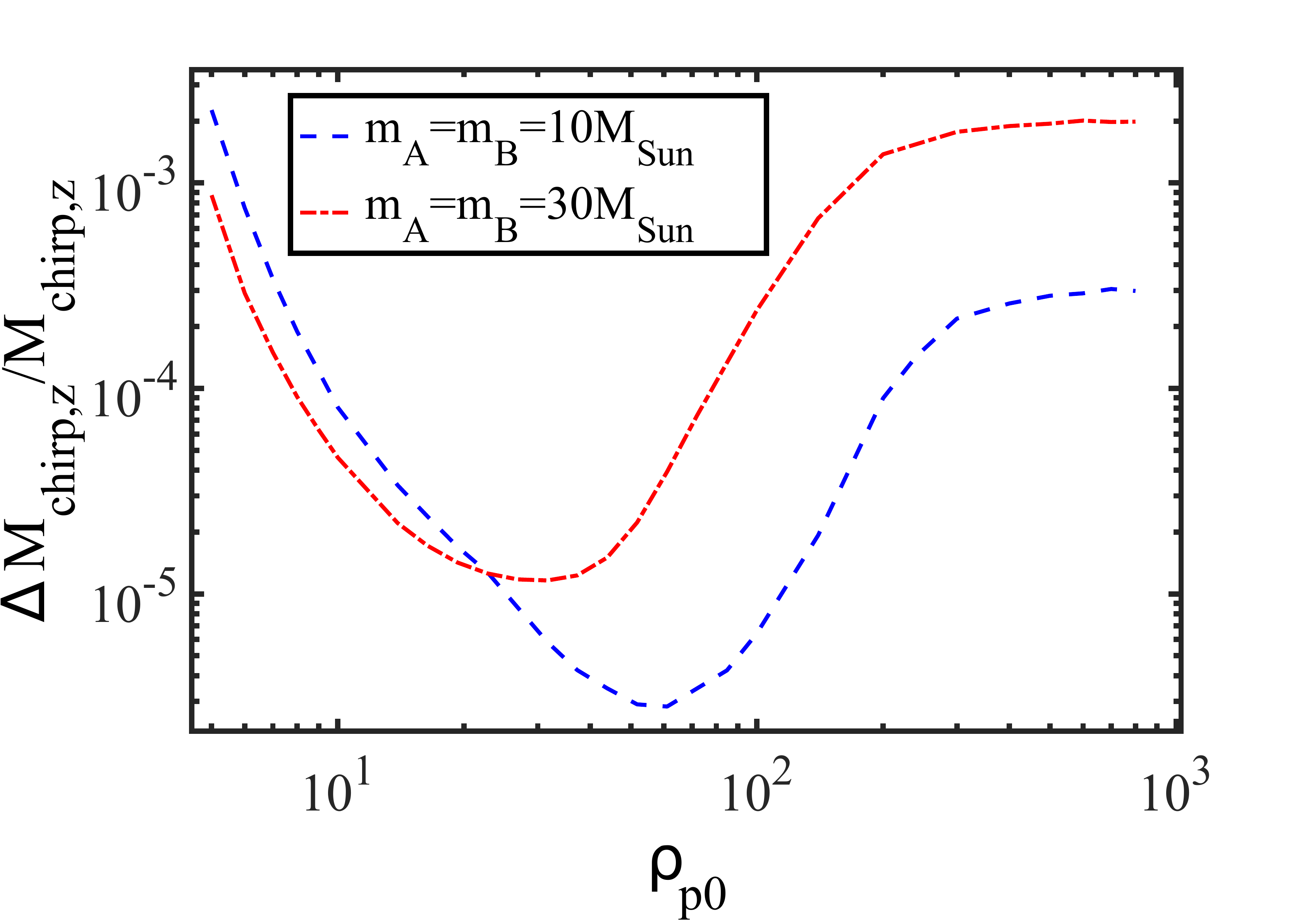} \\
    \includegraphics[width=79mm]{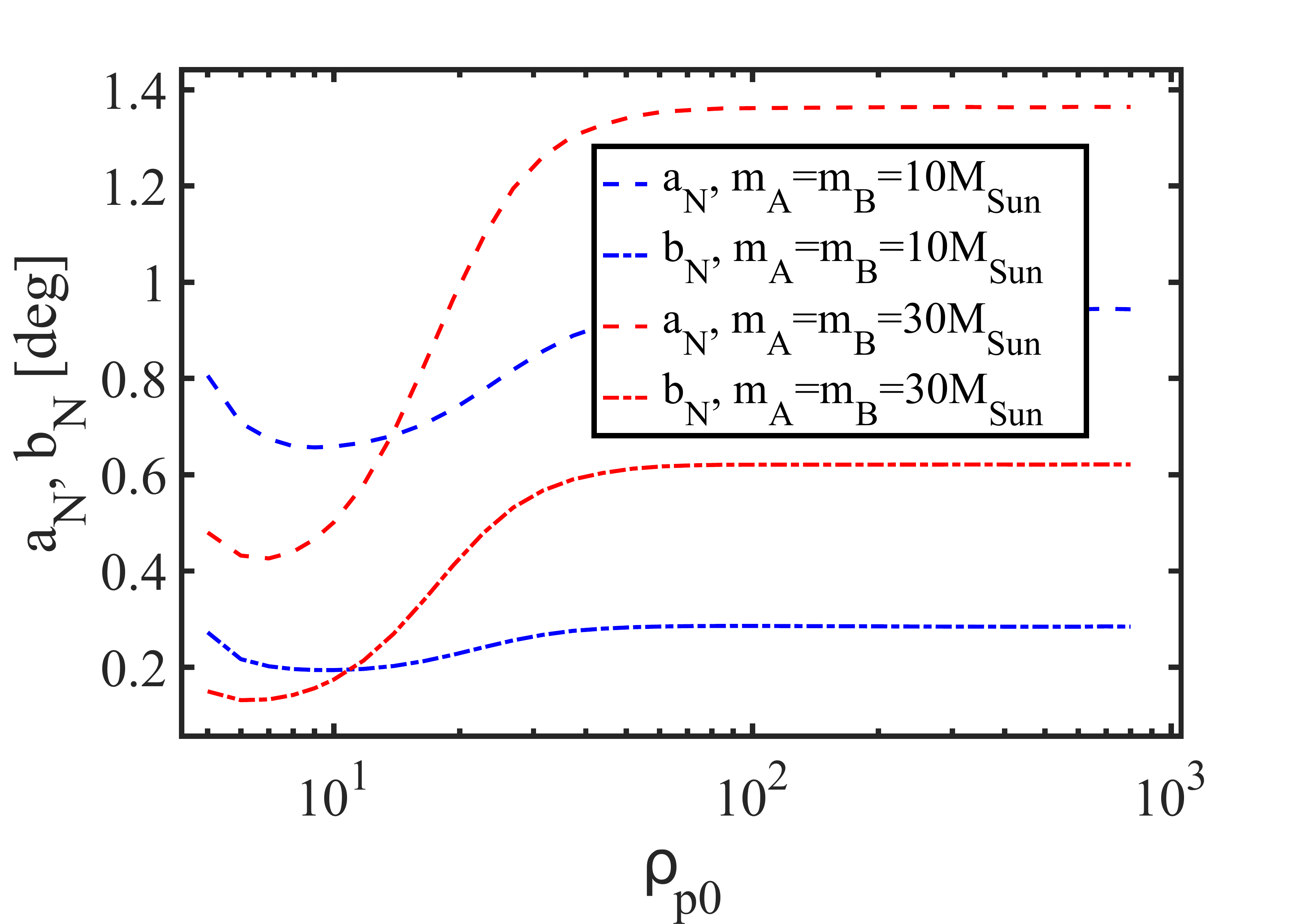}  
    \includegraphics[width=79mm]{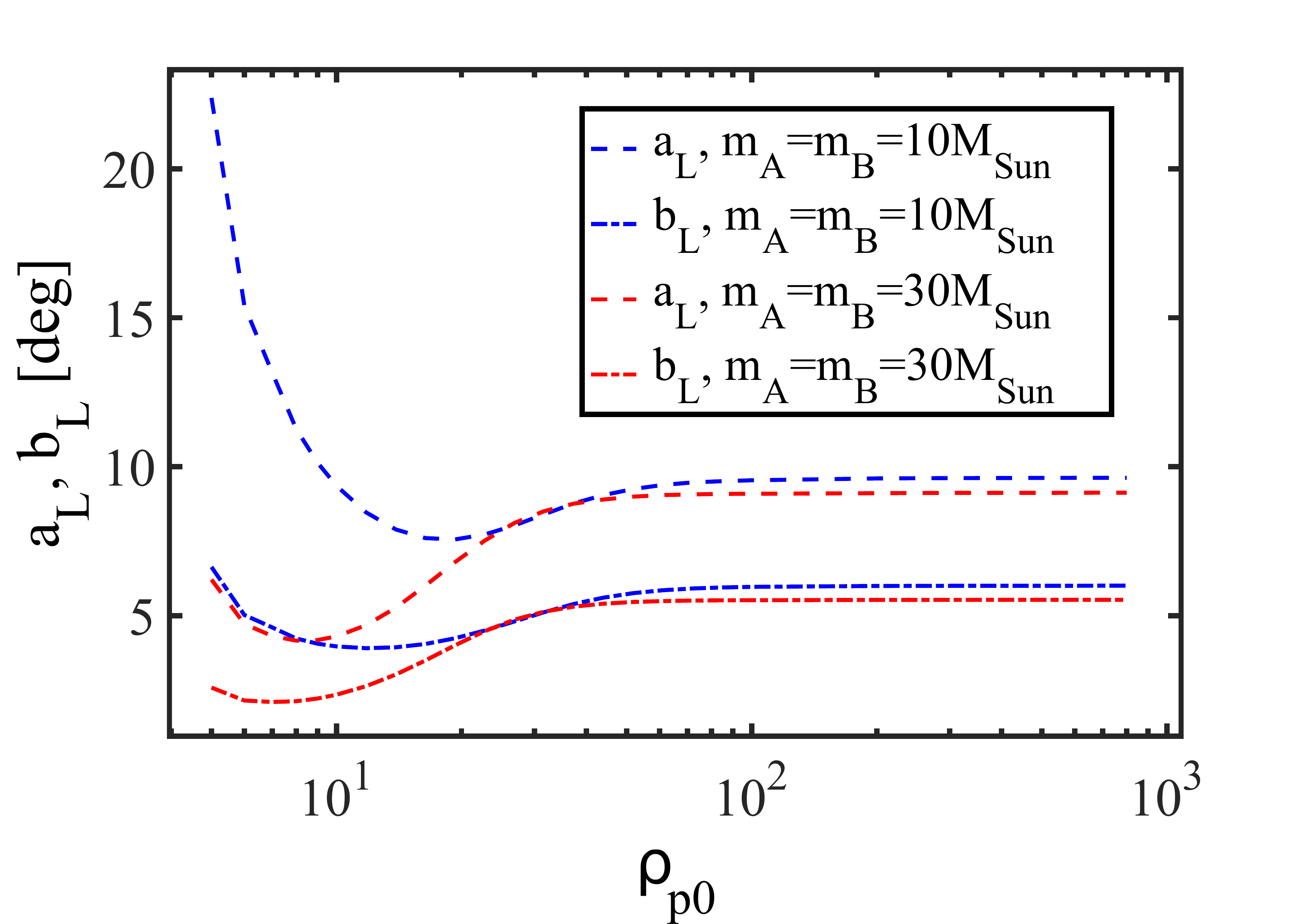} \\
    \includegraphics[width=79mm]{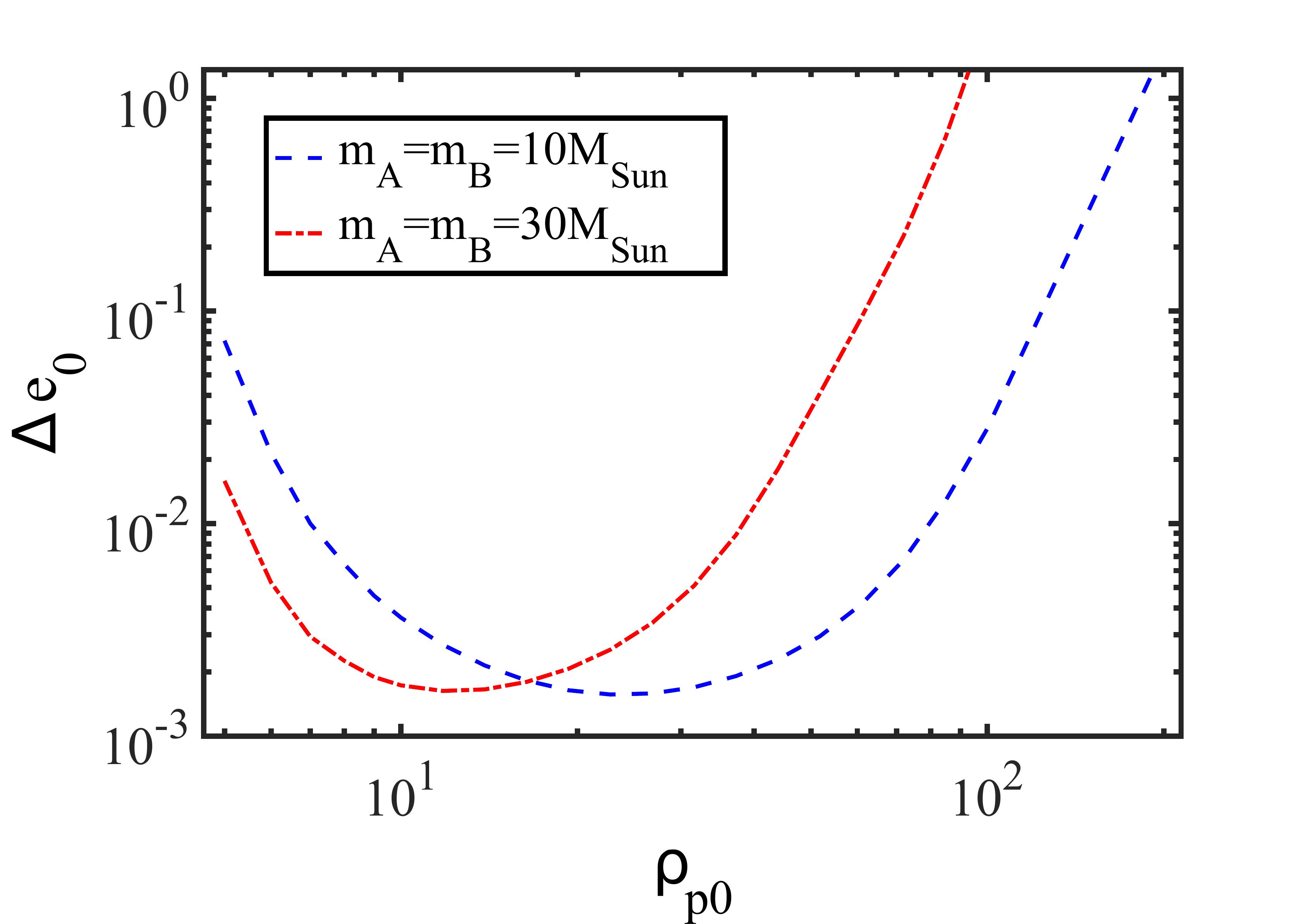}  
    \includegraphics[width=79mm]{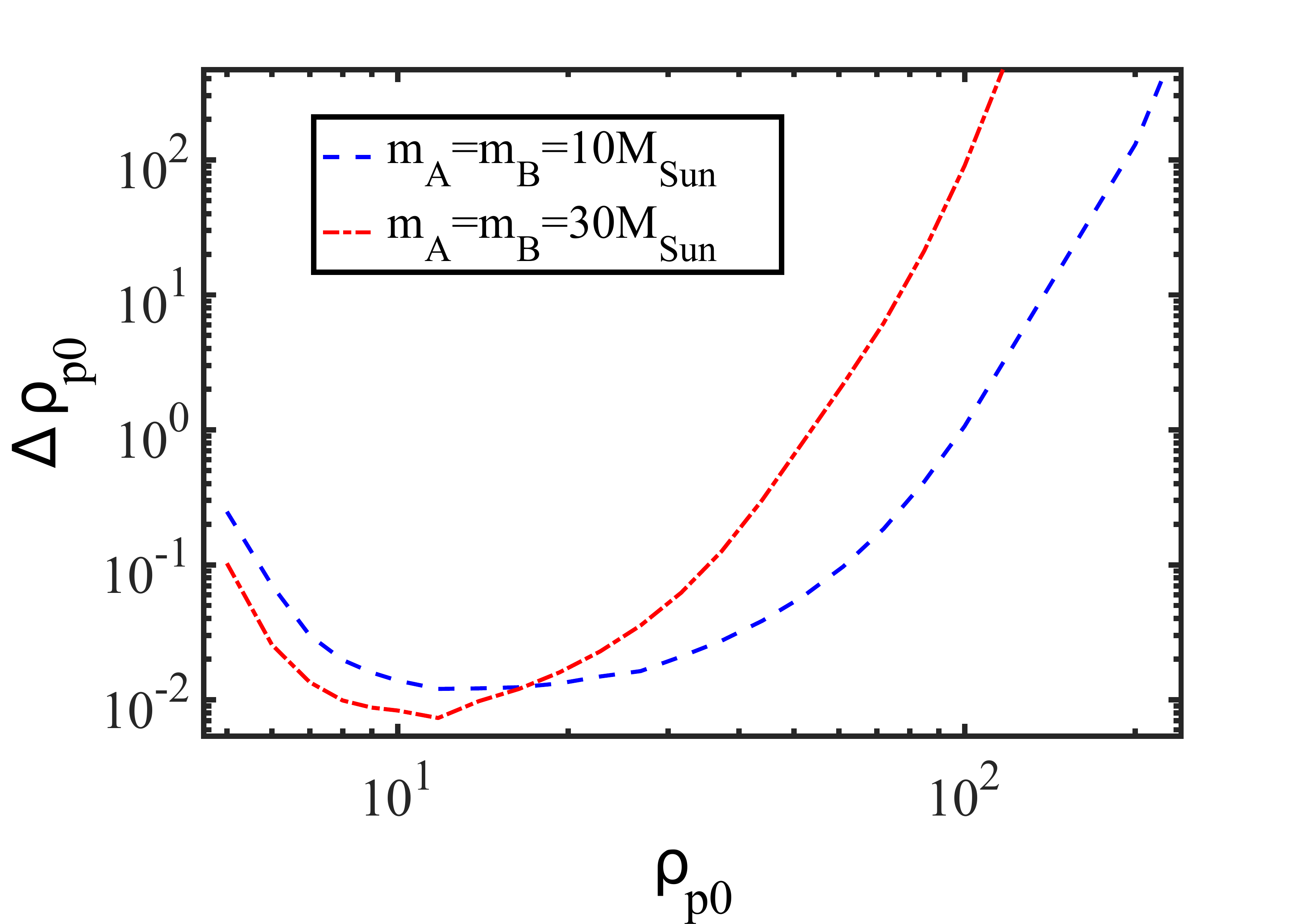} \\
    \includegraphics[width=79mm]{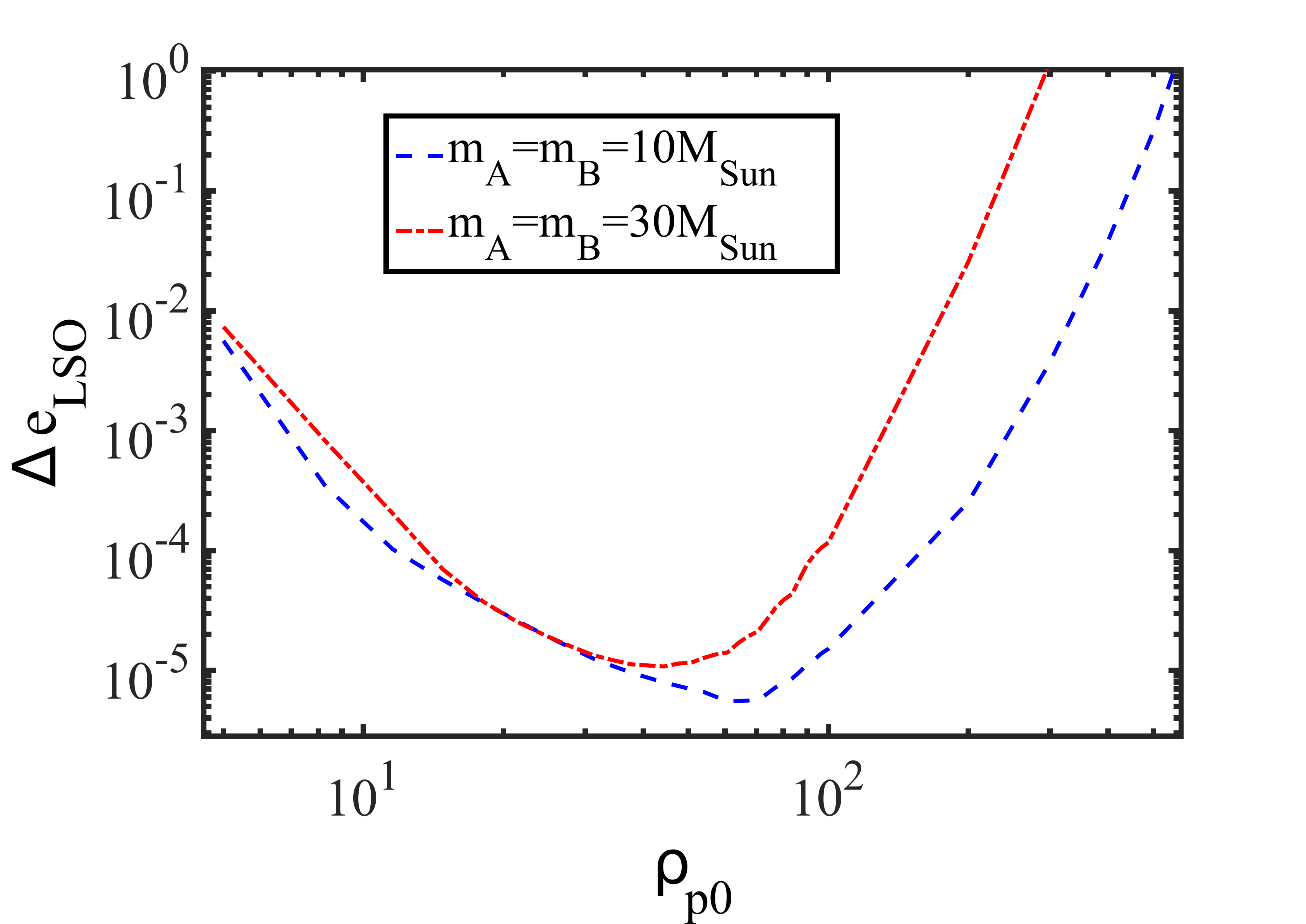}
\caption{ The measurement error of source parameters as a function of initial dimensionless pericenter 
 distance with all other binary parameters fixed as in Figure \ref{fig:SNRtot_rho}. \emph{First row left:}
 luminosity distance $\Delta D_\mathrm{L}/ D_\mathrm{L} = \Delta (\mathrm{ln} D_\mathrm{L})$. 
 \emph{First row right:} redshifted chirp mass, $\Delta \mathcal{M}_z/\mathcal{M}_z= \Delta 
 (\mathrm{ln} \mathcal{M}_z)$. \emph{Second row left:} Semi-major and semi-minor axes of the sky 
 position error ellipse $a_N$ and $b_N$. \emph{Second row right:} Semi-major and semi-minor axes of
 the error ellipse for the binary orbital plane normal vector direction, $a_L$ and $b_L$. \emph{Third 
 row left:} Initial orbital eccentricity, $\Delta e_0$. \emph{Third row right:} Initial dimensionless 
 pericenter distance, $\Delta \rho_{\rm p0}$. \emph{Fourth row:} Eccentricity at the last stable orbit, 
 $\Delta e_{\rm LSO}$. The measurement error of parameters ($\mathrm{ln} (D_\mathrm{L}), \mathrm{ln} 
 (\mathcal{M}_z), a_N, b_N, a_L, b_L$) converge asymptotically to the value of precessing eccentric
 binaries in the circular limit for high $\rho_\mathrm{p0}$, and the measurement error of parameters 
 ($\Delta e_0, \Delta \rho_{\mathrm{p}0}, \Delta e_{\rm LSO}$) increase rapidly with $\rho_{\mathrm{p}0}$
 for high $\rho_{\mathrm{p}0}$. We find similar trends with $\rho_{\mathrm{p}0}$ for other random 
 choices of binary direction and orientation (not shown). Note that the measurement error of $\Delta 
 e_{\rm LSO}$ is undetermined for high $\rho_{\mathrm{p}0}$ because the Fisher matrix algorithm becomes
 invalid for this parameter in this regime, see Section \ref{subsec:ResParamDist} for details. }
 \label{fig:Delta_rho}
\end{figure*} 

\begin{figure}
    \centering 
    \includegraphics[width=78mm]{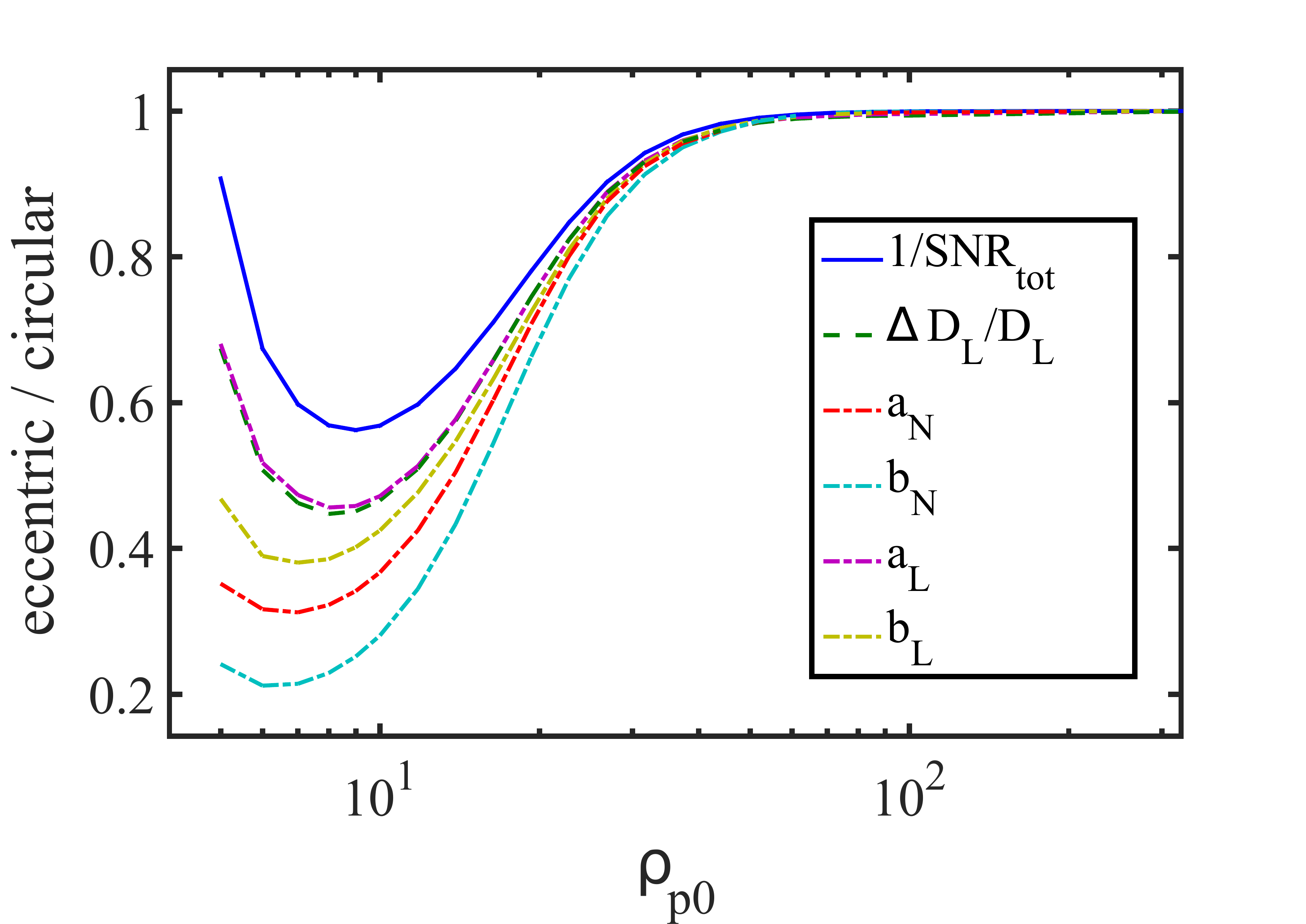}  
\caption{ The initial dimensionless pericenter distance dependence of $\Delta D_\mathrm{L} / 
 D_\mathrm{L}$, the principal axes of the sky localization and binary orientation error ellipses 
 $(a_N,b_N)$ and $(a_L,b_L)$, and $1 / \mathrm{SNR}_\mathrm{tot}$ for $30 \, \Msun-30 \, \Msun$ 
 precessing highly eccentric BH binaries with all other binary parameters fixed as in Figure 
 \ref{fig:SNRtot_rho}. The values are with respect to those of circular binaries and match those 
 in Figure \ref{fig:Delta_rho}. Note that we find similar trends with $\rho_{\mathrm{p}0}$ and 
 $e_0$ for other random choices of binary direction and orientation (not shown). }
 \label{fig:ecc_per_circ}
\end{figure}

 We calculate the parameter measurement errors for precessing highly eccentric BH binaries as a 
 function of $\rho_{\mathrm{p}0}$ for some arbitrarily fixed binary direction and orientations. 
 For one such binary direction and orientation Figure \ref{fig:Delta_rho} shows the 
 $\rho_{\mathrm{p}0}$ dependence of $\Delta D_{\rm L} / D_{\rm L}$, $\Delta \mathcal{M}_z / 
 \mathcal{M}_z$, $\Delta e_0$, $\Delta \rho_ {\mathrm{p}0}$, $\Delta e_{\rm LSO}$, semi-major 
 and semi-minor axes of the sky position error ellipse $(a_N,b_N)$, and semi-major and semi-minor
 axes of the error ellipse for the binary orbital plane normal vector direction $(a_L,b_L)$. Not
 e that we find similar trends with $\rho_{\mathrm{p}0}$ for other random choices of binary direction
 and orientation. We find that measurement errors systematically decrease with decreasing $\rho_
 {\mathrm{p}0}$ for precessing highly eccentric binaries relative to similar binaries in the circular
 limit (Appendix \ref{sec:circlim}), and the errors have a minimum in the range $8 < \rho_{\mathrm{p}0} 
 < 80$ and deteriorate rapidly for $\rho_{\mathrm{p}0} < 8$. The latter is due to the rapid decrease of 
 the $\mathrm{SNR}_\mathrm{tot}$ in that range. The $\rho_{\mathrm{p}0}$ dependence of 
 $\Delta D_\mathrm{L} / D_\mathrm{L}$ and the principal axes of the sky position and binary orientation
 error ellipses $(a_N,b_N)$ and $(a_L,b_L)$ (i.e. quantities derived from slow parameters, see Section 
 \ref{sec:ParamSpace}) are qualitatively similar to that of $1/\mathrm{SNR}_\mathrm{tot}$ in the complete
 range of $\rho_{\mathrm{p}0}$ (i.e. they decrease rapidly with $\rho_{\mathrm{p}0}$ for low $\rho_
 {\mathrm{p}0}$, have a minimum at moderate $\rho_{\mathrm{p}0}$, and converge asymptotically to the 
 value of precessing highly eccentric binaries in the circular limit), see Figure \ref{fig:ecc_per_circ} 
 for details. However, Figure \ref{fig:Delta_rho} shows that the chirp mass errors have a minimum at 
 much higher $\rho_{\mathrm{p}0}$, i.e. between $50 - 60$ and $20 - 40$ for $10 \,\Msun - 10 \, \Msun$ 
 and $30 \, \Msun -30 \, \Msun$ precessing highly eccentric BH binaries, respectively. The main reason 
 for the different behavior of the chirp mass from the distance and angular errors is the fact that the 
 chirp mass is a fast parameter, while the distance and angular parameters are slow parameters. Slow 
 parameters are insensitive to the GW phase perturbations, and depend on the GW amplitude, which is set
 by the $\mathrm{SNR}_\mathrm{tot}$. The SNR of the early part of the waveform near the low-frequency 
 noise wall of the detector is small. However, fast parameters depend sensitively on the GW phase, and the 
 GW phase accumulates mostly at low frequencies, since the residence time (i.e. $\nu/\dot \nu$) is largest
 at low orbital frequencies. Thus, the fast parameters' errors are minimized for binaries which form with 
 $e_0 \sim 1$ with a $\rho_{\rm p0}$ value for which the GW characteristic frequency is near the detectors' 
 minimum frequency. The peak of the spectrum is initially at $f_{\min}$ if $\rho_{\rm p0} = 40.9 [M_{\rm tot}
 /(20 \, \Msun)]^{-2/3} (f_{\min} / 10\,{\rm Hz})^{-2/3}$ \citep{Gondanetal2017}. A slightly lower value of 
 $f_{\min} \sim 7 \, {\rm Hz}$ leads to values that represent the minimum of the fast parameters. This also 
 leads to the result observed in Figure \ref{fig:Dist1} that these parameters have higher errors for $\rho_
 {\mathrm{p}0} = 10$ than for $\rho_{\mathrm{p}0} = 20$. 
 
 In Figure \ref{fig:Delta_rho}, note that $\Delta e_{\mathrm{LSO}}$ errors are relatively small for 
 relatively high $\rho_{\mathrm{p}0}$ up to $\rho_{\mathrm{p}0} \sim 200$. At high $\rho_{\mathrm{p}0}$, 
 the orbital eccentricity approaches zero when it enters the aLIGO band, and $\Delta e_{\mathrm{LSO}}$ 
 increases. We note that the posterior probability distribution function of $e_{\mathrm{LSO}}$ is 
 well-defined even in the circular limit $\rho_{\mathrm{p}0} \rightarrow \infty$, and $\Delta e_
 {\mathrm{LSO}}$ is finite for a given confidence region. However, the Fisher matrix algorithm becomes
 invalid in this regime as the signal is not approximated well by its linear Taylor expansion with respect
 to the $\Delta e_{\mathrm{LSO}}$ parameter, since its first $e_{\mathrm{LSO}}$ derivative vanishes in the 
 circular limit. Therefore the true asymptotic value of $\Delta e_{\mathrm{LSO}}$ for high $\rho_{\mathrm{p}0}$
 cannot be recovered with the Fisher matrix technique used in this paper. Further, note that $\Delta e_0$ 
 and $\Delta \rho_{\mathrm{p}0}$ also increase rapidly with $\rho_{\mathrm{p}0}$ for high $\rho_{\mathrm{p}0}$.
 This is due to the fact that for these parameters the binary forms with a pericenter frequency smaller than 
 the minimum frequency of the detector network, and the information on $e_0$ and $\rho_{\mathrm{p}0}$ is 
 limited to higher harmonics with small power. Thus, these parameters indeed have a very high error and 
 become indeterminate in the circular limit. The fact that the relative error of $e_0$ and $\rho_{\mathrm{p}0}$ 
 can be less than $\sim 5\%$ percent in the range $5 < \rho_{\mathrm{p}0} < 50$ ($ 6 < \rho_{\mathrm{p}0} < 
 100$) for $30 \, \Msun - 30 \, \Msun$ ($10 \, \Msun - 10 \, \Msun$) precessing highly eccentric BH binaries 
 implies that the GW detections might have the potential to constrain the formation environment of these system 
 \citep{OLearyetal2009,Cholisetal2016,Rodriguezetal2016a,Rodriguezetal2016b,Chatterjeeetal2017,Gondanetal2017,Kocsisetal2017,SamsingRamirezRuiz2017,SilsbeeTremaine2017}. 
 
 Furthermore, we found from numerical investigations that $\Delta e_0$ does not correlate
 significantly with other parameters' errors, which is due to the fact that $e_0$ is measured from 
 the truncation of the signal for $e>e_0$ at the start of the waveform, while other parameters of 
 a precessing eccentric binary are measured from the inspiral rate (Section \ref{sec:ParamSpace}). 
 However, $\Delta \rho_{\mathrm{p}0}$ behaves differently from $\Delta e_0$ in this regard, which 
 is due to the fact that $\rho_{\mathrm{p}0}$ is determined by $e_{\mathrm{LSO}}$ in Equation 
 (\ref{eq:deltarhop0}), and $e_{\mathrm{LSO}}$ depends on the mass parameters.

\subsection{Comparison with previous results}
\label{sebsec:CompWithPrevRes}
 
 In this paper, we have determined the SNR and the expected accuracy with which the 
 aLIGO-AdV-KAGRA detector network may determine the parameters that describe highly eccentric BH 
 binaries, and investigated how these quantities depend on the initial pericenter distance $\rho_
 {\mathrm{p}0}$ and initial eccentricity $e_0$. There are some previous studies that also made 
 similar investigations for eccentric compact binaries with significant differences 
 \citep{Yunesetal2009,KyutokuSeto2014,Sunetal2015,Maetal2017}. They considered different detector 
 networks, applied different waveform models, and used different definitions for $e_0$ and $\rho_
 {\mathrm{p}0}$. As a consequence, only a qualitative comparison is possible with those results, 
 which we discuss in this section. At the end of this section, we compare our results for the 
 measurement errors in the circular limit with those presented in previous studies.
 
 We first compare our results with a previous study for the $\rho_{\mathrm{p}0}$ dependence 
 of the $\mathrm{SNR}_\mathrm{tot}$. Our result for the $\rho_{\mathrm{p}0}$ dependence of the $\mathrm{SNR}_
 \mathrm{tot}$ (Figure \ref{fig:SNRtot_rho}) is qualitatively in agreement with the result of Figure
 2 in \citet{KyutokuSeto2014}, i.e. the $\mathrm{SNR}_\mathrm{tot}$ increases rapidly with $\rho_{\mathrm{p}0}$
 for low $\rho_{\mathrm{p}0}$, peaks at a moderate $\rho_{\mathrm{p}0}$, and converges asymptotically 
 to the value of highly eccentric binaries in the circular limit for high $\rho_{\mathrm{p}0}$. 
       
\begin{figure}
    \center
    \includegraphics[width=78mm]{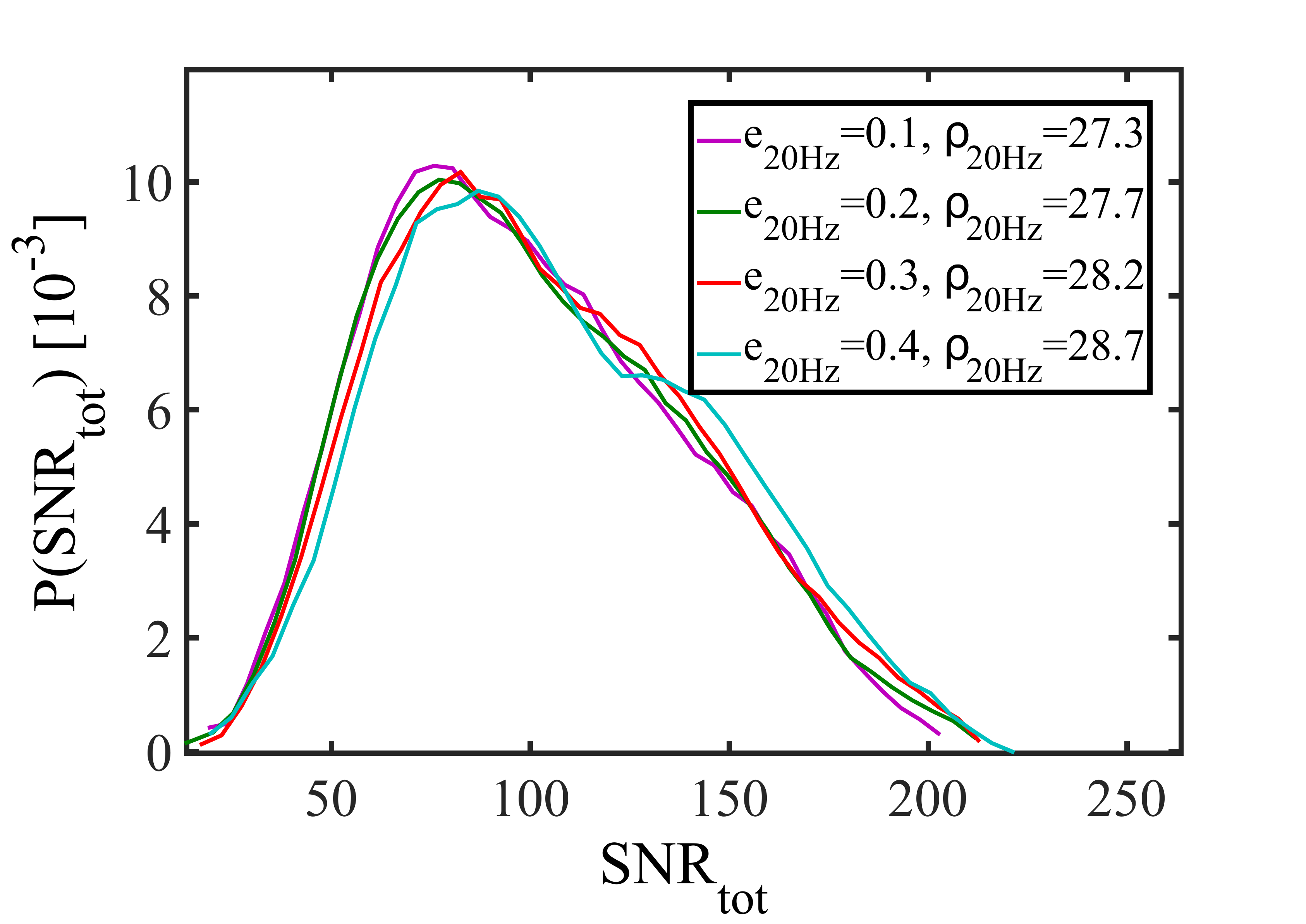}  \\
    \includegraphics[width=78mm]{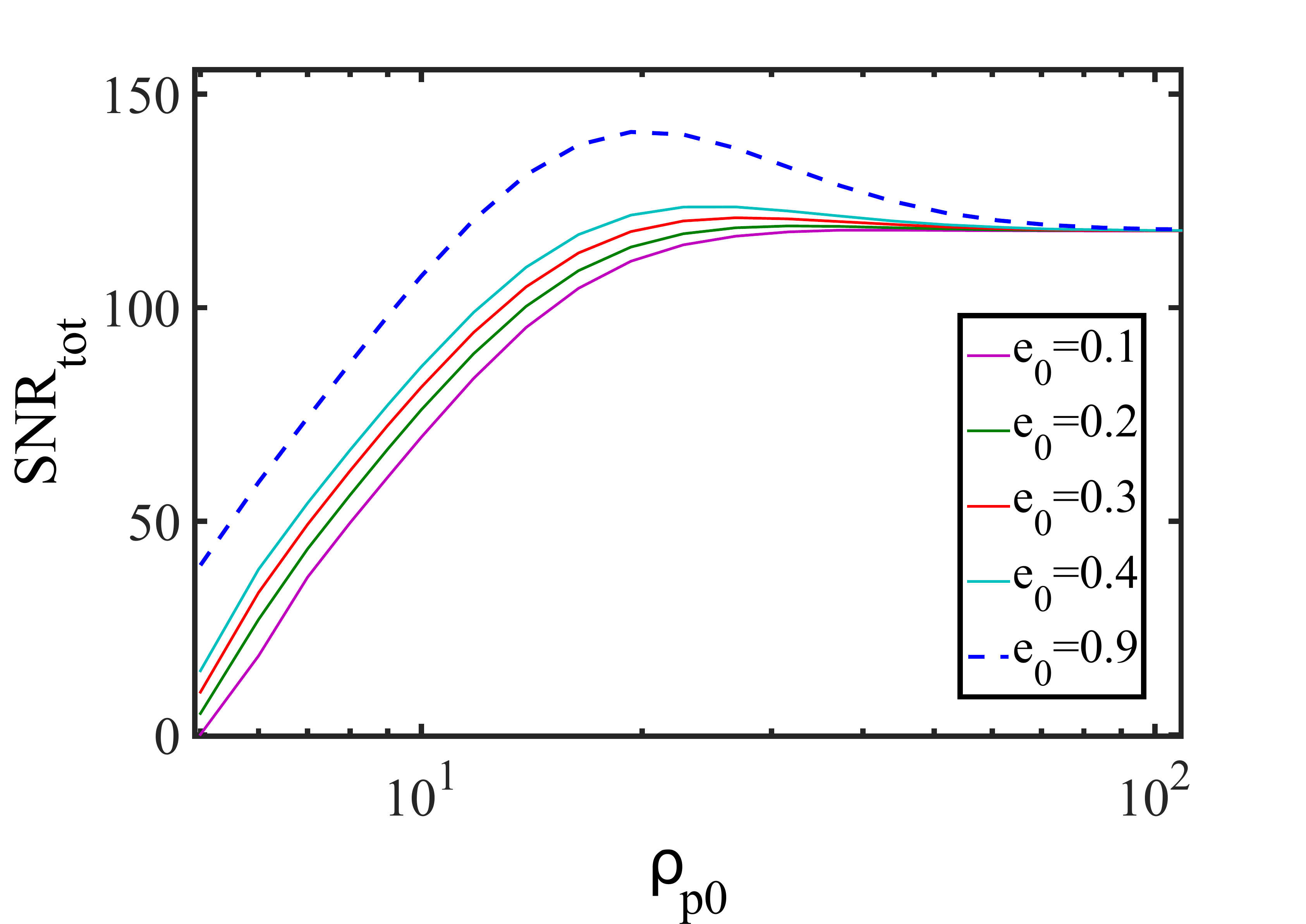}
\caption{ 
     \emph{Top panel}: Smoothed probability density function of the $\mathrm{SNR}_\mathrm{tot}$
     for $10 \, \Msun-10 \, \Msun$ precessing eccentric BH binaries. In these calculations we set the 
     lower bound of the detectors' sensitive frequency band to $20 \, \mathrm{Hz}$. Here $e_{20\Hz}$ 
     and $\rho_{20\Hz}$ represent the initial orbital eccentricity and initial dimensionless pericenter 
     distance at which the peak GW frequency \citep{Wen2003} of the binary is $f_\mathrm{GW} = 20 \, 
     \mathrm{Hz}$. Other details of the calculations are the same as in Figure \ref{fig:SNR_Prec}. 
     \emph{Bottom panel:} The same as in Figure \ref{fig:SNRtot_rho} but for $10 \, \Msun -10 \, \Msun$
     precessing eccentric BH binaries with initial eccentricities $e_0= (0.1, 0.2, 0.3, 0.4, 0.9)$ as a 
     function of $\rho_{\mathrm{p}0}$. We also represent the $e_0 = 0.9$ curve in this plot in order to 
     show that $\mathrm{SNR}_\mathrm{tot}$ is a strictly monotonically increasing function of $e_0$ over 
     the full range of $\rho_{\mathrm{p}0}$. We find similar trends with $\rho_{\mathrm{p}0}$ 
     and $e_0$ for other random choices of binary direction and orientation (not shown). }   
     \label{fig:SNRtot_Comps}
\end{figure}

\begin{figure*}
    \centering 
    \includegraphics[width=78mm]{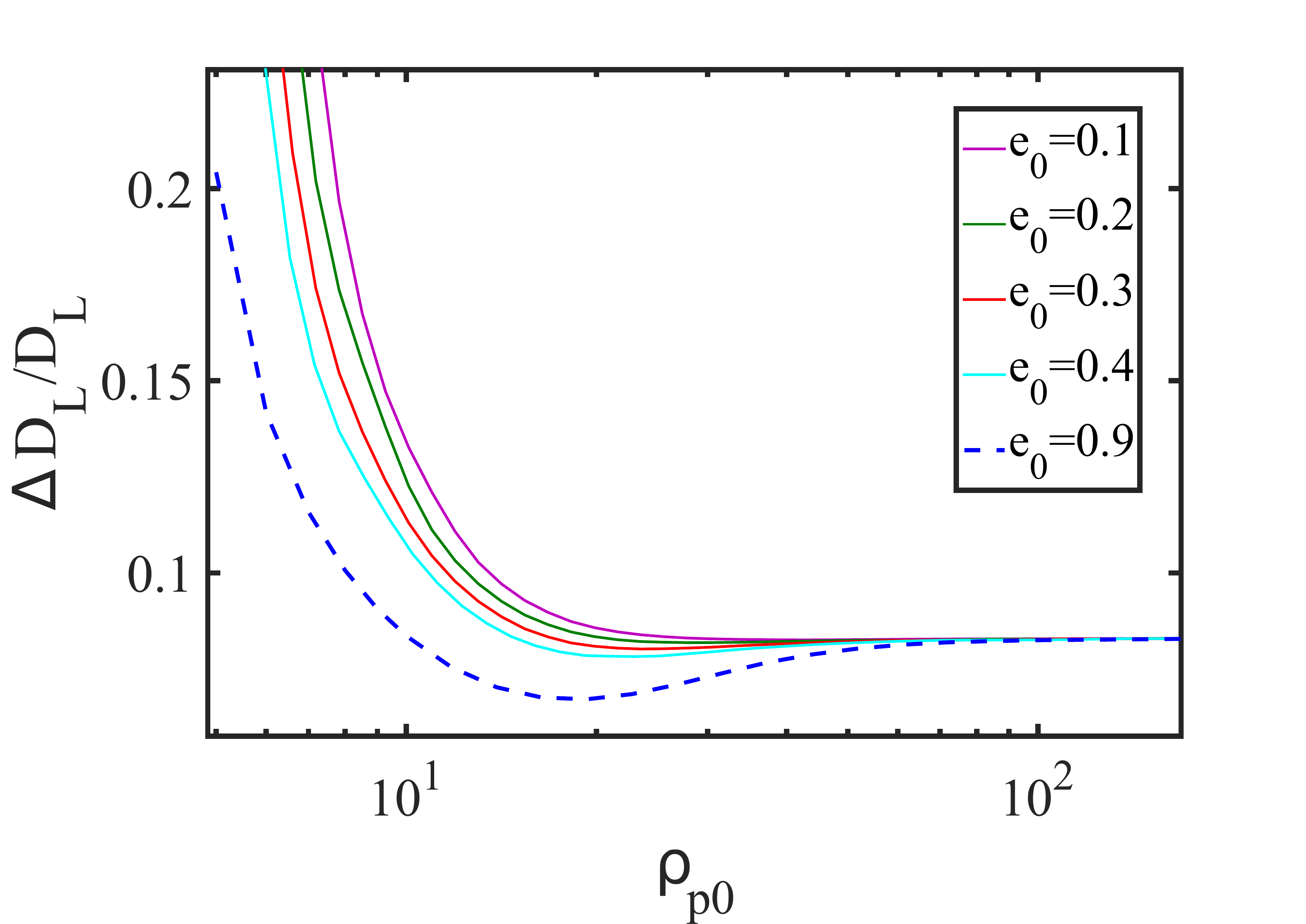} 
    \includegraphics[width=78mm]{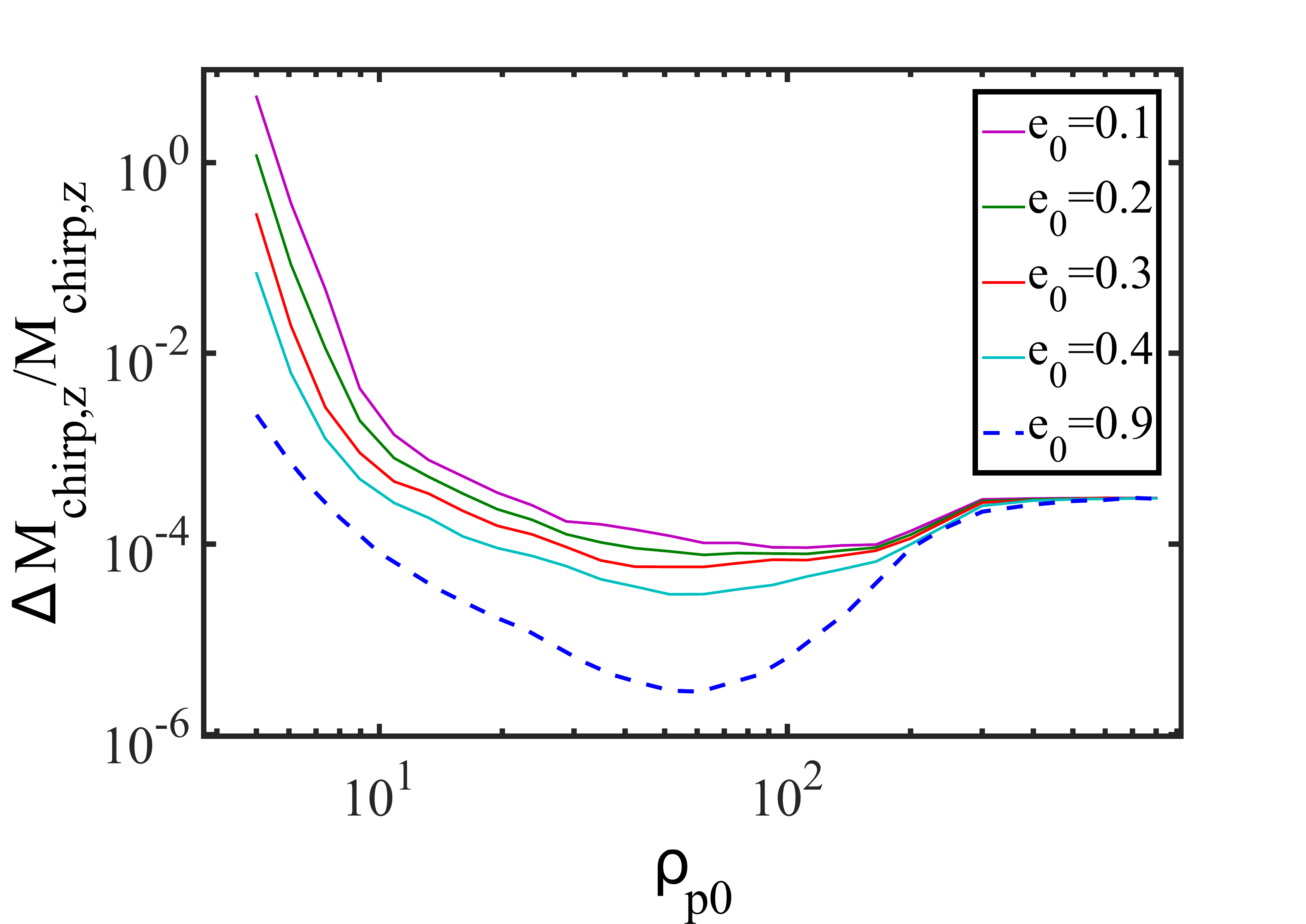} \\
    \includegraphics[width=78mm]{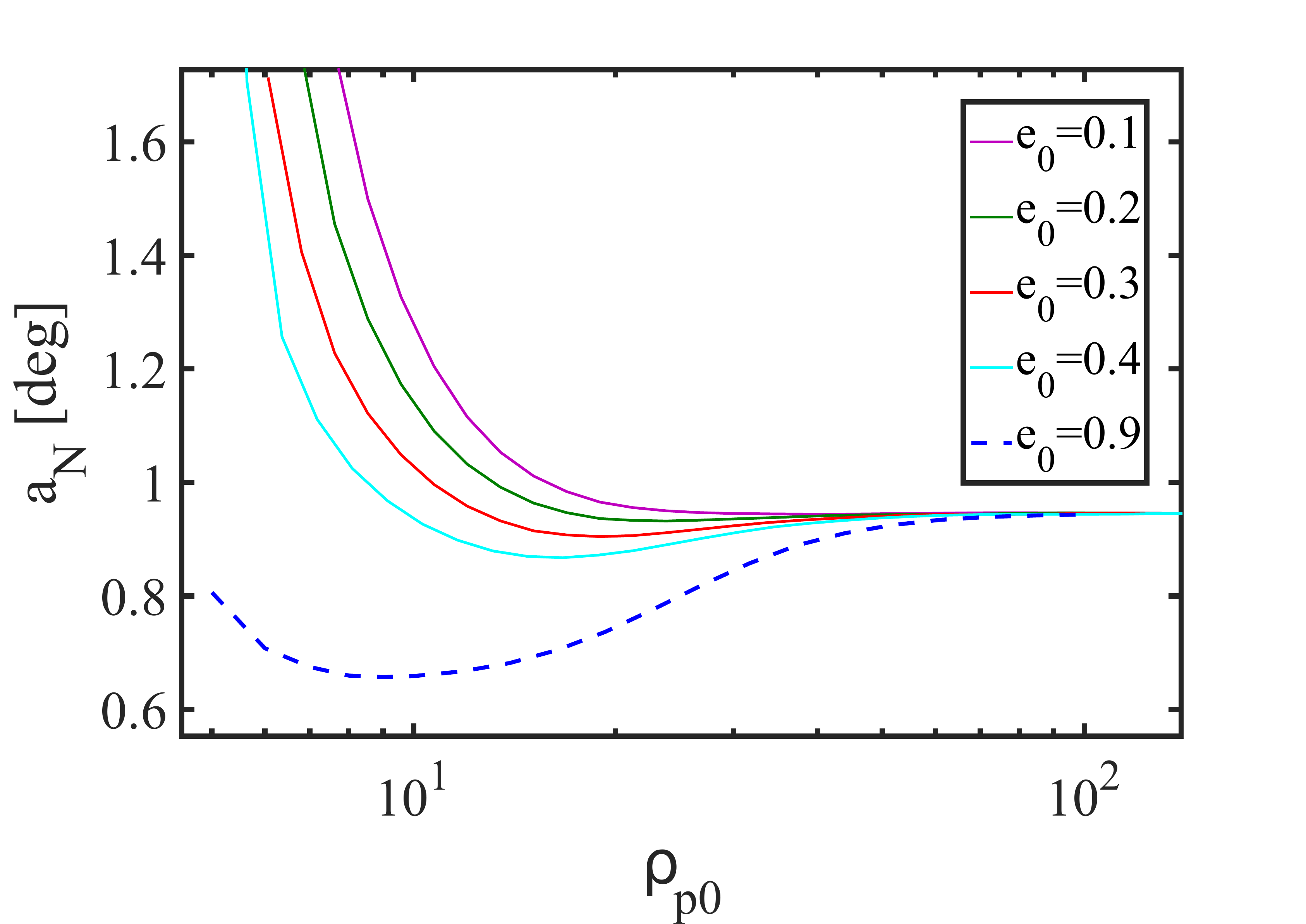} 
    \includegraphics[width=78mm]{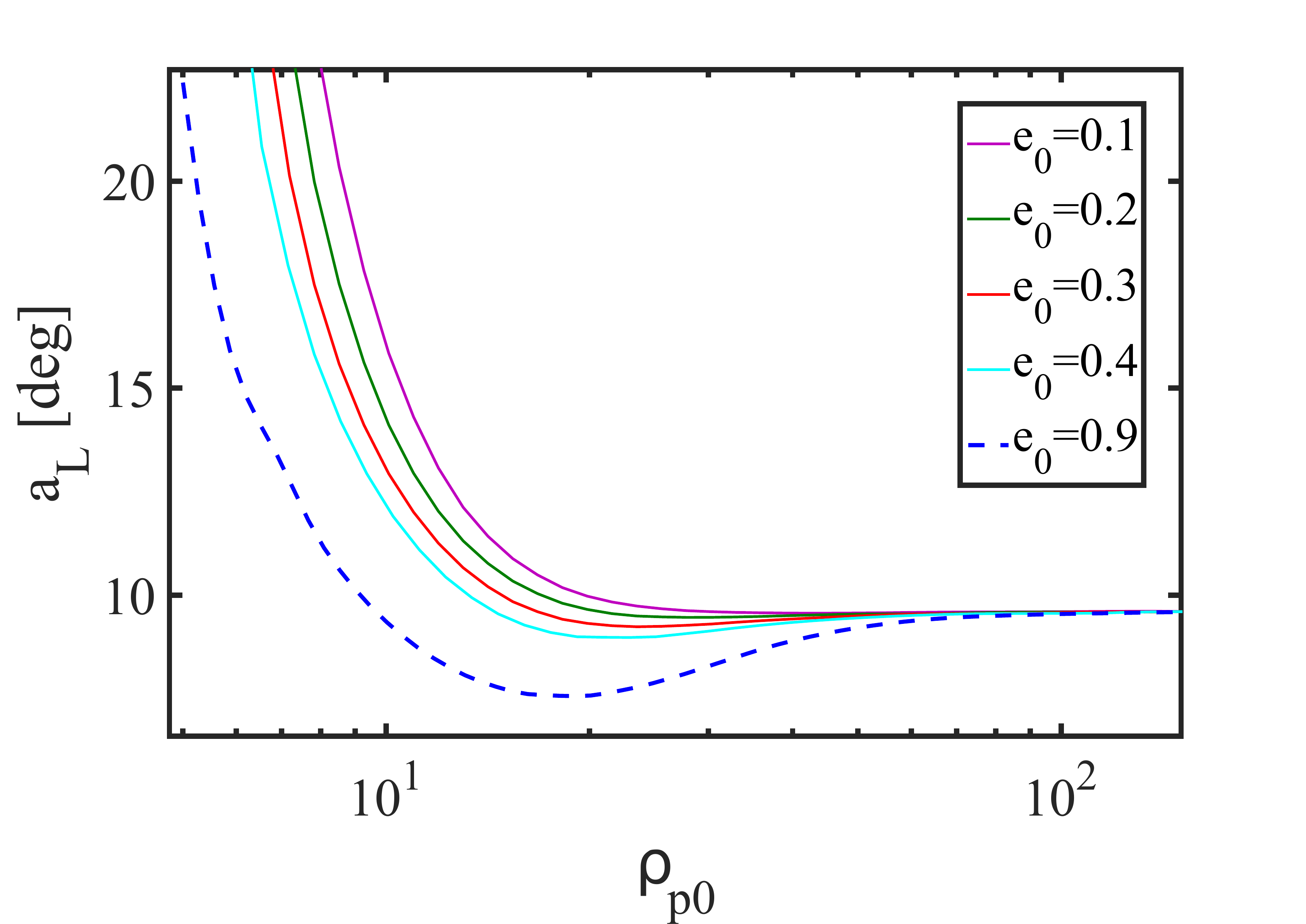} 
\caption{ The initial pericenter distance and eccentricity dependence of the measurement errors
 of various parameters for $10 \, \Msun-10 \, \Msun$ precessing eccentric BH binaries for a specific 
 sky position and inclination chosen as in Figure \ref{fig:SNRtot_rho}. Different lines show different 
 initial eccentricities as labeled. \emph{Top left:} luminosity distance $\Delta D_\mathrm{L}/ D_\mathrm{L} 
 = \Delta (\mathrm{ln} D_\mathrm{L})$. \emph{Top right:} redshifted chirp mass, $\Delta \mathcal{M}_z/
 \mathcal{M}_z= \Delta (\mathrm{ln} \mathcal{M}_z)$. \emph{Bottom left:} Semi-major axis of the sky 
 position error ellipse, $a_N$. \emph{Bottom right:} Semi-major axis of the error ellipse for the 
 binary orbital plane normal vector direction, $a_L$. We find similar trends with $\rho_{\mathrm{p}0}$ 
 and $e_0$ for other random choices of binary direction and orientation (not shown). We also show
 the $e_0 = 0.9$ curve in this plot to illustrate that measurement errors of binary parameters are a
 strictly monotonically increasing functions of $e_0$ over the full range of $\rho_{\mathrm{p}0}$. }
 \label{fig:Err_LowEccDep}
\end{figure*}

 In order to compare our results for the $e_0$ dependence of the $\mathrm{SNR}_\mathrm{tot}$ with
 \citet{Yunesetal2009} and \citet{Sunetal2015}, we also set the lower bound of advanced GW detectors' 
 sensitive frequency band to $f_{\min} =20 \, \mathrm{Hz}$. We define $e_{20 \, \Hz}$ to be the eccentricity
 at which the peak GW frequency of the binary defined in \citet{Wen2003} is $f_\mathrm{GW} = 20 \, 
 \mathrm{Hz}$, and evaluate $\rho_{\mathrm{p}0}$ corresponding to $e_0=e_{\mathrm{20\Hz}}$ from Equation
 (37) in \citet{Wen2003} as
 \begin{equation}  \label{eq:rho_GW}
   \rho_{20\Hz} = \left[ (1+e_{20\Hz})^{0.3046} f_\mathrm{GW} \pi M_{{\rm tot},z} \right]^{-2/3} \, .
 \end{equation} 
 We recalculate the distribution of the $\mathrm{SNR}_\mathrm{tot}$ for $10 \, \Msun-10 \, \Msun$ 
 precessing eccentric compact binaries with $e_{20\Hz}= (0.1, 0.2, 0.3, 0.4)$. The top panel of Figure
 \ref{fig:SNRtot_Comps} shows that the $\mathrm{SNR}_\mathrm{tot}$ is roughly the same for different $e_
 {20\Hz}$, which is consistent with results presented in Figure 2 in \citet{Sunetal2015} and in the 
 left panel of Figure 8 in \citet{Yunesetal2009}. Moreover, we find that the $\mathrm{SNR}_\mathrm{tot}$
 increases weakly with $e_\mathrm{20\Hz}$, which is in agreement with results in the left panel of 
 Figure 8 in \citet{Yunesetal2009}. Note that this result disagrees with Table 5 in \citet{Sunetal2015}. 
 The $\mathrm{SNR}_\mathrm{tot}$ does not depend significantly on $e_{20\Hz}$ in the range of $[0.1,0.4]$ 
 for $10 \, \Msun-10 \, \Msun$ for $\rho_{p0}=\rho_{20\Hz}\sim 28$ as seen in the bottom panel of Figure
 \ref{fig:SNRtot_Comps}.\footnote{For binaries with relatively high $\rho_{\mathrm{p}0}$, binaries are 
 well-circularized by the time their peak GW frequency enters the sensitive frequency band of advanced
 ground-based GW detectors, thus the information about the initial eccentricity vanishes from the detectable
 part of the waveform. This explains the very weak $e_0$ dependence of the $\mathrm{SNR}_\mathrm{tot}$ 
 and of parameter measurement errors for high $\rho_{\mathrm{p}0}$ in the bottom panel of Figure 
 \ref{fig:SNRtot_Comps} and in Figure \ref{fig:Err_LowEccDep}. } For $10 \, \Msun - 10 \, \Msun$ binaries
 with $e_0<0.4$ and $\rho_{\mathrm{p}0} \sim 28$, we find that $\rho_{20\Hz}$ is high enough to fall into
 the range of $\rho_{\mathrm{p}0}$ where the $\mathrm{SNR}_\mathrm{tot}$ depends on $e_0$ at the $\sim 10\%$
 level for $e_{20\Hz}<0.4$. The influence of $e_{20\Hz}$ on the $\mathrm{SNR}_\mathrm{tot}$ increases with
 $M_\mathrm{tot}$ since in this case $\rho_{20\Hz}$ is lower as shown by Equation (\ref{eq:rho_GW}). Thus, 
 the influence of $e_{20\Hz}$ on the distribution of $\mathrm{SNR}_\mathrm{tot}$ is more significant for 
 higher-mass low-eccentricity binaries. Examples for this characteristic of the $\mathrm{SNR}_\mathrm{tot}$ 
 are seen in Figure 8 in \citet{Yunesetal2009}.

 Finally, we compare our results with previous studies for the $e_0$ dependence of measurement
 errors of parameters describing eccentric binaries. \citealt{Sunetal2015,Maetal2017} set the initial 
 orbital parameters to be $e_{20\Hz}$ ($e_{10\Hz}$) and $\rho_{20\Hz}$ ($\rho_{10\Hz}$). For various 
 values of $e_{20\Hz}$ ($e_{10\Hz}$) in the range $[0.1, 0.2, 0.3, 0.4]$ and the corresponding values
 of $\rho_{20\Hz}$ ($\rho_{10\Hz}$), they determined the measurement accuracies for various parameters
 of eccentric binaries. Since they applied different waveform models and different parameters 
 describing the eccentric binaries, we resort to a qualitative comparison. We repeated the analysis of 
 Figure \ref{fig:Delta_rho} for $e_0= (0.1, 0.2, 0.3, 0.4)$ and determined the measurement error of 
 parameters as a function of $\rho_{\mathrm{p}0}$ as shown in Figure \ref{fig:Err_LowEccDep}. We find 
 qualitative agreement with \citet{Sunetal2015,Maetal2017}. The measurement accuracies of parameters 
 increase strictly monotonically with $e_0$. 
 
 Previous papers have investigated the $M_\mathrm{tot}$ dependency of the measurement errors for $\lbrace 
 t_c, \Phi_c, \mathrm{ln}(\mathcal{M}_{z}), \mathrm{ln}(\eta) \rbrace$ by using different PN order waveform 
 models for non-spinning inspiraling binaries for a fixed SNR in a single aLIGO type detector. Previous 
 results showed that the measurement accuracy of these parameters decreases with increasing $M_\mathrm{tot}$  
 for $2.8 \, \Msun \leq M_\mathrm{tot} \leq 20 \, \Msun$ provided that the SNR accumulated in one GW detector 
 is fixed, see Table 1 in \citet{Arunetal2005} and references therein. Therefore, we determined the measurement 
 errors in the circular limit for $\lbrace t_c, \Phi_c, \mathrm{ln}(\mathcal{M}_{z}), \mathrm{ln}(\eta) \rbrace$
 for a qualitative comparison.\footnote{A quantitative agreement is not expected since our precessing waveform 
 model differs from the waveform models in those studies.} To calculate the measurement error of the $\mathrm{ln}
 (\eta)$ parameter we use the fact that $\eta= (\mathcal{M}_z M_{{\rm tot},z}^{-1})^{5/3}$ and so 
\begin{equation}  \label{eq:Delta_eta} 
  \frac{\langle\Delta \eta^2\rangle}{\eta^2}  = \frac{25}{9} \frac{\langle\Delta \mathcal{M}_z^2\rangle}{
  \mathcal{M}_z^2}  +\frac{25}{9} \frac{ \langle\Delta M_{{\rm tot},z}^2\rangle }{ M_\mathrm{tot,z}^2 } 
  + \frac{50}{9} \frac{ \langle \Delta \mathcal{M}_z \Delta M_{{\rm tot},z} \rangle }{ 
  \mathcal{M}_z M_{{\rm tot},z} } \, .
\end{equation}
 In agreement with the 1PN order case in \citet{Arunetal2005}, we find that $\Delta t_c$, $\Delta \Phi_c$,
 $\Delta \mathcal{M}_{z} / \mathcal{M}_z$, and $\Delta \eta / \eta$ increase with $M_\mathrm{tot}$ for fixed 
 $\mathrm{SNR}_\mathrm{tot}$ (Table \ref{Table:CircLimParams}). Such a qualitative agreement is expected since 
 the adopted precessing eccentric waveform approximates the full 1PN waveform in its most important features, and 
 the $M_\mathrm{tot}$-dependent trends of error distributions do not depend on the number of detectors or on the 
 sky position or angular momentum unit vectors of the source. 
 
\begin{table} 
\centering 
   \begin{tabular}{@{}ccccc} 
      $m_A-m_B$  &  $\Delta t_c$  &  $\Delta \Phi_c $  &  $\Delta \mathcal{M}_{z} / \mathcal{M}_z$ 
      &  $\Delta \eta / \eta$  \\ 
     \hline\hline 
   $10 \, \Msun-10 \, \Msun$  &  $0.30$  &  $0.32$  &  $3.6 \times 10^{-4}$  &  $3.4 \times 10^{-3}$  \\
   \hline
   $15 \, \Msun-15 \, \Msun$  &  $0.55$  &  $0.39$  &  $8.7 \times 10^{-4}$  &  $6.5 \times 10^{-3}$  \\
   \hline
   $20 \, \Msun-20 \, \Msun$  &  $0.94$  &  $0.49$  &  $1.5 \times 10^{-3}$  &  $10^{-2}$  \\
   \hline
   $25 \, \Msun-25 \, \Msun$  &  $1.38$  &  $0.57$  &  $2.2 \times 10^{-3}$  &  $1.3 \times 10^{-2}$  \\
   \hline
   $30 \, \Msun-30 \, \Msun$  &  $1.84$  &  $0.62$  &  $2.8 \times 10^{-3}$  &  $1.5 \times 10^{-2}$  \\
   \hline\hline 
   \end{tabular} 
   \caption{ Errors in $t_c$ (msec), $\Phi_c$ (rad) and for the relative errors
   in $\mathcal{M}_z$ and $ \eta $ in the circular limit for equal-mass binaries 
   for a specific sky position and inclination
   $\theta_N=\pi/2$, $\phi_N=\pi/3$, $\theta_L=\pi/4$, and $\phi_L=\pi/5$ in
   each case, we have assumed detection with the detector network introduced in Table 
   \ref{tab:DetCord}, and errors correspond to a fixed $\mathrm{SNR}_\mathrm{tot}=100$. 
   We find similar trends for other random choices of binary directions and orientations (not shown). }  
   \label{Table:CircLimParams}
\end{table}

\section{Summary and Conclusion} 
\label{sec:Conc} 

 We carried out a Fisher-matrix-type study to determine the accuracy with which the parameters of highly 
 eccentric BH binaries may be measured using the aLIGO-AdV-KAGRA GW detector network. Eccentricity changes 
 the GWs of binaries compared to circular binaries, in several ways. In time-domain, the gravitational 
 waveform of eccentric binaries is quasiperiodic but not sinusoidal. Relativistic precession adds a slow 
 amplitude modulation to the waveform for each polarization. Eccentricity also changes the inspiral rate 
 at which the binary separation and period shrink. We take all of these effects into account using the 
 stationary phase approximation \citep{MorenoGarridoetal1994,Mikoczietal2012}. In contrast to circular 
 binaries, the waveform of eccentric binaries includes several prominent orbital frequency harmonics, 
 general relativistic precession causes each harmonic to split into three frequencies for both GW 
 polarizations respectively, and the eccentric inspiral creates a spectrum, which is different from the
 $\tilde{h} \propto f^{-7/6}$ waveform of circular inspiral sources for each harmonic. These features in 
 the waveform make it possible to accurately determine the eccentricity and angle of periapsis, and the 
 modulated inspiral rate improves the measurement accuracy of mass parameters for eccentric inspirals. 

 The main parameters that describe eccentric inspiraling binaries are the initial pericenter distance 
 $\rho_{\mathrm{p}0}$ when the eccentricity is close to unity and the final eccentricity at the last 
 stable orbit $e_\mathrm{LSO}$. These parameters are systematically different for different formation 
 channels. Thus their measurement may have important implications on the astrophysical origin of the 
 sources
 \citep{OLearyetal2009,Cholisetal2016,Rodriguezetal2016a,Rodriguezetal2016b,Chatterjeeetal2017,Gondanetal2017,SamsingRamirezRuiz2017,SilsbeeTremaine2017}. 
 Based on a survey with $10 \, \Msun - 10 \, \Msun$ and $30 \, \Msun - 30 \, \Msun$ precessing 
 highly eccentric BH binaries at $100\, \mathrm{Mpc}$ using the planned aLIGO-AdV-KAGRA detector network,
 our results are be summarized as follows.
 \begin{enumerate}
 \item  The $\mathrm{SNR}_\mathrm{tot}$ improves by a factor of $\sim 1- 1.7$ (depending on 
 $\rho_{\mathrm{p}0}$, the component masses, and the sky position and binary orientation angles, see 
 Figure \ref{fig:SNR_Prec}) for $30 \, \Msun - 30 \, \Msun$ precessing highly eccentric BH binaries 
 compared to similar binaries in the circular limit \footnote{We adopted the leading order stationary 
 phase approximation waveform for circular sources.} with the same masses and  distance. The volume in 
 the Universe for a fixed maximum $S/N$ is \mbox{$\langle (S/N)^3\rangle \sim 5 \times$ ($2\times$)} 
 larger for eccentric inspiraling binaries with $\rho_{\mathrm{p}0}=10$ ($\rho_{\mathrm{p}0} 
 = 20$) than for similar binaries in the circular limit.
 
\item We determined how the parameters' measurement accuracies depend on the initial dimensionless
 pericenter distance ($\rho_{\mathrm{p}0}$) for precessing highly eccentric BH binaries. The smallest 
 errors are obtained for small $\rho_{\mathrm{p}0} < 10$ values for the sky position and angular 
 momentum and $\rho_{\mathrm{p}0} < 20$ for the luminosity distance and $\rho_{\mathrm{p}0}$. However, 
 the errors for fast parameters, which are sensitive to the GW phase, like the chirp mass, the initial 
 eccentricity, and the eccentricity at the last stable orbit improve most significantly for a higher 
 $\rho_{\mathrm{p}0}$ between $10$ and $80$ (Figure \ref{fig:Delta_rho}). 
 
 \item The parameter estimation errors can improve significantly for highly eccentric precessing BH binaries 
 compared to similar binaries in the circular limit by a factor of (depending on $\rho_{\mathrm{p}0}$,  
 the component masses, and the sky position and binary orientation angles, 
 see Figures \ref{fig:Dist1} and \ref{fig:Delta_rho})
\begin{itemize}
 \item  $\sim 1-200$ for the mass errors, 
 \item  $\sim 1-4.5$ for the semi-major and semi-minor axes of the sky localization ellipse, 
 \item  $\sim 1-2$ for the distance errors, 
 \item  $\sim 1-3$ for the semi-major and semi-minor axes of the error ellipse for the binary 
 orientation.
\end{itemize}

 \item \label{i:ecc} For initially highly eccentric BH binaries at $D_{\rm L}=100\,\rm Mpc$, 
 the measurement errors for parameters specific to precessing highly eccentric BH binary sources may 
 be as low as of order (depending on $\rho_{\mathrm{p}0}$, the component masses, and the sky position 
 and binary orientation angles, see Figures \ref{fig:Dist2} and \ref{fig:Delta_rho})
\begin{itemize}
 \item $10^{-5}$ for the final eccentricity errors at LSO, 
 \item $10^{-4}$ for the initial eccentricity errors,
 \item  $10^{-3}$ for the initial pericenter distance.
\end{itemize}

 \end{enumerate}

 For initially moderately eccentric to low eccentricity binaries, the parameter measurement
 errors and SNRs improve by a smaller amount for low to intermediate $\rho_{\mathrm{p}0}$ (Figure 
 \ref{fig:Err_LowEccDep}).

 Note that the eccentricity errors are remarkably low, which is not surprising given 
 that eccentricity is encoded in several measurable features of the waveform including the orbital 
 harmonics, the splitting of each frequency harmonic into triplets, the frequency evolution 
 of harmonics (the ``chirp''), the frequency evolution of the distance between the spectral 
 triplets, the low frequency cutoff of the signal at the initial pericenter frequency, and 
 the eccentricity dependence of the last stable orbit where the inspiral transitions into a
 rapid coalescence. 
 
 However, there are several factors which may significantly increase the measurement errors in
 more typical cases. First, more typical sources are expected to be at much larger distances 
 than $100\,\mathrm{Mpc}$. Assuming crudely that the eccentricity errors scale with $D_\mathrm{L}$,
 the median measurement errors for a $30 \, \Msun - 30 \, \Msun$ precessing highly eccentric BH
 binary at $\sim 410\, \mathrm{Mpc}$ (similar to GW150914) are expected to be $\Delta e_\mathrm{LSO} \sim 8.8 \times 
 10^{-4}$ ($1.3 \times 10^{-4}$) for the final eccentricity at the last stable orbit if $\rho_
 {\mathrm{p}0} = 10$ ($20$)
 for the design sensitivity of the aLIGO/AdV/KAGRA instruments. In these cases, the expected median initial eccentricity error is 
 $\Delta e_0 \sim 8.9 \times 10^{-3}$ $(1.2 \times 10^{-2})$, and the median initial pericenter 
 distance is $\Delta \rho_{\mathrm{p}0} \sim 4.4 \times 10^{-2}$ ($9.4 \times 10^{-2}$; see Table 
 \ref{tab:PrecBin}). 
 
 Another important simplifying assumption, which may have skewed the errors to lower values, was 
 to neglect higher order post-Newtonian (PN) corrections which depend on the spin of the merging
 objects. The spins of the two binary components introduce $6$ additional parameters, which may 
 become partially degenerate with all other parameters thereby increasing their errors. On the 
 other hand, spin precession breaks degeneracies between the binary orientation and other slow 
 parameters \citep{LangHughes2006,Kocsisetal2007,Chatziioannouetal2014}. However, the 
 eccentricity-induced orbital harmonics enter at the Newtonian order, the frequency-triplets 
 due to GR precession enter at the low $1$PN order. Therefore, the eccentricity-related spectral
 features are dominant already at the early stages of the inspiral when higher order PN corrections
 are negligible. For this reason, the estimated $\Delta e_0$ and $\Delta \rho_{\mathrm{p}0}$ errors
 are expected to be robust. On the other hand, most of the SNR accumulates at late times for 
 stellar-mass BH binaries where the high order PN corrections are significant. We leave an 
 estimate of parameter estimation errors for spinning binaries to future work. 
 
 Finally, an important simplification is the Fisher matrix method itself, which is valid only 
 if the waveform model is a faithful representation of the GW signal and the noise is Gaussian 
 and sufficiently small that the SNR is sufficiently large that the parameter error region is 
 an ellipsoid in parameter space and when the parameter derivate of the waveform is non-vanishing. 
 For smaller SNR, the parameter error region geometry is more complex and the uncertainties are 
 generally higher \citep[see][and reference therein]{CornishLittenberg2015}. The parameter 
 estimation errors may also be affected by theoretical uncertainties of the waveform model 
 \citep{CutlerVallisneri2007}, which may be especially important for highly eccentric binaries, 
 where the post-Newtonian expansion is slowly convergent \citep{KocsisLevin2012}. 
 
 Such low eccentricity errors may give the aLIGO-AdV-KAGRA GW detector network the capability 
 to distinguish among different astrophysical formation channels. In a companion study 
 \citep{Gondanetal2017}, we illustrate the expected distribution of eccentricities and other physical
 parameters for single-single GW capture sources in galactic nuclei. Similar studies for other 
 astrophysical formation channels are underway.

 Future multi-waveband searches of eccentric inspiral sources with \textit{LISA} and aLIGO-AdV-KAGRA
 \citep{KocsisLevin2012,Sesana2016} have good prospects for even more accurate measurements of the
 physical parameters well beyond the level reported here. In that case the GW frequency range is 
 much wider. Since eccentricity decreases due to GW emission, eccentricity may be expected to be 
 much higher at lower frequencies in the \textit{LISA} band. This leads to a much larger total GW 
 phase shift caused by eccentricity. The better measurement of relativistic precession may more 
 efficiently break degeneracies between mass and other parameters. The modulation caused by the 
 orbit of the instrument around the Sun and Earth's spin can help break degeneracies among source
 direction, orientation, and other parameters. Accounting for eccentricity for third generation 
 Earth based (e.g. Einstein telescope), deci-Hertz to mHz space-based instruments  will be essential 
 \citep{ChenAmaroSeoane17}.

\section*{Acknowledgment}
 We thank the anonymous referee for constructive comments, which helped improve the quality of the 
 paper. This project has received funding from the European Research Council (ERC) under the European
 Union's Horizon 2020 research and innovation programme under grant agreement No 638435 (GalNUC) and 
 by the Hungarian National Research, Development, and Innovation Office grant NKFIH KH-125675. This 
 work was performed in part at the Aspen Center for Physics, which is supported by National Science 
 Foundation grant PHY-1607761. The calculations were carried out on the NIIF HPC cluster at the
 University of Debrecen, Hungary.

\appendix

\section{The circular limit} 
\label{sec:circlim} 
 
 The circular limit follows from the limit $e_0 \rightarrow 0$ for arbitrary $\rho_{\mathrm{p}0}$ or 
 $\rho_{\mathrm{p}0} \rightarrow \infty$ for arbitrary $e_0 \leq 1$ (see Figure 5 in \citet{OLearyetal2009}). 
 In practice we set $e_0 = 10^{-4}$, $\rho_{\mathrm{p}0} = 1000$, and omit the parameters from the
 Fisher matrix that become degenerate or unconstrained in the circular limit: $\gamma_c$, $e_0$, and 
 $e_\mathrm{LSO}$. Thus, the parameters in the circular limit are
 \begin{align} \label{eq:lambdacirc2}
   \bm{\lambda}_\mathrm{Prec,circ} & =  \lbrace t_c,\Phi_c,\mathrm{ln} (D_\mathrm{L}), \mathrm{ln} 
  (\mathcal{M}_z),  \mathrm{ln} (M_{\mathrm{tot},z}),\theta_N,\phi_N,\theta_L,\phi_L\rbrace 
  \, , 
\end{align}  
 where $M_{\mathrm{tot},z}$ arises due to precession even in the circular limit (see Equation 
 (\ref{eq:dgammadt})). In the $\rho_{\mathrm{p}0} \rightarrow \infty$ limit, $\Delta \gamma_0 
 \rightarrow \infty$ and $\Delta e_0 \rightarrow \infty$, however the Fisher matrix algorithm 
 becomes invalid for $\Delta e_\mathrm{LSO}$ in this limit (Section \ref{subsec:ResParamDist}).
 
 We have also examined how the measurement errors depend on the total mass of the binary in the
 circular limit, and qualitatively compared our results to those of previous parameter estimation
 studies in Section \ref{sebsec:CompWithPrevRes}. We have found an excellent match between our 
 results and the results of previous studies.

\section{Response of an individual ground-based detector} 
\label{sec:RoGBD}
 Here we describe the measured signal of individual ground-based detectors, and introduce the adopted
 coordinate system.
 
 We define the Cartesian coordinate system with basis vectors $\mathbf{i}$, $\mathbf{j}$, $\mathbf{k}$
 and a spherical coordinate system ($\theta, \phi$) fixed relative to the center of the Earth, such that 
 $\mathbf{k}$ and $\theta=0$ is along the North geographic pole and $\phi=0$ is along the prime meridian.
 We denote the unit vector pointing from the center of Earth to the binary's sky position as $\mathbf{N}$,
 and define $\mathbf{L}$ to be the normal vector parallel with the binary's orbital angular momentum, 
\begin{align} \label{eq:nvec}
  \mathbf{N} &= \sin \theta_N \cos \phi_N \, \mathbf{i} + \sin \theta_N 
  \sin \phi_N \, \mathbf{j} + \cos \theta_N \, \mathbf{k} \, ,
  \\\label{eq:Lvec}
  \mathbf{L} &= \sin \theta_L \cos \phi_L \, \mathbf{i} + \sin \theta_L 
  \sin \phi_L  \, \mathbf{j} + \cos \theta_L \, \mathbf{k} \, .
\end{align} 

 We denote the unit vectors parallel with the arms of the $k$-th detector as $\mathbf{x}_k$ and 
 $\mathbf{y}_k$, and set $\mathbf{z}_k = \mathbf{x}_k \times \mathbf{y}_k$. As $\mathbf{x}_k$ and
 $\mathbf{y}_k$ are parallel with the surface of Earth for all detectors, $\mathbf{z}_k$ points 
 from the center of Earth toward the geographical location of the $k^{\mathrm{th}}$ detector. Let 
 the coordinates ($\theta_k, \phi_k$) denote the location of the $k^{\mathrm{th}}$ detector, thus 
 the unit vectors along the arms can be expressed as
\begin{align} \label{eq:xarm}
 \nonumber \mathbf{x}_k & = \left( \cos \psi_k \sin \phi_k  - \sin \psi_k \cos \phi_k 
   \cos \theta_k \right) \mathbf{i} \\ \nonumber
   & + \left( -\cos \psi_k \cos \phi_k -\sin \psi_k \sin \phi_k \cos \theta_k
   \right) \mathbf{j} \\ 
   & + \left( \sin \psi_k \sin \theta_k \right) \mathbf{k} \, , 
\end{align}
\begin{align} \label{eq:yarm}
\nonumber  \mathbf{y}_k & = \left( -\sin \psi_k \sin \phi_k - \cos \psi_k \cos \phi_k 
  \cos \theta_k \right) \mathbf{i} \\ \nonumber
  & + \left( \sin \psi_k \cos \phi_k - \cos \psi_k \sin \phi_k \cos \theta_k 
   \right) \mathbf{j} \\  
  & + \left( \cos \psi_k \sin \theta_k \right) \mathbf{k} \, ,
\end{align}
\begin{equation} \label{eq:zxarm}
 \mathbf{z}_k = \sin \theta_k \cos \phi_k \, \mathbf{i} + \sin \theta_k 
  \sin \phi_k \, \mathbf{j} + \cos \theta_k \, \mathbf{k} \, 
\end{equation}  
 \citep{CreightonAnderson2011}, where the orientation angle of the $k^{\mathrm{th}}$ detector, 
 $\psi_k$, is defined in Section \ref{sec:GWdetectors}. 
 
 These vectors define the response tensor for the $k^{\mathrm{th}}$ detector:
\begin{equation}  \label{eq:DetResp} 
  D_k^{ij} = \frac{1}{2} \left( x_k^i x_k^j - y_k^i y_k^j \right) \, 
\end{equation} 
 \citep{FinnChernoff1993}, where $x_k^i$ and $y_k^i$ are the $i^\mathrm{th}$ Cartesian components 
 of $\mathbf{x}_k$ and $\mathbf{y}_k$. 
 
 We adopt the basis vectors following the conventions of previous studies 
 \citep{FinnChernoff1993,CutlerFlanagan1994,Andersonetal2001,Dalaletal2006,Nissankeetal2010} 
\begin{equation}  \label{eq:polbasis} 
 \mathbf{X} = \frac{\mathbf{N} \times \mathbf{L}}{\arrowvert \mathbf{N} \times \mathbf{L} \arrowvert} 
 \, , \quad
 \mathbf{Y} = \frac{\mathbf{X} \times \mathbf{N}}{\arrowvert \mathbf{X} \times \mathbf{N} \arrowvert} 
 \, 
\end{equation}
 with preferred polarization basis tensors 
\begin{align}
 e^+_{ij}= X_i X_j - Y_i Y_j \, , \label{eq:ePtensor} \\
 e^{\times}_{ij} = X_i Y_j + Y_i X_j \, ,  \label{eq:eXtensor}
\end{align}
 where $i$ and $j$ are Cartesian components. Thus, the transverse-traceless metric perturbation describing
 the GW is written as 
\begin{equation} \label{eq:TTgauge} 
  h_{ij} = h_+ e^{+}_{ij} + h_{\times} e^{\times}_{ij} \, , 
\end{equation} 
 where $h_+$ and $h_{\times}$ are given in Equations (\ref{eq:hplusBessel}) and (\ref{eq:hxBessel}). 
 
The response of the $k^{\mathrm{th}}$ detector to a GW with frequency $f$ can be given in time-domain by
\begin{equation} \label{eq:ha} 
  h_k = e^{i \Delta \Phi_k} D_k^{ij} h_{ij} 
  = e^{i \Delta \Phi_k}\left( h_+ F_{+,k} + h_{\times} F_{\times,k} \right) \,  
\end{equation} 
 \citep{Nissankeetal2010}, where $\mathbf{r}_k = R_{\oplus} \mathbf{z}_{k}$ is the position of the 
 $k^{\mathrm{th}}$ detector, the factor $-\mathbf{N} \cdot\mathbf{r}_k$ measures the light travel time
 between the $k^{\mathrm{th}}$ detector and the coordinate origin, thus the factor $\Delta \Phi_k = - 2
 \pi f \mathbf{N} \cdot \mathbf{r}_k$ measures the phase shift between the $k^{\mathrm{th}}$ detector 
 and the coordinate origin. In Equation (\ref{eq:ha}), $F_{+,k}$ and $F_{\times,k}$ are the antenna factors 
\begin{equation}  \label{eq:FxFp}
  F_{+,k} = e^{+}_{ij} D_k^{ij} \, , \quad F_{\times,k} = e^{\times}_{ij} D_k^{ij} \, .
\end{equation}
 In our calculations, Earth is taken to be a sphere with radius of $R_{\oplus} = 6,370 \, \mathrm{km}$. 
 
 If the time that the GW signal spends in the detectors' sensitive frequency band is negligible compared 
 to the rotation period of Earth, then the measured waveform in frequency domain for the $k^{\mathrm{th}}$ 
 detector is
\begin{equation}   \label{eq:Fourierha}
 \tilde{h}_k (f) = \left[ F_{+,k}\tilde{h}_+(f)  
 +  F_{\times,k}\tilde{h}_{\times} (f) \right] e^{- 2\pi i f \mathbf{N} \cdot \mathbf{r}_k}\, ,
\end{equation}
 where $\tilde{h}_+(f)$ and $\tilde{h}_{\times} (f)$ are the Fourier-transformed expressions of 
 $h_+$ and $h_{\times}$ at Earth's center, and $F_{+,k}$ and $F_{\times,k}$ are given by the 
 (practically time-independent) orientation of the detectors shown in Table \ref{tab:DetCord}.
 
 Similarly, using the frequency harmonic triplets for eccentric precessing inspiraling binaries in 
 the stationary phase approximation, the measured waveform for the $k^{\mathrm{th}}$ detector is
\begin{equation} \label{eq:haGen}
  \tilde{h}_k(\mathbf{f}) =  F_{+,k}\tilde{h}_{+,k} (\mathbf{f}) +
   F_{\times,k} \tilde{h}_{\times,k} (\mathbf{f}) \, ,
\end{equation}
 where $\tilde{h}_{+,k} (\mathbf{f})$ and $\tilde{h}_{\times,k} (\mathbf{f})$ can be derived from
 $\tilde{h}_{+} (\mathbf{f})$ and $\tilde{h}_{\times} (\mathbf{f})$ by multiplying each term of
 $\mathbf{f}$ with the phase shift factors $e^{ -2\pi i f_n \mathbf{N} \cdot \mathbf{r}_k}$ and 
 $e^{ -2\pi i f_n^{\pm} \mathbf{N} \cdot\mathbf{r}_k}$ for each harmonic, respectively. More 
 specifically, the measured signal's Fourier phase in Equations (\ref{eq:PsiN}) and (\ref{eq:Psipm}) 
 in the $k^{\mathrm{th}}$ detector is shifted respectively according to
 \begin{align}
 \Psi_{n,k} &= \Psi_{n} -  2\pi i f_n \mathbf{N} \cdot \mathbf{r}_k\,,\\
  \Psi_{n,k}^{\pm} &= \Psi_{n}^{\pm} -  2\pi i f_n^{\pm} \mathbf{N} \cdot \mathbf{r}_k\,.
 \end{align}

\section{Time evolution of the orbit}
\label{sec:TimeinDetBand}
 
 In this section, we derive the time evolution of different harmonics in the detectors' sensitive frequency
 band, and for each orbital harmonic we determine the eccentricity at which the signal enters the detectors' 
 sensitive frequency band. These formulae will be utilized in Appendix \ref{sec:NumEffSNRFisher}.
 
 The time-dependent GW signal of a precessing eccentric BH binary as measured by the $k^{\mathrm{th}}$ 
 detector can be given as
\begin{align}  \label{eq:hkt}
  h_k(t) & = h_+(t) F_{+,k}\left[\alpha_N(t), \beta_N, \alpha_L, \beta_L \right] 
  \nonumber  \\ 
 & + h_{\times}(t) F_{\times,k}\left[\alpha_N(t), \beta_N, \alpha_L, \beta_L \right] \, ,
\end{align}
 where $h_+(t)$ and $h_{\times}(t)$ are given in Equations (\ref{eq:hplusBessel}) and (\ref{eq:hxBessel}), 
 and $F_{+,k}$ and $ F_{\times,k}$ are quantified by Equation (\ref{eq:FxFp}). We neglect spins in this 
 study, therefore the angular momentum vector direction $(\alpha_L, \beta_L)$ is conserved during the 
 eccentric inspiral \citep{CutlerFlanagan1994}. The polar angle of the source $\alpha_N$ relative to 
 the detector depends on the rotation phase of Earth during the day. We neglect Earth's rotation, since 
 the total duration of an eccentric inspiral from $e=0.9$ to merger is of order $[(\rho_{\mathrm{p}0} / 
 40)^4 (4\eta)^{-1} M_{ \rm tot}/(20 \, \Msun) \, \mathrm{min}]$ as shown in Figure 3 in 
 \citet{OLearyetal2009}. \footnote{\label{fn:Earthrotate}Earth's rotation may be relevant for highly 
 eccentric low mass compact objects with large $\rho_{\mathrm{p}0} \gtrsim 40$, such as neutron star 
 binaries. For black holes, if $\rho_{p0} \gg 40$, then the signal mostly circularizes before it enters
 the detectors' sensitive frequency band, and the amount of time it spends in the band with a significant
 $\mathrm{SNR}_\mathrm{tot}$ is limited to less than a minute. For an illustration of the accumulation of 
 the $\mathrm{SNR}_\mathrm{tot}$ with time we refer the reader to Figure 10 of \citet{OLearyetal2009} and 
 Figure 7 of \citet{KocsisLevin2012}. } 

 The waveform of an eccentric binary in the stationary phase approximation is a sum over harmonics
 $n$ for each component of the frequency triplet ($f_n, f_n^+, f_n^-$) with different reference 
 times $(t_n, t_n^+, t_n^-)$  (see Section \ref{subsec:FreqDom}). During the evolution $e(t)$
 (and $\rho_p(t)$) shrinks strictly monotonically in time \citep{Peters1964}, therefore its inverse 
 function $t(e)$ is well-defined and determines $t_n$, $t_n^+$, and $t_n^-$. For inspiraling 
 circular binaries, time $t$ can be expressed using the frequency of the emitted GW signal $f=2\nu$ as
\begin{equation}  \label{eq:defte}
  t(f) = t_c - \int_f^{\infty} \frac{df'}{\dot{f}'} = t_c - 5 (8 \pi f)^{-8/3} \mathcal{M}^{-5/3}
\end{equation} 
 (see equations (2.13) and (2.19) in \citet{CutlerFlanagan1994} for details), where the constant
 of integration, $t_c$, is defined by the requirement that $t \rightarrow t_c$ as $f \rightarrow 
 \infty$. We generalize Equation (\ref{eq:defte}) for eccentric inspirals by changing the integration
 variable from $f$ to $e$ in Equation (\ref{eq:defte}), 
\begin{equation}  \label{eq:te}
 t(e) = t_c + \int_0^e \frac{de'}{\dot{e}(e')} = t_c - \tau I_t(e) \, ,
\end{equation} 
 where $\dot{e}$ is given by Equation (\ref{eq:dedt}), and we introduced
\begin{align}  \label{eq:tau}
\tau & = \frac{15}{304} \mathcal{M}^{-5/3} (2 \pi \CC)^{-8/3}  \nonumber  \\
 & = \frac{15}{304} \frac{M_{\mathrm{tot}}^{8/3}}{\mathcal{M}^{5/3}} \frac{[(1-e_\mathrm{LSO})
 \rho_\mathrm{pLSO}(e_\mathrm{LSO})]^{4}}{H(e_\mathrm{LSO})^{8/3}}  \, ,
\end{align} 
 and substituted $\CC$ using Equation (\ref{eq:CC}). Here $\rho_{\mathrm{pLSO}}(e_{\mathrm{LSO}})$ and 
 $H(e_{\mathrm{LSO}})$ are given by Equations (\ref{eq:defeLSO}) and (\ref{eq:He}), and $I_t(e)$ in 
 Equation (\ref{eq:te}) is of form
\begin{equation}  \label{eq:Ite}
  I_t(e) = \int_0^e \frac{x^{29/19} 
 \left( 1+\frac{121}{304} x^2\right)^{\frac{1181}{2299}}} {(1-x^2)^{3/2} } \, dx 
\end{equation} 
 \citep[see][for an analytic result]{Mikoczietal2012}. Using Equation (\ref{eq:te}) we get the 
 total duration of the $n^{\mathrm{th}}$ harmonic in the detector's sensitive frequency band
\begin{align}\label{eq:Tn1}
  T_{n} & = t(e_\mathrm{min,n})-t(e_\mathrm{max,n}) = [ I_1(e_{\mathrm{max},n}) - I_1 (e_{
  \mathrm{min},n}) ]\tau  \, ,
  \\ 
  T_n^+ & = t(e_\mathrm{min,n}^+)-t(e_\mathrm{max,n}^+) 
  =  [ I_1(e_{\mathrm{max},n}^+) - I_1 (e_{\mathrm{min},n}^+) ]\tau
  \, ,
  \\
  T_n^- & = t(e_\mathrm{min,n}^-)-t(e_\mathrm{max,n}^-) 
  =  [ I_1(e_{\mathrm{max},n}^-) - I_1 (e_{\mathrm{min},n}^-) ]\tau\, .\label{eq:Tn3}
\end{align}
 Here $e_\mathrm{min,n}$ ($e_\mathrm{min,n}^+,e_\mathrm{min,n}^-$) refers to the eccentricity at which the 
 harmonic $f_n$ ($f_n^+,f_n^-$) reaches the LSO or when it exists the detectable highest frequency for the 
 given detector, and $e_\mathrm{max,n}$ ($e_\mathrm{max,n}^+,e_\mathrm{max,n}^-$) refers to the eccentricity
 at which the signal related to $f_n$ ($f_n^+,f_n^-$) first enters the detector's sensitive frequency band 
 or when it forms within the band. Thus, 
 \begin{align}  
 \label{eq:emin} e_{\mathrm{min},n}  & = 
  \mathrm{max}(e_\mathrm{LSO}, e^{\mathrm{min}}_{\mathrm{det},n}) \,, 
    \\  
  \label{eq:emax} e_{\mathrm{max},n} &  = 
 \mathrm{min}(e_0, e^{\mathrm{max}}_{\mathrm{det},n}) \, , 
 \end{align} 
  where
  \begin{align}  \label{eq:eminmaxdet}
    e^{\mathrm{min}}_{\mathrm{det},n} & = \nu_n^{-1}( f_{\mathrm{det,max}})\, , 
    \\ 
    e^{\mathrm{max}}_{\mathrm{det},n} & = \nu_n^{-1}( f_\mathrm{det,min}) \, ,
    \\
    \label{eq:nune} \nu_n (e) & = n \nu(e) \, .
  \end{align}
  Here $\nu(e)$ is given analytically by Equation (\ref{eq:nue}), $\nu_n^{-1}(\cdot)$ denotes the
  inverse function of $\nu_n(e)$, $f_\mathrm{det,min}$ and $f_\mathrm{det,max}$ are the lower and
  upper limits of the detector's sensitive frequency band (typically $10 \, \mathrm{Hz}$ and $10^4
  \, \mathrm{Hz}$, respectively), and $e_\mathrm{LSO}$ is determined by Equation (\ref{eq:defeLSO}).
  Similarly, we define the parameters corresponding to $f_n^{+}$ and $f_n^{-}$ as
  \begin{align} 
 \label{eq:efirst} e_{\mathrm{min},n}^+ & = \mathrm{max}(e_{\mathrm{LSO}},
  e^{\mathrm{min}}_{\mathrm{det},n+} ) \, , \\ 
  e_{\mathrm{max},n}^+ & = \mathrm{min}(e_0, e^{\mathrm{max}}_{\mathrm{det},n+} )
  \, , \\ 
  e_{\mathrm{min},n}^- & = \mathrm{max}(e_{\mathrm{LSO}}, e^{\mathrm{min}}_{\mathrm{det},n-} ) \, , \\  
  e_{\mathrm{max},n}^- & = \mathrm{min}(e_0,  
 \label{eq:last} e^{\mathrm{max}}_{\mathrm{det},n-} ) \, ,
  \end{align} 
  where 
  \begin{align} 
  e^{\mathrm{min}}_{\mathrm{det},n+} & = \nu_{n+}^{-1}( f_\mathrm{det,max}) \, ,
  \\ 
  e^{\mathrm{max}}_{\mathrm{det},n+} & = \nu_{n+}^{-1}( f_\mathrm{det,min}) \, , 
  \\ 
  e^{\mathrm{min}}_{\mathrm{det},n-} & = \nu_{n-}^{-1}( f_\mathrm{det,max}) \, ,
  \\ 
  e^{\mathrm{max}}_{\mathrm{det},n-} & = \nu_{n-}^{-1}( f_\mathrm{det,min}) \, ,
  \\
  \nu_{n\pm}(e) & = n\nu (e) \pm \frac{\dot{\gamma}(e)}{\pi} \, , \label{eq:nupme}
  \end{align} 
 where $\dot\gamma(e)$ is given by Equation (\ref{eq:dgammadt}), and $\nu^{-1}_{n\pm}(\cdot)$ is the 
 inverse function of $\nu_{n\pm}(e)$ given by Equation (\ref{eq:nupme}).

\begin{figure}
    \centering
    \includegraphics[width=78mm]{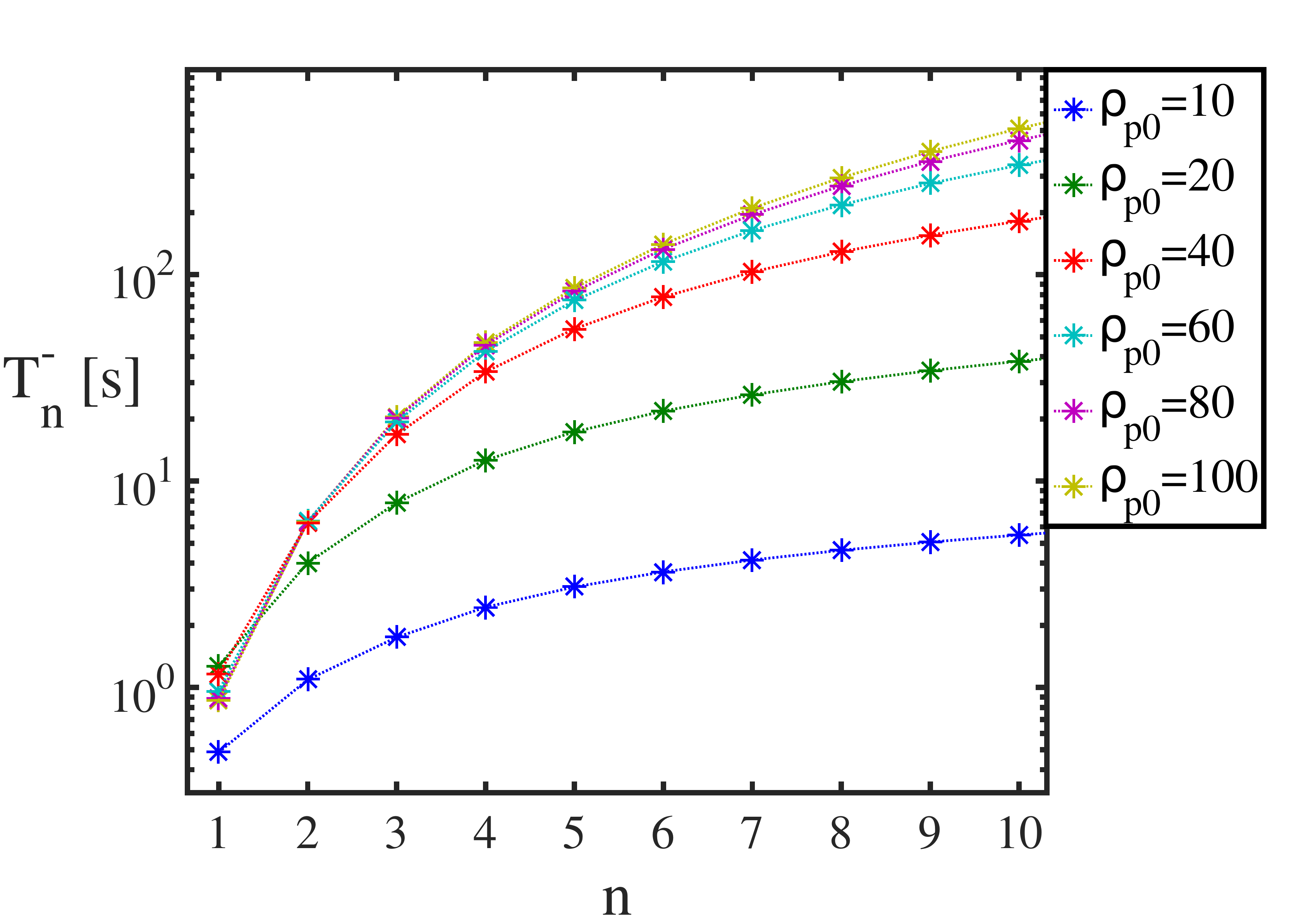} \\
    \includegraphics[width=78mm]{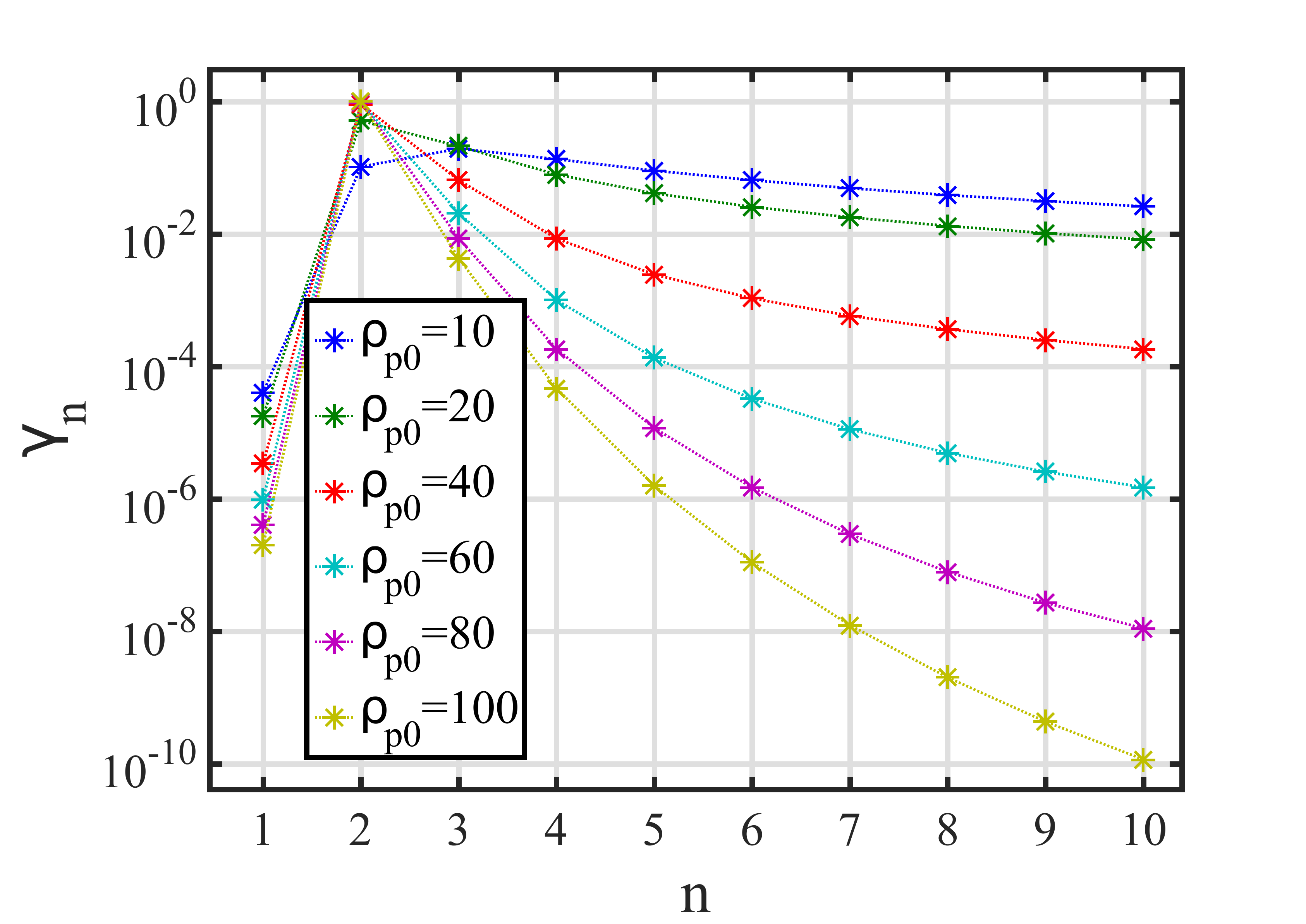}
\caption{ \emph{Top panel}: The time duration that the first 10 harmonics (specifically $f_n^-$ 
 here) spend in an aLIGO type detector's sensitive frequency band for \mbox{$30 \, \Msun-30 \, \Msun$
 precessing eccentric BH binaries with} initial eccentricity $e_0=0.9$ for various $\rho_{\mathrm{p}0}$ 
 values between 10 and 100 as labeled. For this choice of masses $T^-_n$ varies within $10 \%$ of its
 value shown for $e_0=0.9$ for $0.9 \leq e_0< 1$ for $n \in \{1,2,\dots,10\}$. \emph{Bottom panel}: The
 fraction of squared signal-to-noise ratio  \mbox{($\gamma_{n}=(\mathrm{SNR}_n^-)^2 \cdot (\mathrm{SNR}_
 \mathrm{tot})^{-2}$)} in the first ten harmonics corresponding to $f_n^-$ frequencies in aLIGO for 
 \mbox{$30 \, \Msun-30 \, \Msun$} precessing eccentric BH binaries with initial eccentricity $e_0=0.9$ 
 for different $\rho_{\mathrm{p}0}$ values as labeled. We show results for the $f_n^-$ components
 of the frequency triplet $(f_n, f_n^+, f_n^-)$ because the dominant fraction of the $\mathrm{SNR}_
 \mathrm{tot}$ accumulates in these frequencies. For any $n \in \{1,2,\dots,10\}$ and \mbox{$0.9 
 \leq e_0\leq 0.99$}, $\gamma_n$ varies by less than $60 \%$ of its value shown for $e_0=0.9$. } 
 \label{fig:time}
\end{figure}

 In practice, the second term is negligible in  Equations (\ref{eq:Tn1})-(\ref{eq:Tn3}).
 We find that $T_n^{+}$ and $T_n^{-}$ are within $\leq 20 \%$ of $T_n$ for any fixed $n$. The top
 panel of Figure \ref{fig:time} shows the total time the GW signal spends in an aLIGO type detector's
 sensitive frequency band for different harmonics. Higher harmonics enter the aLIGO band earlier and 
 that depending on $\rho_{\mathrm{p}0}$ the first 10 orbital harmonics spend between seconds to minutes
 in the detector's sensitive frequency band for a $30 \, \Msun - 30 \, \Msun$ precessing highly eccentric 
 BH binary. 
 
 The bottom panel of Figure \ref{fig:time} shows the fraction of the squared SNR that accumulates in
 different orbital harmonics for various $\rho_{\mathrm{p}0}$ for aLIGO. For high $\rho_{\mathrm{p}0}$,
 the signal effectively circularizes by the time it enters the detector's sensitive frequency band 
 and the $n=2$ harmonic dominates. However, the contribution of $n\neq 2$ is significant for $\rho_
 {\mathrm{p}0} \lesssim 20$ for a $30 \, \Msun - 30 \, \Msun$ precessing highly eccentric BH binary. 

\section{Calculating the signal-to-noise Ratio and the Fisher matrix} 
\label{sec:NumEffSNRFisher} 
 
 In this section, we derive numerically efficient formulae to calculate the SNR and the Fisher matrix 
 for individual detectors. We first neglect pericenter precession, then extend the calculations for 
 precessing eccentric sources.

\subsection{Signal-to-noise ratio}
\label{subsec:ExpSNR} 
 
\subsubsection{Eccentric inspirals without precession}  \label{subsubsec:ExpSNRNoPrec}
 
 The \emph{NoPrec} signal measured by a detector at position $\bm{r}$ is given in Fourier space from 
 Equations (\ref{eq:SPAp}) and (\ref{eq:SPAx}) as
\begin{equation}  \label{eq:hPPN} 
  \tilde{h}_\mathrm{NoPrec} = \sum_{n=1}^{\infty} L_n(e, f_n) \Theta_{\mathrm{H}}(e_0-e) 
  e^{i \Psi_n(e, f_n) } \, , 
\end{equation} 
 where $\Psi_n$ is the Fourier phase at the origin of the coordinate system set to the Earth's center, 
 given by Equation (\ref{eq:PsiN}), we set $\dot{\gamma} \equiv 0$, $\gamma\equiv \gamma_c$, $\Theta_
 {\mathrm{H}}(\cdot)$ denotes the Heaviside function, which is zero and unity for negative and positive 
 arguments\footnote{More precisely, we assume a smoothed truncation of the signal as 
 \begin{equation}\label{eq:ThetaH}
 \Theta_\mathrm{H}(e_0-e) = \left\{
 \begin{array}{ll}
 0 &{\rm if~~} e > e_0 \\
 \frac{e_0-e}{\delta e_0} &{\rm if~~} e_0 - \delta e_0 < e \leq e_0\\
 1 &{\rm if~~} e \leq e_0 - \delta e_0
 \end{array}
 \right.
 \end{equation}
 where $\delta e_0$ is the absolute change of the eccentricity during the first orbit, which from 
 Equations (\ref{eq:nuKepler}) and (\ref{eq:dedt}) is 
 \begin{equation}  \label{eq:deltae0}
 \delta e_0 = 2\pi\left|\frac{\dot{e}}{\nu}\right|_{e_0} = \frac{ 1216 \pi^2 }{ 15 } \frac{\eta}{
 \rho_{p0}^{5/2}}\frac{ e_0 }{(1+e_0)^{5/2}}\left(1+\frac{121}{304}e_0^2\right) \, ,
 \end{equation}
 where $\eta= (\mathcal{M}_z/M_{{\rm tot},z})^{5/3}$.}, respectively, and 
\begin{align}  \label{eq:Ln} 
 \nonumber L_n(e,f_n) = & - \left[\frac{  A_n \sin^2 \Theta}{4} +\frac{(1+\cos^2 \Theta)}{ 4 } 
  (B_n^+e^{2 i \gamma_c} - B_n ^- e^{-2 i \gamma_c} )\right]  \nonumber \\
 & \times h_0 F_+ \Lambda_n e^{i( \Delta \Phi_n - \pi/4)}  \nonumber \\
 & - \frac{i h_0 F_{\times} \Lambda_n \cos\Theta}{ 2 } 
  (B_n^+ e^{2 i \gamma_c} + B_n ^- e^{-2 i \gamma_c}) e^{i(\Delta \Phi_n+\pi/4)} \, , 
\end{align} 
 where $h_0$ and $\Lambda_n$ are given by Equations (\ref{eq:h0}) and (\ref{eq:Lambda}), $A_n$ and $B_n
 ^{\pm}$ are given by Equations (\ref{eq:AnBn}) and (\ref{eq:Cn}), $\gamma_c$ specifies the argument of 
 pericenter, which is assumed to be fixed here, $F_+$ and $F_\times$ are the antenna factors given by 
 Equation (\ref{eq:FxFp}). The factor \mbox{$\Delta \Phi_{n} =-2 \pi f_n \mathbf{N} \cdot \mathbf{r}$} 
 gives the phase shift of the measured signal between the position of the detector $\bm{r}$ and the 
 origin of the coordinate system for the $n^{\mathrm{th}}$ harmonic (Appendix \ref{sec:RoGBD}). $L_n$ 
 depends on $f_n$ implicitly through $\Delta \Phi_n$ and $h_0$. 
 
 In Equation (\ref{eq:hPPN}), $\Theta_{\mathrm{H}}(e_0-e)$ accounts for the start of the waveform 
 when the binary forms with initial eccentricity\footnote{The initial eccentricity $e_0$ does not 
 enter the waveform anywhere else, $L_n$ and $\Psi_n$ are independent of $e_0$. Due to this term, 
 $e_0$ and $e_{\mathrm{LSO}}$ may be measured independently, and $\Delta \rho_{\mathrm{p}0}$ follows
 from Equation (\ref{eq:deltarhop0}). } $e_0$. Along the same lines, a similar term $\Theta_
 {\mathrm{H}} (e-e_\mathrm{LSO})$ could be incorporated to account for the end of the eccentric 
 inspiral where the waveform transitions to a plunge and ringdown phase. However, we conservatively 
 do not account for such a term, since the waveform near the end of the inspiral is sensitive to 
 higher order post-Newtonian corrections, which are not known and neglected here 
 \citep{KocsisLevin2012,LoutrelYunes2017}. Nevertheless, the inspiral rate is sensitive to $e_\mathrm{LSO}$ 
 in Equations (\ref{eq:nue}) and (\ref{eq:He}) and (\ref{eq:CC}), which affects $L_n$ and $\Psi_n$. 
 
 For each detector, the square of the SNR for the \emph{NoPrec} waveform, $\mathrm{SNR}^2
 _\mathrm{NoPrec}$, can be obtained by substituting $\tilde{h}_\mathrm{NoPrec}$ into Equation 
 (\ref{eq:SNR2_1det}). We find that the product of sums in $\tilde{h}_\mathrm{NoPrec} \, \tilde{h}
 _\mathrm{NoPrec}^*$ is dominated by the elements such that \footnote{Numerically we confirm that 
 cross-terms proportional to $L_n \, L_m^* \, \exp(i\Psi_n- i\Psi_m)$ have a negligible contribution 
 for $n\neq m$.} 
\begin{equation}  \label{eq:SNR2PPNfn} 
 \left( \mathrm{SNR}_\mathrm{NoPrec} \right)^2 \approx 4 \sum_{n=1}^{\infty} \int_{f_{\mathrm{min},n}}
 ^{f_{\mathrm{max},n}} \frac{|L_n (e(f_n), f_n)|^2}{S_{n}(f_n)} df_n \, , 
\end{equation} 
 where $f_{\mathrm{min},n}$ is the frequency at which the $n^{\rm th}$ harmonic first enters the detector's
 sensitive frequency band or when it forms in the band, and similarly $f_{\mathrm{max},n}$ is the frequency
 at which the signal exits the detector's sensitive frequency band or when it reaches the LSO, 
 \begin{align}  \label{eq:fmaxn}
  f_{\mathrm{max},n} &= \mathrm{min} ( \nu_n(e_\mathrm{LSO}), f_\mathrm{det,max}) \, ,\\
   \label{eq:fminn}
   f_\mathrm{min,n} &= \mathrm{max}(\nu_n(e_0), f_\mathrm{det,min})  \, .
\end{align}
 
 Computationally it is practical to change the integration variable from $f_n$ to $e$ as
\begin{gather} \label{eq:dfnde} 
  df_n = n\left\arrowvert \frac{d\nu}{de} \right\arrowvert \, de \, , 
\end{gather} 
 thus Equation (\ref{eq:SNR2PPNfn}) can be rewritten generally as 
\begin{equation}  \label{eq:SNR2PPNe_Gen}
  \left( \mathrm{SNR}^{\mathrm{Gen}}_\mathrm{NoPrec} \right)^2 \approx 4 \sum_{n=1}^{n_\mathrm{max}(e_0)}
  \int_{e_{\mathrm{min},n}}^{e_{\mathrm{max},n}}  \frac{n|L_n (e)|^2}{S_{n}(n \nu(e))}  \left\arrowvert 
  \frac{d\nu}{de}  \right\arrowvert \, de \, , 
\end{equation} 
 where $\nu(e)$ is given analytically by Equation (\ref{eq:nue}), and $L(e)$ may be obtained from Equation
 (\ref{eq:Ln}) by substituting \mbox{$f_n = n \nu(e)$}. The integration bounds $e_{\mathrm{min},n}$ and 
 $e_{\mathrm{max},n}$ are given by Equations (\ref{eq:emin}) and (\ref{eq:emax}). We truncate the calculation
 beyond a maximum spectral harmonic $n_\mathrm{max}(e)$ defined in Equation (\ref{eq:nmax}). 
 
 $\mathrm{SNR}^{\mathrm{Gen}}_\mathrm{NoPrec}$ accurately recovers Equation (\ref{eq:SNR2PPNfn}) 
 generally for any value of the initial eccentricity in the range \mbox{$0 < e_0 < 1$}, however the number
 of considered harmonics $n_\mathrm{max}(e_0)$ increases rapidly for high $e_0$  (Equation \ref{eq:nmax}), 
 and $n_\mathrm{max}(e_0) \rightarrow \infty$ in the limit $e_0 \rightarrow 1$. Therefore, $\mathrm{SNR}^
 {\mathrm{Gen}}_\mathrm{NoPrec}$ is computationally efficient for low to moderate initial eccentricities 
 ($e_0 \la 0.8$) and it is inefficient for higher $e_0$. In order to make $\mathrm{SNR} ^{\mathrm{Gen}}
 _\mathrm{NoPrec}$ computationally efficient for high initial eccentricities, we reverse the order of 
 the sum and the integral in Equation (\ref{eq:SNR2PPNe_Gen}) and truncate the sum over harmonics at 
 $n_\mathrm{max}(e)$ \citep{OLearyetal2009}.\footnote{In this case the number of considered harmonics 
 reduces significantly.} Thus, we get
\begin{equation}  \label{eq:SNR2PPNe} 
 \left( \mathrm{SNR}^{\mathrm{High}}_\mathrm{NoPrec} \right)^2 \approx 4 \int_{e_{\mathrm{min},n}}
 ^{e_{\mathrm{max},n}} \sum_{n=1}^{n_\mathrm{max}(e)} \frac{n|L_n (e)|^2}{S_{n}(n \nu(e))}  
 \left\arrowvert \frac{d\nu}{de} \right\arrowvert \, de \, .
\end{equation} 
 We use the above introduced trick to derive computationally efficient formulae in the high 
 initial eccentricity limit for Fisher matrix elements in the precession-free case (Appendix
 \ref{subsubsec:ExpFisherNoPrec}) and for the SNR and the Fisher matrix elements in the 
 precessing case (Appendices \ref{subsubsec:ExpSNRPrec} and \ref{subsubsec:ExpFisherWithPrec}).

\subsubsection{Eccentric inspirals with precession} 
\label{subsubsec:ExpSNRPrec}
 
 We derive the SNR of the precessing model in this section. The Fourier-transformed waveform given
 by Equation (\ref{eq:haGen}) can be rewritten as 
\begin{align} \label{eq:hPPC} 
  \tilde{h}_\mathrm{Prec} & = \sum_{n=1}^{\infty} K_n (e, f_n) \Theta_{\mathrm{H}}(e_0-e) 
  e^{i \Psi_n(e, f_n)} 
  \nonumber \\
 & +  \sum _{n=1}^{\infty} K_n^+ (e, f_n^+) \Theta_{\mathrm{H}}(e_0-e) e^{i \Psi_n^+(e, f^+_n)} 
 \nonumber \\ 
 & + \sum_{n=1}^{\infty} K_n^- (e, f_n^-) \Theta_{\mathrm{H}}(e_0-e) e^{i \Psi_n^-(e, f^-_n)} \, , 
\end{align} 
 where the terms $K_n$, $K_n^+$, and $K_n^-$ are defined as 
\begin{align} 
 \label{eq:Cn_Sprec} K_n (e, f_n) & = -\frac{h_0 \sin^2 \Theta}{4} A_n \Lambda_n F_+ 
  e^{i (\Delta \Phi_n -\pi/4) }  \, , \\
  \label{eq:Cnp_Sprec} K_n^+ (e, f_n^+) & =  -\frac{h_0 \cos \Theta}{2} B_n^- \Lambda_n ^+
  F_{\times} e^{i (\Delta \Phi_n^+ + \pi/4) }  \nonumber \\
  & + \frac{h_0 \left( 1+ \cos^2\Theta \right)}{4} 
  B_n ^- \Lambda_n^+ F_+ e^{i( \Delta \Phi_n^+ - \pi/4) } \, , \\
  \label{eq:Cnm_Sprec} K_n^-(e, f_n^-) & = - \frac{h_0 \cos\Theta}{2} B_n^+ \Lambda _n^- 
  F_{\times} e^{i(\Delta \Phi_n^- + \pi/4)}  \nonumber \\ 
  & - \frac{h_0}{4} B_n^+ \Lambda_n^- F_+ \left(1+\cos^2 \Theta \right) e^{i (\Delta \Phi_n^- -\pi/4)} \, .
\end{align} 
 The terms $K_n$ and $K_n^{\pm}$ depend on $\nu$ through $h_0$, which are expressed with $f_n,f_n ^{\pm}$
 using Equations (\ref{eq:fn}) and (\ref{eq:fnpm}). Furthermore these equations depend on $f_n$ and $f_n^
 {\pm}$ through $\Delta \Phi_n$ and  $\Delta \Phi_n^{\pm}$ (e.g. $\Delta \Phi_n = -2 \pi f_n \mathbf{N} 
 \cdot \mathbf{r})$ and $\Delta \Phi_n^{\pm} = -2 \pi f_n^{\pm} \mathbf{N} \cdot \mathbf{r}$).
 
 Next, we substitute this waveform $\tilde{h}_\mathrm{Prec}$ into Equation (\ref{eq:SNR2_1det}).
 Similarly to that of the \emph{NoPrec} signal (Equation \ref{eq:hPPN}), cross terms in the product of sums in 
 $\tilde{h}_\mathrm{Prec} \, \tilde{h}_\mathrm{Prec}^*$ have negligible contributions to $\left( \mathrm{SNR}
 _\mathrm{Prec} \right)^2$, and so 
\begin{align}  \label{eq:SNR2PPCfnpm} 
 \nonumber \left( \mathrm{SNR}_\mathrm{Prec} \right)^2 & \approx 4 \sum_{n=1}^{\infty} 
 \int_{f_\mathrm{min,n}}^{f_\mathrm{max,n}} 
 \frac{\arrowvert K_n (e(f_n), f_n) \arrowvert^2}{S_{h}(f_n)} \, df_n 
 \nonumber \\
 & + 4 \sum_{n=1}^{\infty} \int_{f_\mathrm{min,n}^+}^{f_\mathrm{max,n}^+} 
 \frac{\arrowvert K_n^+ (e(f_n^+), f_n^+) \arrowvert^2 }{S_{h}(f_n^+)} \, df_n^+ 
 \nonumber \\
 & + 4 \sum_{n=1}^{\infty} \int_{f_\mathrm{min,n}^-}^{ f_\mathrm{max,n}^-} 
 \frac{ \arrowvert K_n^-(e(f_n^-), f_n^-) \arrowvert^2 }{S_{h}(f_n^-)} \, df_n^- \, , 
\end{align} 
 where $f_\mathrm{max,n}$ and $f_\mathrm{min,n}$ are defined in Equations (\ref{eq:fmaxn}) and 
 (\ref{eq:fminn}). The integration bounds for the integrals over $f_n^\pm$ are defined similarly 
 to $f_\mathrm{max,n}$ and $f_\mathrm{min,n}$ in Equation (\ref{eq:fmaxn}) and (\ref{eq:fminn}), 
\begin{align} 
  f_{\mathrm{max},n}^{\pm} & = \mathrm{min} (\nu_{n\pm}(e_\mathrm{LSO}),
  f_\mathrm{det,max}) \, , 
  \\
 f_\mathrm{min,n}^{\pm} & = \mathrm{max}(\nu_{n\pm}(e_0), f_\mathrm{det,min})  \, , 
\end{align} 
 where $\nu_{n\pm}$ is defined in Equation (\ref{eq:nupme}).

  Next, we change integration variables from $f_n$ to $e$ using Equation (\ref{eq:dfnde}) 
 and similarly from $f_n^{\pm}$ to $e$ using Equation (\ref{eq:fnpm}) as
\begin{equation} \label{eq:dfnpm}
 df_n^{\pm} = \left\arrowvert n \pm \frac{1}{\pi}\frac{d\dot{\gamma}}{d\nu} 
  \right\arrowvert \left\arrowvert \frac{d\nu}{de}
 \right\arrowvert \, de \, .
\end{equation} 
 Here $d\dot\gamma/d\nu$ is given by Equation (\ref{eq:dgammadt}) as
\begin{align}
 \nonumber \frac{d\dot{\gamma}}{d\nu} & = \frac{\partial\dot{\gamma}(\nu,e)}{\partial \nu} + 
 \frac{\partial\dot{\gamma}(\nu,e)}{\partial e} \frac{1}{d \nu/d e} 
 \\
 & = \left(\frac{5}{3\nu} -\frac{2 e}{1-e^2} \frac{1}{d \nu/d e}\right)\dot\gamma \, .
\end{align}  
 After truncating the sum over the harmonics to the relevant range as in Equation (\ref{eq:SNR2PPNe_Gen}),
 Equation (\ref{eq:SNR2PPCfnpm}) can be written as 
\begin{align}  \label{eq:SNRPPC2e_Gen}
 \left( \mathrm{SNR}^{\mathrm{Gen}}_\mathrm{Prec} \right)^2 & \approx 4 \sum_{n=1}^{n_\mathrm{max}(e_0)}
 \int_{e_{\mathrm{min},n}}^{e_{\mathrm{max},n}} \frac{n \arrowvert K_n (e) \arrowvert^2}{ S_{n}( \nu_n(e)) }
 \left\arrowvert \frac{d\nu}{ de} \right\arrowvert de  
 \nonumber  \\
  & + 4 \sum_{n=1}^{n_\mathrm{max}(e_0)} \int_{e_{\mathrm{min},n}^+}^{e_{\mathrm{max},n}^+}   
 \frac{ \arrowvert K_n^+ (e) \arrowvert^2 }{ S_{n}(\nu_{n+}(e))}
 \left\arrowvert n + \frac{1}{\pi}\frac{d\dot{\gamma}}{d\nu}
  \right\arrowvert \left\arrowvert \frac{d\nu}{de} \right\arrowvert de 
 \nonumber  \\
  & + 4 \sum_{n=1}^{n_\mathrm{max}(e_0)} \int_{ e_{\mathrm{min},n}^-}^{ e_{\mathrm{max},n}^-}   
 \frac{ \arrowvert K_n^- (e) \arrowvert^2 }{ S_{n}( \nu_{n-}(e)) }
  \left\arrowvert n - \frac{1}{\pi}\frac{d\dot{\gamma}}{d\nu}
   \right\arrowvert \left\arrowvert \frac{d\nu}{de} \right\arrowvert de \, ,
\end{align}
 where $K_n(e) \equiv K_n(e, \nu_n(e))$ and $K_n^{\pm}(e)\equiv K_n^{\pm}(e, \nu_{n\pm}(e))$. 
 The integration bounds are given by Equations (\ref{eq:efirst}) and (\ref{eq:last}). 
 
 Similarly to $\mathrm{SNR}^{\mathrm{Gen}}_\mathrm{NoPrec}$, $\mathrm{SNR}^{\mathrm{Gen}}_
 \mathrm{Prec}$ is computationally efficient only for low to moderate initial eccentricities \mbox{($e_0 
 \la 0.8$)}. For high initial eccentricities, the computationally efficient form of $\mathrm{SNR}
 ^{\mathrm{Gen}} _\mathrm{Prec}$, $\mathrm{SNR}^{\mathrm{High}}_\mathrm{Prec}$, can be given by 
 reversing the order of the sum and the integral as
\begin{align}  \label{eq:SNRPPC2e}
 \left( \mathrm{SNR}^{\mathrm{High}}_\mathrm{Prec} \right)^2 & \approx 4 
 \int_{e_{\mathrm{min},n}}^{e_{\mathrm{max},n}} \sum_{n=1}^{n_\mathrm{max}(e)}
 \frac{n \arrowvert K_n (e) \arrowvert^2}{ S_{n}( \nu_n(e)) }
 \left\arrowvert \frac{d\nu}{ de} \right\arrowvert \, de  
 \nonumber  \\
  & + 4 \int_{e_{\mathrm{min},n}^+}^{e_{\mathrm{max},n}^+} \sum_{n=1}^{n_\mathrm{max}(e)}  
 \frac{ \arrowvert K_n^+ (e) \arrowvert^2 }{ S_{n}(\nu_{n+}(e))}
 \left\arrowvert n + \frac{1}{\pi}\frac{d\dot{\gamma}}{d\nu}
  \right\arrowvert \left\arrowvert \frac{d\nu}{de} 
 \right\arrowvert\, de 
 \nonumber  \\
  & + 4 \int_{ e_{\mathrm{min},n}^-}^{ e_{\mathrm{max},n}^-} 
  \sum_{n=1}^{n_\mathrm{max}(e)}  
 \frac{ \arrowvert K_n^- (e) \arrowvert^2 }{ S_{n}( \nu_{n-}(e)) }
  \left\arrowvert n - \frac{1}{\pi}\frac{d\dot{\gamma}}{d\nu}
   \right\arrowvert \left\arrowvert \frac{d\nu}{de} 
  \right\arrowvert \, de \, ,
\end{align}

\subsection{Fisher Matrix}
\label{subsec:ExpFisher}

 Due to the similarity of the equations defining the SNR (see Equation (\ref{eq:SNR2_1det}))
 and the Fisher matrix (see Equation (\ref{eq:Fisher})), we may follow the same procedure to 
 derive numerically efficient formulae for the Fisher matrix in the limit of high initial
 eccentricity. Similarly to Appendix \ref{subsec:ExpSNR}, we start the analysis with the 
 \emph{NoPrec} model and then generalize the calculation to the \emph{Prec} model, which 
 accounts for the precessing case.

 \subsubsection{Eccentric inspirals without precession} 
 \label{subsubsec:ExpFisherNoPrec} 
 
 Let us substitute Equation (\ref{eq:hPPN}) into Equation (\ref{eq:Fisher}). Similarly to the 
 product of sums of orbital harmonics $\tilde{h}_\mathrm{NoPrec}\tilde{h}_\mathrm{NoPrec}^*$,
 we find numerically that the cross-terms in $\partial_j \tilde{h}_\mathrm{NoPrec} \, \partial_k 
 \tilde{h} _\mathrm{NoPrec}^*$ with $j\neq k$ have a negligible contribution to $\Gamma_{jk}$. 
 Thus, we find that the stationary phase approximation is applicable if we drop the cross-terms,
 and thus in Equation (\ref{eq:Fisher}) we may use
\begin{align} \label{eq:PPNhhPPN}
  \partial_j \tilde{h}_\mathrm{NoPrec} \, \partial_k \tilde{h}_\mathrm{NoPrec}^* & = 
  \sum_{n=1}^{\infty} \tilde{L}_{n,j}(e) \, \tilde{L}_{n,k}^*(e) \Theta_{H}(e_0-e) 
  \nonumber  \\
  & + \delta_{e_0,j} \sum_{n=1}^{\infty} \tilde{L}_n(e) \, \tilde{L}_{n,k}^*(e) \delta(e-e_0)
  \nonumber  \\
  & + \delta_{e_0,k} \sum_{n=1}^{\infty}\tilde{L}_{n,j}(e) \, \tilde{L}_{n}^*(e) \delta(e-e_0)
  \nonumber  \\
 & + \delta_{e_0,k} \delta_{e_0,j} \sum_{n=1}^{\infty} | L_n(e)|^2 
   \left[ \partial_{e_0} \Theta_{H}(e_0-e) \right]^2 \, . 
\end{align}
 Here
\begin{equation} \label{eq:tildeL}
 \tilde{L}_{n}(e) = L_n(f_n(e),e) \, e^{i \Psi_n(f_n(e),e)}  \, ,
\end{equation}
 and 
\begin{equation}  \label{eq:Lnpartdiff}
  \tilde{L}_{n,j}(e) = \partial_j \left[ L_{n}(f_n) e^{i \Psi_n(f_n)} \right](e) \, , 
\end{equation}
 where
\begin{equation}
 L_{n}(f_n) \equiv L_{n}(e(f_n),f_n) \, , \quad  \Psi_n(f_n) \equiv \Psi_n(e(f_n),f_n) \, ,
\end{equation}
 and $L_{n}(e,f_n)$ and $\Psi_n(e,f_n)$ are given by Equations (\ref{eq:Ln}) and (\ref{eq:PsiN}).
 We first differentiate the expressions in the bracket $[\,]$ in Equation (\ref{eq:Lnpartdiff}) 
 with respect to $\lambda_j$, then change the variable from $f_n$ back to $e$. Note that for 
 all $n$ and $e$, $\tilde{L}_n(e)$ is independent of $e_0$, and so $\tilde{L}_{n,j}(e)=0$ for
 $\lambda_j=e_0$. In Equation (\ref{eq:PPNhhPPN}), $\delta_{a,b}$ in the second, third, and fourth
 terms denote the Kronecker-$\delta$, defined to be unity if $a=b$ and zero otherwise. In Equation 
 (\ref{eq:PPNhhPPN}), the second and third terms arise due to the Heaviside function in the waveform 
 in Equation (\ref{eq:hPPN}), which represents the start of the waveform with eccentricity $e_0$.
 The $e_0$-derivative of this function is $\delta(e_0-e)$, which denotes the Dirac-$\delta$ function.
 Note that we use a smoothed version of $\Theta_\mathrm{H}(e_0-e)$ over a scale $\delta e_0$, which 
 is given in Equation (\ref{eq:ThetaH}), whose derivative is approximately\footnote{We neglect the
 partial derivatives of $\delta e_0$ with respect to the physical parameters.}
 \begin{align}
 [\partial_{e_0} \Theta_{H}(e_0-e)]^2 &\approx \frac{1}{(\delta e_0)^2} {~\rm if~~} e_0-\delta e_0 
 \leq e < e_0
 \nonumber\\
 & \approx \frac{\delta(e_0-e)}{\delta e_0} \, .
 \end{align}
 To avoid confusion note that the numerator denotes the Dirac-$\delta$ function, which has a 
 unit integral over $e\approx e_0$, and $\delta e_0$ in the denominator is the quantity given 
 by Equation (\ref{eq:deltae0}).

 Furthermore, we note that 
\begin{equation}
  \Re \left( \tilde L_{n}(e_0)\tilde L^*_{n,k}(e_0) \right)  = \frac12 \partial_k \left\arrowvert
  L_n(e_0) \right\arrowvert ^2
\end{equation}  
 in Equation (\ref{eq:PPNhhPPN}). In these equations $f_n$ enters when substituting $f_n/n$ for 
 $\nu$. By substituting Equation (\ref{eq:PPNhhPPN}) into Equation (\ref{eq:Fisher}), changing
 the integration variable from $f_n$ to $e$ respectively for each harmonic using\footnote{Note that
 $\nu(e)$ depends on $e_{\rm LSO}$ as seen in Equations (\ref{eq:nue}) and (\ref{eq:CC}).} $f_n=n
 \nu(e)$ and Equations (\ref{eq:nue}) and (\ref{eq:CC}), and truncating the sum over the harmonics 
 to the relevant range, the Fisher matrix becomes\footnote{We label this general expression with ``Gen'' to distinguish from an approximation ``High'' used below for high eccentricities.}
\begin{align} \label{eq:FisherPPN_Gen}
 \Gamma_{jk}^{\mathrm{NoPrec,Gen}} & \approx 4 \sum_{n=1}^{n_\mathrm{max}(e_0)} 
 \int_{e_{\mathrm{min},n}}^{e_{\mathrm{max},n}} \frac{ \Re \left( \tilde{L}_{n,j}(e) 
 \tilde{L}_{n,k}^*(e) \right) }{S_n( \nu_n(e))} \left| n \frac{d\nu}{de} \right| \, de 
 \nonumber  \\
 & + 2 \delta_{k,e_0} \sum_{n}  \left| n\frac{d\nu}{de} \right|_{e_0} 
 \frac{ \partial_j | L_n(e_0)|^2 }{S_n[\nu_n(e_0)]}
 \nonumber  \\ 
 & + 2 \delta_{j,e_0} \sum_{n} \left| n\frac{d\nu}{de} \right|_{e_0} 
 \frac{ \partial_k | L_n(e_0)|^2 }{S_n[\nu_n(e_0)]} 
 \nonumber  \\ 
 & + 4 \frac{\delta_{j,e_0} \delta_{k,e_0} }{\delta e_0} \sum_{n} 
  \left| n \frac{d\nu}{de} \right|_{e_0} \frac{ |L_n(e_0)|^2 }{S_n( \nu_n(e_0))} \, .
\end{align} 
 The limits of integration in Equation (\ref{eq:FisherPPN_Gen}) are defined by Equations (\ref{eq:emin})
 and (\ref{eq:emax}). Here the four terms correspond respectively to the four terms in Equation 
 (\ref{eq:PPNhhPPN}). The first term is directly analogous to that appearing in the SNR (see Equation 
 \ref{eq:SNR2PPNe}). Note that in particular, the elements corresponding to $j=e_{\rm LSO}$ and 
 $k=e_{\rm LSO}$ terms are nonzero. The $e_{\rm LSO}$-dependence enters in $\nu(e)$ as shown in 
 Equations (\ref{eq:nue}) and (\ref{eq:CC}). However, the first term in Equation (\ref{eq:FisherPPN_Gen}) 
 is zero for the $j=e_0$ and $k=e_0$ elements. If the binary forms in the detector's sensitive frequency
 band the second, the third, and the fourth terms in Equation (\ref{eq:PPNhhPPN}) contribute to this 
 element of the Fisher matrix. The eccentricity integral in the Fisher matrix may be carried analytically 
 over the $\delta$ function, which yields the second and third terms in Equation (\ref{eq:FisherPPN_Gen}). 
 There $\delta_{j, e_0}$ is the Kronecker-$\delta$, which is zero unless $j$ corresponds to the parameter
 $e_0$, and similarly for $\delta_{k,e_0}$. Note further that only harmonics with \mbox{$f_\mathrm{\max,
 det}/\nu(e_0) \geq n \geq f_\mathrm{min,det}/\nu(e_0)$} contribute to these boundary terms, since 
 otherwise $S_n(\nu_n(e_0)) = \infty$. 
 
 Similarly to the $\mathrm{SNR}^{\mathrm{Gen}}_\mathrm{NoPrec}$, $\Gamma_{jk}^{\mathrm{NoPrec}}$ is 
 generally valid for any initial eccentricity in the range $0 < e_0 < 1$, but it is computationally 
 efficient only for low to moderate initial eccentricities ($e_0 \la 0.8$). In order to make the calculation 
 computationally efficient for high initial eccentricities, we reverse the order of the sum and the 
 integral in the first term in Equation (\ref{eq:FisherPPN_Gen}), and truncate the sum over harmonics
 at $n_\mathrm{max}(e)$ as in Appendix \ref{subsubsec:ExpSNRNoPrec}. We get 
 \begin{align} \label{eq:FisherPPN}
 \Gamma_{jk}^{\mathrm{NoPrec,High}} & \approx 4 \int_{e_{\mathrm{min},n}}^{e_{\mathrm{max},n}} 
 \sum_{n=1}^{n_\mathrm{max}(e)} \frac{ \Re \left( \tilde{L}_{n,j}(e) \tilde{L}_{n,k}^*(e) 
 \right) }{S_n( \nu_n(e))} \left| n \frac{d\nu}{de} \right| \, de 
 \nonumber  \\
 & + 2 \delta_{k,e_0} \sum_{n}  \left| n\frac{d\nu}{de} \right|_{e_0} 
 \frac{ \partial_j | L_n(e_0)|^2 }{S_n[\nu_n(e_0)]}
 \nonumber  \\ 
 & + 2 \delta_{j,e_0} \sum_{n} \left| n\frac{d\nu}{de} \right|_{e_0} 
 \frac{ \partial_k | L_n(e_0)|^2 }{S_n[\nu_n(e_0)]} 
 \nonumber  \\ 
 & + 4 \frac{\delta_{j,e_0} \delta_{k,e_0} }{\delta e_0} \sum_{n} 
  \left| n \frac{d\nu}{de} \right|_{e_0} \frac{ |L_n(e_0)|^2 }{S_n( \nu_n(e_0))} \, .
\end{align}

\subsubsection{Eccentric inspirals with precession}
\label{subsubsec:ExpFisherWithPrec}

 Following the steps of Appendix \ref{subsubsec:ExpFisherNoPrec} for the precession-free model, 
 we may generalize the calculation of the Fisher matrix to include precession similar to Appendix 
 \ref{subsubsec:ExpSNRPrec}. The Fisher matrix, which is computationally efficient for low to moderate
 initial eccentricities \mbox{($e_0 \la 0.8$)}, can be given as 
\begin{equation}\label{eq:GammaWP}
\Gamma _{jk}^{\mathrm{Prec,Gen}} = \Gamma_{jk}^{\mathrm{Gen},n} + \Gamma _{jk}^{\mathrm{Gen},n+} 
 + \Gamma _{jk}^{\mathrm{Gen},n-} \, ,
\end{equation}
 where 
\begin{align}  \label{eq:FisherPrecn_Gen}
 \nonumber 
 \Gamma _{jk}^{\mathrm{Gen},n} & \approx 4 \sum_{n=1}^{n_\mathrm{max}(e_0) }
 \int_{e_{\mathrm{min},n}}^{e_{\mathrm{max},n}}  \frac{ \Re \left( \tilde{K}_{n,j}(e) 
 \tilde{K}_{n,k}^*(e) \right) }{ S_{n}(\nu_n(e)) } \left|n \frac{d\nu}{de} \right| \, de 
 \nonumber  \\  
  & + 2 \delta_{j,e_0} \sum_{n} 
  \left| n \frac{d\nu}{de} \right|_{e_0}
  \frac{ \partial_k \left| K_{n}(e_0)\right|^2}{ S_{n}(\nu_n(e_0)) }
  \nonumber  \\ 
 & + 2 \delta_{k,e_0} \sum_{n} 
 \left| n \frac{d\nu}{de} \right|_{e_0}
  \frac{ \partial_j \left| K_{n}(e_0)\right|^2}{ S_{n}(\nu_n(e_0)) }
  \nonumber  \\
 & + 4 \frac{\delta_{j,e_0} \delta_{k,e_0} }{\delta e_0} \sum_{n} 
  \left| n \frac{d\nu}{de} \right|_{e_0} \frac{ |K_n(e_0)|^2 }{S_n( \nu_n(e_0))} \, ,
\end{align}
 and
\begin{align}  \label{eq:FisherPrecpm_Gen}
\Gamma _{jk}^{\mathrm{Gen},n\pm} & \approx  4 \, \sum_{n=1}^{n_\mathrm{max}(e_0) }
  \int_{e_{\mathrm{min},n}^{\pm}}^{e_{\mathrm{max},n}^{\pm}} de \,  
   \frac{ \Re \left( \tilde{K}_{n,j}^{\pm}(e) \tilde{K}_{n,k}^{\pm,*}(e) \right) }{ S_{n}(\nu_n^{\pm}(e)) }
  \nonumber \\ 
 & \times \left| n \pm \frac{1}{\pi}\frac{d\dot{\gamma}}{d\nu} \right| \left| \frac{d\nu}{de} \right| 
  \nonumber \\
 & + 2 \delta_{j,e_0}  \sum_{n} 
 \left| n \pm \frac{1}{\pi}\frac{d\dot{\gamma}}{d\nu}  \right|_{e_0} \left| \frac{d\nu}{de} \right|_{e_0}
 \frac{\partial_k \left| K_{n}^{\pm}(e_0)\right|^2}{S_{n}(\nu_n^{\pm}(e_0))}
    \nonumber \\ 
 & + 2 \delta_{k,e_0} \sum_{n} 
 \left| n \pm  \frac{1}{\pi}\frac{d\dot{\gamma}}{d\nu} \right|_{e_0} \left| \frac{d\nu}{de} \right|_{e_0}
 \frac{\partial_j\left| K_{n}^{\pm}(e_0)\right|^2}{S_{n}(\nu_n^{\pm}(e_0))}
 \nonumber \\
 & + 4 \frac{\delta_{j,e_0} \delta_{k,e_0} }{\delta e_0} \sum_{n} 
   \left| n \pm  \frac{1}{\pi}\frac{d\dot{\gamma}}{d\nu} \right|_{e_0}
  \left| \frac{d\nu}{de} \right|_{e_0}  \frac{ |K_{n}^{\pm}(e_0)|^2 }{S_n( \nu_n^{\pm}(e_0))} \, ,
\end{align}
 where the integration bounds are given by Equations (\ref{eq:efirst})-(\ref{eq:last}), and 
\begin{align}
  \label{eq:Kn} \tilde{K}_{n,j}(e) & = \partial_j \left[ K_n (f_n) \, e^{i \Psi_n(f_n)} 
  \right] (e) \, , \\ 
  \label{eq:Knpm}  \tilde{K}_{n,j}^{\pm}(e) & = \partial_j \left[ K_n^{\pm} (f_n^{\pm}) 
  \, e^{i \Psi_n^{\pm}(f_n^{\pm})} \right] (e) \, .
\end{align} 
 Similar to the precession-free case, here 
\begin{align}
  K_n(f_n) & \equiv K_n (e(f_n), f_n), \quad  \Psi_n(f_n) \equiv \Psi_n(e(f_n), f_n) \, , \\
  K_n^{\pm}(f_n) & \equiv K_n^{\pm} (e(f_n^{\pm}), f_n^{\pm}), \quad \Psi_n^{\pm}(f_n^{\pm})
  \equiv  \Psi_n^{\pm}(e(f_n^{\pm}), f_n^{\pm}) \, ,
\end{align}  
 where $K_n (e, f_n)$ and $K_n^{\pm} (e, f_n^{\pm})$ are given by Equations
 \mbox{(\ref{eq:Cn_Sprec})-(\ref{eq:Cnm_Sprec})}, $\Psi_n(e, f_n)$ and $\Psi_n^{\pm}(e, f_n^{\pm})$
 are expressed by Equations (\ref{eq:PsiN}) and (\ref{eq:Psipm}), and we first differentiate the 
 expressions in the bracket $[\,]$ in Equations (\ref{eq:Kn}) and (\ref{eq:Knpm}) with respect 
 to $\lambda_j$, then change variables from $f_n$ and $f_n^{\pm}$ back to $e$. Similar to the
 precession-free case, only harmonics with \mbox{$f_\mathrm{max,det}/\nu(e_0) \geq n \geq f_\mathrm{min,det}
 /\nu(e_0)$} and \mbox{$f_\mathrm{max,det}/\nu^{\pm}(e_0) \geq n \geq f_\mathrm{min,det}/\nu^{\pm}(e_0)$} 
 contribute to the boundary terms in Equations (\ref{eq:FisherPrecn_Gen}) and (\ref{eq:FisherPrecpm_Gen}), 
 since otherwise $S_n(\nu_n(e_0)) = \infty$ and $S_n(\nu_n^{\pm}(e_0)) = \infty$. 
 
 For high initial eccentricities, the computationally efficient form of $\Gamma _{jk}^
 {\mathrm{Prec,Gen}}$, $\Gamma _{jk}^{\mathrm{Prec,High}}$, can be derived by reversing the order 
 of the sum and the integral in the first term in Equations (\ref{eq:FisherPrecn_Gen}) and 
 (\ref{eq:FisherPrecpm_Gen}), and truncating the sum over harmonics at $n_\mathrm{max}(e)$ as in
 Appendix \ref{subsubsec:ExpSNRNoPrec}. Thus, $\Gamma _{jk}^{\mathrm{Prec,High}}$ can be given as 
 \begin{equation}\label{eq:GammaWP}
 \Gamma _{jk}^{\mathrm{Prec,High}} = \Gamma_{jk}^{\mathrm{High},n} + \Gamma _{jk}^{\mathrm{High},n+}
 + \Gamma _{jk}^{\mathrm{High},n-} \, ,
\end{equation}
 where 
\begin{align}  \label{eq:FisherPrecn}
 \nonumber 
 \Gamma _{jk}^{\mathrm{High},n} & \approx 4 \int_{e_{\mathrm{min},n}}^{e_{\mathrm{max},n}} 
 \sum_{n=1}^{n_\mathrm{max}(e) } \frac{ \Re \left( \tilde{K}_{n,j}(e) 
 \tilde{K}_{n,k}^*(e) \right) }{ S_{n}(\nu_n(e)) } \left|n \frac{d\nu}{de} \right| \, de 
 \nonumber  \\  
  & + 2 \delta_{j,e_0} \sum_{n} 
  \left| n \frac{d\nu}{de} \right|_{e_0}
  \frac{ \partial_k \left| K_{n}(e_0)\right|^2}{ S_{n}(\nu_n(e_0)) }
  \nonumber  \\ 
 & + 2 \delta_{k,e_0} \sum_{n} 
 \left| n \frac{d\nu}{de} \right|_{e_0}
  \frac{ \partial_j \left| K_{n}(e_0)\right|^2}{ S_{n}(\nu_n(e_0)) }
  \nonumber  \\
 & + 4 \frac{\delta_{j,e_0} \delta_{k,e_0} }{\delta e_0} \sum_{n} 
  \left| n \frac{d\nu}{de} \right|_{e_0} \frac{ |K_n(e_0)|^2 }{S_n( \nu_n(e_0))} \, ,
\end{align}
and
\begin{align}  \label{eq:FisherPrecpm}
\Gamma _{jk}^{\mathrm{High},n\pm} & \approx  4 \int_{e_{\mathrm{min},n}^{\pm}}^{e_{\mathrm{max},n}^{\pm}} 
  de \, \sum_{n=1}^{n_\mathrm{max}(e) } 
   \frac{ \Re \left( \tilde{K}_{n,j}^{\pm}(e) \tilde{K}_{n,k}^{\pm,*}(e) \right) }{ S_{n}(\nu_n^{\pm}(e)) }
  \nonumber \\ 
 & \times \left| n \pm \frac{1}{\pi}\frac{d\dot{\gamma}}{d\nu} \right| \left| \frac{d\nu}{de} \right| 
  \nonumber \\
 & + 2 \delta_{j,e_0}  \sum_{n} 
 \left| n \pm \frac{1}{\pi}\frac{d\dot{\gamma}}{d\nu}  \right|_{e_0} \left| \frac{d\nu}{de} \right|_{e_0}
 \frac{\partial_k \left| K_{n}^{\pm}(e_0)\right|^2}{S_{n}(\nu_n^{\pm}(e_0))}
    \nonumber \\ 
 & + 2 \delta_{k,e_0} \sum_{n} 
 \left| n \pm  \frac{1}{\pi}\frac{d\dot{\gamma}}{d\nu} \right|_{e_0} \left| \frac{d\nu}{de} \right|_{e_0}
 \frac{\partial_j\left| K_{n}^{\pm}(e_0)\right|^2}{S_{n}(\nu_n^{\pm}(e_0))}
 \nonumber \\
 & + 4 \frac{\delta_{j,e_0} \delta_{k,e_0} }{\delta e_0} \sum_{n} 
   \left| n \pm  \frac{1}{\pi}\frac{d\dot{\gamma}}{d\nu} \right|_{e_0}
  \left| \frac{d\nu}{de} \right|_{e_0}  \frac{ |K_{n}^{\pm}(e_0)|^2 }{S_n( \nu_n^{\pm}(e_0))} \, .
\end{align}

\section{Validation of Codes}
\label{sec:TestCodes}
 
 \subsection{Analytic circular limit without precession}

 First, we study the circular limit of the eccentric waveform, $\tilde{h}_{\times}(\mathbf{f})$ 
 and $\tilde{h}_+ (\mathbf{f})$, defined by Equations (\ref{eq:SPAp}) and (\ref{eq:SPAx}) for the
 \emph{NoPrec} model. For $e\rightarrow 0$, $B_n^+=\delta_{n,2}$ and $\Psi_n^{\pm}=\Psi_n \mp 2 
 \gamma_c$ implying that $\gamma_c$ is degenerate with $\Phi_c$, which we henceforth omit. After
 integrating the term $\nu'/\dot{\nu}(\nu')$ over $\nu'$ in Equation (\ref{eq:Phint}) according 
 to previous considerations, substituting $\nu$ with $f/2$, and expanding the expression of $h_0
 \Lambda_2$, the polarization components become
\begin{align} 
 \label{eq:circhxee}  h_{\times} (f) & = - 2 i \sqrt{\frac{5}{96}} 
 \frac{\mathcal{M}_z^{5/6} f^{-7/6} e^{i \Psi} }{\pi^{2/3} D_{\rm L}} \cos \Theta
 \, , \\
 \label{eq:circhpee}  h_+ (f) & = -  \sqrt{ \frac{5}{96} } \frac{\mathcal{M}_z^{5/6}
 f^{-7/6} e^{i\Psi} }{\pi^{2/3} D_{\rm L}} (1 + \cos ^2 \Theta) \, ,
\end{align}
 where the phase function $\Psi$ can be given as 
\begin{equation}
 \Psi = 2 \pi f t_c - \Phi_c - \pi/4 + \frac{3}{4} 
 \left(8 \pi \mathcal{M}_z f \right)^{-5/3} \, .
\end{equation} 
 These are indeed the well-known frequency domain polarization components of circular binaries 
 in leading order \citep{CutlerFlanagan1994}. The parameter set characterizing this waveform is
\begin{equation}\label{eq:lambdacirc}
  \bm{\lambda}_\mathrm{circ} = \lbrace \mathrm{ln} (D_\mathrm{L}), \mathrm{ln} (\mathcal{M}_z),
  \theta_N, \phi_N,\theta_L,\phi_L , t_c,\Phi_c\rbrace \, . 
\end{equation}
 
 For validation tests, we calculate the SNR of circular binaries, $\mathrm{SNR}_\mathrm{ circ}$,  
 for a single aLIGO detector. $\mathrm{SNR}_\mathrm{circ}$ is calculated by substituting Equations (\ref{eq:circhxee}) 
 and (\ref{eq:circhpee}) into Equation (\ref{eq:Fourierha}) and then using Equation (\ref{eq:SNR2_1det}). 
 The Fisher matrix of circular binaries for the parameter set $\bm{\lambda} _\mathrm{circ}$ for each detector 
 $\Gamma _{jk}^{\mathrm{circ}}$ is calculated by substituting Equations (\ref{eq:circhxee}) and (\ref{eq:circhpee}) 
 into Equation (\ref{eq:Fourierha}) and then using Equations (\ref{eq:Fisher}).

\subsection{Eccentric inspiral without pericenter precession} 
\label{subsec:ValidatePPN} 

 Next, we discuss validation tests performed for the codes using the \emph{NoPrec} waveform model. 
 In this case, the parameters are
\begin{equation}\label{eq:lambdaNoPrec}
 \bm{\lambda}_\mathrm{NoPrec} = \lbrace \mathrm{ln} (D_\mathrm{L}), \mathrm{ln} (\mathcal{M}_z),
 \theta_N, \phi_N, \theta_L, \phi_L , t_c, \Phi_c, c_0, e_0, \gamma_c \rbrace \, .   
\end{equation}
 Compared to $\bm{\lambda}_\mathrm{circ}$, $\bm{\lambda}_\mathrm{NoPrec}$ includes $c_0$ (set by 
 $M_{{\rm tot},z}$ and $e_\mathrm{LSO}$, see Equation (\ref{eq:CC})) and $e_0$.

\subsubsection{Signal-to-noise ratio}
\label{subsec:validTest_NoPrec_SNR}

 First, we generate a set of source parameters for comparison $(m_A, m_B, D_\mathrm{L}, 
 \theta_N, \phi_N, \theta_L, \phi_L)$, and compare the output of $\mathrm{SNR}_\mathrm{ circ}$ 
 with the output of $\mathrm{SNR}^{\mathrm{Gen}}_\mathrm{NoPrec}$ in the circular limit for a single
 aLIGO detector (Table \ref{tab:DetCord}). Here and in further validation test, the set
 of fiducial source parameters generated for comparison are $(m_A, m_B, D_\mathrm{L}, \theta_N, \phi_N, 
 \theta_L, \phi_L)$, where $\cos \theta_N$ and $\cos \theta_L$ are drawn from a uniform distribution 
 between $[-1, 1]$, the set of $\phi_N$ and $\phi_L$ are drawn from a uniform distribution between $[0, 
 2 \pi]$, $m_A$ and $m_B$ are drawn from a uniform distribution between $[5 \, \Msun, 100 \, \Msun]$,
 and $D_\mathrm{L}$ is drawn from a uniform distribution between $[100 \, \mathrm{Mpc}, 1000 \, 
 \mathrm{Mpc}]$. The generation of other source parameters are described in details in the corresponding
 paragraph. We assume the fiducial value $t_c = \Phi_c = \gamma_c = 0$ for each binary, and
 in practice we set $e_0 = 10^{-4}$ and $\rho_\mathrm{p0} = 1000$ when considering the circular 
 limit (Appendix \ref{sec:circlim}). We find that the relative discrepancy between $\mathrm{SNR}_
 \mathrm{ circ}$ and $\mathrm{SNR}^{\mathrm{Gen}}_\mathrm{NoPrec}$ is less than $10^{-3}$ in all 
 cases. 

 Next, we examine if the output of $\mathrm{SNR} ^{\mathrm{Gen}} _\mathrm{NoPrec}$ 
 agrees with Figure 11 in \citet{OLearyetal2009}, which shows the source sky position- and 
 orientation-averaged RMS SNR for a single aLIGO detector as a function of 
 $\rho_\mathrm{p0}$ and $M_\mathrm{tot}$. \citet{OLearyetal2009} used the Fourier domain 
 orbit-averaged leading order waveform \citep{PetersMathews1963}, which corresponds to our 
 \emph{NoPrec} model.  We set the sensitivity curve to that used in \citet{OLearyetal2009},
 set $e = 0.95$, and find the results for several $\rho_\mathrm{p0}$ and $M_\mathrm{tot}$ 
 in the range $[5,100]$ and of $[10 \, \Msun, 1000 \, \Msun]$, respectively. For each $\rho
 _\mathrm{p0}$ and $M_\mathrm{tot}$, we generate random Monte Carlo samples of source sky 
 location ($\theta_N, \phi_N$) and binary orientation ($\theta_L, \phi_L$) as introduced 
 above in this section. We find that the RMS of the $\left( \mathrm{SNR} ^{\mathrm{Gen}} 
 _\mathrm{NoPrec} \right) ^2$ distributions are in agreement with the results of Figure 11
 in \citet{OLearyetal2009}.
 
 Finally, we generate a set of source parameters for comparison, and compare the output
 of $\mathrm{SNR} ^{\mathrm{Gen}}_\mathrm{NoPrec}$ and $\mathrm{SNR} ^{\mathrm{High}} _\mathrm{
 NoPrec}$ for several high $e_0$ and $\rho_\mathrm{p0}$ and for a single aLIGO detector. In particular, 
 $e_0$ and $\rho_\mathrm{p0}$ are drawn from a uniform distribution between $]0.9, 1[$ and $[5, 1000]$ 
 in these calculations, respectively. The relative discrepancy between $\mathrm{SNR} ^{\mathrm{Gen}} 
 _\mathrm{NoPrec}$ and $\mathrm{SNR}^{\mathrm{High}} _\mathrm{NoPrec}$ is less than $10^{-3}$ 
 in all cases.

\subsubsection{Fisher matrix}
\label{subsec:validTest_NoPrec_Fisher}
 
 Since the parameter set of the leading order circular and eccentric binaries, $\bm{\lambda}
 _\mathrm{circ}$ and $\bm{\lambda}_\mathrm{NoPrec}$ differ, see Equations (\ref{eq:lambdacirc}) and 
 (\ref{eq:lambdaNoPrec}), we cannot simply compare the output of $\Gamma _{jk}^{\mathrm{circ}}$ with 
 the output of $\Gamma_{jk}^{\mathrm{NoPrec,Gen}}$ in the circular limit. Therefore, we first restrict
 $\bm{\lambda}_\mathrm{NoPrec}$ to the parameters set $\bm{\lambda}_\mathrm{circ}$. Next, we generate 
 a set of source parameters for comparison, and compare the output of $\Gamma _{jk}^{\mathrm{circ}}$ 
 with the output of $\Gamma_{jk}^{\mathrm{NoPrec,Gen}}$ in the circular limit for the Fisher matrix 
 elements corresponding to the circular parameters $\bm{\lambda}_\mathrm{circ}$ and for a single aLIGO 
 detector. The relative discrepancy between $\Gamma _{jk} ^{\mathrm{circ}}$ and $\Gamma_{jk} 
 ^{\mathrm{NoPrec,Gen}}$ is less than $10^{-2}$ in all cases. 
 
 Finally, we generate a set of source parameters for comparison, and compare the output 
 of $\Gamma_{jk} ^{\mathrm{NoPrec,Gen}}$ and $\Gamma_{jk}^{\mathrm{NoPrec,High}}$ for the Fisher
 matrix elements corresponding to the parameter set $\bm{\lambda}_\mathrm{NoPrec}$ in the high 
 eccentricity limit for several $\rho_\mathrm{p0}$ and for a single aLIGO detector. Similarly to 
 Appendix \ref{subsec:validTest_NoPrec_SNR}, $e_0$ and $\rho_\mathrm{p0}$ are drawn from a 
 uniform distribution between $]0.9, 1[$ and $[5, 1000]$ in these calculations, respectively. 
 The relative discrepancy between Fisher matrix elements is less than $10^{-3}$ in all cases.

\subsection{Eccentric pericenter-precessing binary waveforms}
\label{subsec:ValidatePPC}

\subsubsection{Signal-to-noise ratio} 
\label{subsec:validTest_WithPrec_SNR}

 We test the numerical accuracy of the precessing waveform using the following theorem for
 leading order post-Newtonian binary inspirals. The amount of energy radiated in GWs is equal
 to the loss of mechanical energy of the binary, which is the same for the \emph{NoPrec} and 
 for the \emph{Prec} models. This is due to the fact that $a$ determines the mechanical energy 
 of the binary in the Newtonian approximation, and the orbital elements $a$ and $e$ are not 
 affected by pericenter precession. This also implies that the amount of SNR must be equal 
 for the \emph{NoPrec} and our precessing models for white noise. 
 
 Thus, assuming white noise, we generate a set of source parameters, and compare the output
 of $\mathrm{SNR}^{\mathrm{Gen}}_\mathrm{NoPrec}$ with the output of $\mathrm{SNR}^{\mathrm{Gen}}
 _\mathrm{Prec}$ for several $e_0$ and $\rho_\mathrm{p0}$, where $e_0$ and $\rho_\mathrm{p0}$ are 
 drawn from a uniform distribution between $]0, 1[$ and $[5, 1000]$ in these calculations, respectively. 
 The relative discrepancy between $\mathrm{SNR}^ {\mathrm{Gen}}_\mathrm{NoPrec}$ and $\mathrm{SNR}^
 {\mathrm{Gen}} _\mathrm{Prec}$ is less than $10^{-3}$ in all cases. 
 
 Furthermore, we repeat the same analysis for $\mathrm{SNR}^{\mathrm{High}}_\mathrm{NoPrec}$ 
 and $\mathrm{SNR}^{\mathrm{High}} _\mathrm{Prec}$, where $e_0$ is drawn from a uniform distribution 
 between $]0.9, 1[$. We find that the discrepancy between SNR values is less than $10^{-3}$ in all 
 cases. 
 
 Finally, we generate a set of source parameters for comparison, and compare the output 
 of $\mathrm{SNR}^{\mathrm{Gen}} _\mathrm{Prec}$ with the output of $\mathrm{SNR}^{\mathrm{High}}
 _\mathrm{Prec}$ for several high $e_0$ and $\rho_\mathrm{p0}$ and assuming white noise. Similarly 
 to the \emph{NoPrec} model, $e_0$ and $\rho_\mathrm{p0}$ are drawn from a uniform distribution 
 between $]0.9, 1[$ and $[5, 1000]$ in these calculations, respectively. We find that the relative 
 discrepancy between $\mathrm{SNR}^{\mathrm{Gen}} _\mathrm{Prec}$ and $\mathrm{SNR} ^{\mathrm{High}}
 _\mathrm{Prec}$ is less than $10^{-3}$ in all cases.

\subsubsection{Fisher matrix}
\label{subsubsec:validTest_WithPrec_Fisher}

 $\Gamma _{jk}^{\mathrm{Prec,High}}$ can be validated by following the procedure for $\Gamma _{jk}
 ^{\mathrm{Prec,Gen}}$, but considering $\Gamma_{jk}^{\mathrm{NoPrec,High}}$ and high $e_0$ values 
 ($e_0 \geq 0.9$). We find that the relative discrepancy between Fisher matrix elements are
 less than $10^{-2}$ in all cases. 

 Finally, after generating a set of source parameters for comparison, we compare the output of 
 $\Gamma _{jk}^{\mathrm{Prec,High}}$ and $\Gamma _{jk}^{\mathrm{Prec,Gen}}$ for the Fisher matrix 
 elements corresponding to the parameter set $\bm{\lambda}_\mathrm{Prec}$ in the high eccentricity 
 limit for several $\rho_\mathrm{p0}$ in the range $[5, 1000]$ and for a single aLIGO detector. The 
 discrepancy between Fisher matrix elements is less than $10^{-3}$ in all cases.

\bibliographystyle{yahapj}
\bibliography{refs}

\end{document}